\documentclass[preprint]{aastex631}
\usepackage{mathtools}
\usepackage{graphicx}
\usepackage{upgreek}
\usepackage{gensymb}

\received{...}
\revised{...}
\accepted{...}

\shorttitle{B2CAT Sources}
\shortauthors{Paggi et al.}

\begin{document}
	
\title{The Multi-Wavelength Environment of Second Bologna Catalog Sources}

\correspondingauthor{Alessandro Paggi}
\email{alessandro.paggi@unito.it}

\author[0000-0002-5646-2410]{A. Paggi}
\affiliation{Dipartimento di Fisica, Universit\`{a} degli Studi di Torino, via Pietro Giuria 1, I-10125 Torino, Italy}
\affiliation{Istituto Nazionale di Fisica Nucleare, Sezione di Torino, via Pietro Giuria 1, I-10125 Torino, Italy}

\author[0000-0002-1704-9850]{F. Massaro}
\affiliation{Dipartimento di Fisica, Universit\`{a} degli Studi di Torino, via Pietro Giuria 1, I-10125 Torino, Italy}
\affiliation{Istituto Nazionale di Fisica Nucleare, Sezione di Torino, via Pietro Giuria 1, I-10125 Torino, Italy}
\affiliation{INAF - Osservatorio Astrofisico di Torino, via Osservatorio 20, I-10025 Pino Torinese, Italy}
\affiliation{Consorzio Interuniversitario per la Fisica Spaziale (CIFS), via Pietro Giuria 1, 10125 Torino, Italy}

\author[0000-0003-0032-9538]{H. Pen\~a-Herazo}
\affiliation{Dipartimento di Fisica, Universit\`{a} degli Studi di Torino, via Pietro Giuria 1, I-10125 Torino, Italy}
\affiliation{Instituto Nacional de Astrof\'{i}sica, \'{O}ptica y Electr\'{o}nica, Tonantzintla, Puebla 72840, Mexico}
\affiliation{East Asian Observatory, Hilo, HI 96720, USA}

\author[0000-0001-8382-3229]{V. Missaglia}
\affiliation{Institute of Astrophysics, Foundation for Research and Technology - Hellas, Voutes, 7110 Heraklion, Greece}
\affiliation{Istituto Nazionale di Fisica Nucleare, Sezione di Torino, via Pietro Giuria 1, I-10125 Torino, Italy}
\affiliation{INAF - Osservatorio Astrofisico di Torino, via Osservatorio 20, I-10025 Pino Torinese, Italy}

\author[0000-0003-4413-7722]{A. Jimenez-Gallardo}
\affiliation{Dipartimento di Fisica e Astronomia, Universit\`{a} di Bologna, Via Gobetti 93/2, 40122 Bologna, Italy}

\author[0000-0001-5742-5980]{F. Ricci}
\affiliation{Dipartimento di Matematica e Fisica, Universit\`{a} Roma Tre, via della Vasca Navale 84, I-00146 Roma, Italy}
\affiliation{INAF - Osservatorio Astronomico di Roma, Via Frascati 33, I-00040 Monte Porzio Catone, Italy}

\author[0000-0003-4117-8617]{S. Ettori}
\affiliation{INAF - Osservatorio di Astrofisica e Scienza dello Spazio di Bologna, via Gobetti 93/3, 40129 Bologna, Italy}
\affiliation{Istituto Nazionale di Fisica Nucleare, Sezione di Bologna, viale Berti Pichat 6/2, 40127 Bologna, Italy}

\author[0000-0003-4916-6362]{G. Giovannini}
\affiliation{Dipartimento di Fisica e Astronomia, Universit\`{a} di Bologna, Via Gobetti 93/2, 40122 Bologna, Italy}
\affiliation{INAF - Istituto di Radioastronomia (IRA), Via P. Gobetti 101, 40129 Bologna, Italy}

\author[0000-0002-2831-7603]{F. Govoni}
\affiliation{INAF - Osservatorio Astronomico di Cagliari, Via della Scienza 5, 09047 Selargius, Italy}

\author[0000-0002-1824-0411]{R. D. Baldi}
\affiliation{INAF - Istituto di Radioastronomia (IRA), Via P. Gobetti 101, 40129 Bologna, Italy}

\author[0000-0001-5649-938X]{B. Mingo}
\affiliation{School of Physical Sciences, The Open University, Walton Hall, Milton Keynes, MK7 6AA, UK}

\author[0000-0002-4800-0806]{M. Murgia}
\affiliation{INAF - Osservatorio Astronomico di Cagliari, Via della Scienza 5, 09047 Selargius, Italy}

\author[0000-0003-0995-5201]{E. Liuzzo}
\affiliation{INAF - Istituto di Radioastronomia (IRA), Via P. Gobetti 101, 40129 Bologna, Italy}
\affiliation{Italian Alma Regional Center (ARC), Via Piero Gobetti 101, 40129 Bologna, Italy}

\author{F. Galati}
\affiliation{Dipartimento di Fisica, Universit\`{a} degli Studi di Torino, via Pietro Giuria 1, I-10125 Torino, Italy}

\begin{abstract}
We present the first results of the \textit{Chandra} Cool Targets (CCT) survey of the Second Bologna Catalog (B2CAT) of powerful radio sources, aimed at investigating the extended X-ray emission surrounding these sources. For the first 33 sources observed in the B2CAT CCT survey, we performed both imaging and spectral X-ray analysis, producing multi-band \textit{Chandra} images, and compared them with radio observations. To evaluate the presence of extended emission in the X-rays, we extracted surface flux profiles comparing them with simulated ACIS Point Spread Functions. We detected X-ray nuclear emission for 28 sources. In addition, we detected 8 regions of increased X-ray flux originating from radio hot-spots or jet knots, and a region of decreased flux, possibly associated with an X-ray cavity. We performed X-ray spectral analysis for 15 nuclei and found intrinsic absorption significantly larger than the Galactic values in four of them. We detected significant extended X-ray emission in five sources, and fitted their spectra with thermal models with gas temperatures \(\sim 2 \text{ keV}\). In the case of B2.1 0742+31, the surrounding hot gas is compatible with the ICM of low luminosity clusters of galaxies, while the X-ray diffuse emission surrounding the highly disturbed WAT B2.3 2254+35 features a luminosity similar to those of relatively bright galaxy groups, although its temperature is similar to those of low luminosity galaxy clusters. These results highlight the power of the low-frequency radio selection, combined with short \textit{Chandra} snapshot observations, to investigate the properties of the X-ray emission from radio sources.
\end{abstract}


\section{Introduction}\label{sec:intro}

Diffuse X-ray emission associated with radio sources, extending well beyond their host galaxies up to hundred of kpc scales, is known and observed since the first X-ray \textit{Uhuru} \citep[e.g.,][]{1971ApJ...167L..81G} and \textit{Einstein} \citep[e.g.,][]{1979ApJ...234L..21J} missions, and more recently with \textit{XMM-Newton} \citep[e.g.,][]{2011A&A...526A.133G} and \textit{Chandra} \citep[e.g.,][]{2018A&A...619A..75M} telescopes \citep[see also][]{2003ApJ...596..105S, 2003MNRAS.341..729F, 2006MNRAS.371...29E, 2006ApJ...642...96E}.

In the last decades, \textit{Chandra} telescope has extensively studied the X-ray emission of high-redshift radio galaxies, often used as tracers of galaxy clusters with poor or moderately rich environments \citep[see e.g.,][]{2002NewAR..46..121W, 2003MNRAS.339.1163C, 2009A&ARv..17....1W, 2021ApJ...907...65G}, since extended X-ray emission from these sources can be due to the thermal radiation arising from the hot gas trapped by the gravitational attraction of giant galaxies or permeating the intergalactic medium \citep[see e.g.,][]{2001MNRAS.322L..11F, 2013ApJ...770..136I, 2015MNRAS.453.2682I}.

Alternatively, when the extended X-ray emission in these sources shows a general alignment with the radio axis and/or is spatially coincident with radio structures, a significant contribution to its flux is expected to come from non-thermal processes, and in particular from inverse Compton (IC) scattering of the radio-emitting electrons. In radio lobes, the X-ray emission is generally interpreted as due to IC of these electrons on Cosmic Microwave Background (CMB) photons permeating the radio lobes \citep[IC/CMB][]{1965Natur.208..111H, 1967NCimB..52..495B, 1972MNRAS.160..339O, 1973MNRAS.165..413O, 1979MNRAS.188...25H, 2000ApJ...540L..69S, 2000ApJ...544L..23T, 2019ApJ...883L...2M, 2023MNRAS.518.3222B}, while in radio hot spots the X-ray emission is believed to be dominated by synchrotron self Compton radiation \citep[SSC,][]{2005ApJ...622..797K}, that is, IC scattering of synchrotron photons by electrons in the radio jets that emitted the synchrotron photons in the first place \citep{2004ApJ...612..729H}. Finally, X-ray emission in radio galaxies on \(\sim 100-200 \text{ kpc}\) scale can also be significantly contributed by IC from far-IR photons in galactic starbursts \citep{2009ApJ...702L.114S, 2012ApJ...760..132S}.

To investigate the nature of the extended X-ray emission surrounding radio sources and study their evolution, in the last decade we carried out the \textit{Chandra} snapshot survey of the Third Cambridge catalog \citep[3CR,][]{1962MNRAS.125...75B, 1985PASP...97..932S} to obtain X-ray coverage of the entire 3CR catalog \citep{2010ApJ...714..589M, 2012ApJS..203...31M, 2015ApJS..220....5M}. Through this observational program, we found X-ray emission associated with radio jets \citep[see e.g.,][]{2009ApJ...696..980M}, hotspots \citep[see e.g.,][]{2011ApJS..197...24M, 2012MNRAS.419.2338O} as well as diffuse X-ray emission from hot atmospheres and intra-cluster medium (ICM) in galaxy clusters \citep[see e.g.,][]{2010MNRAS.401.2697H, 2012MNRAS.424.1774H, 2016MNRAS.458..681D, 2018ApJ...867...35R, 2021ApJS..255...18M}, extended X-ray emission aligned with the radio axis of several moderate and high redshift radio galaxies \citep[see e.g.,][]{2013ApJS..206....7M, 2018ApJS..234....7M, 2018ApJS..235...32S, 2020ApJS..250....7J, 2021A&A...647A..79P, 2021ApJS..252...31J}, and the presence of extended X-ray emission spatially associated with optical emission line regions not coincident with radio structures \citep{2009ApJ...692L.123M, 2012A&A...545A.143B, 2022ApJ...941..114J}.

In this work we present the first results from a \textit{Chandra} snapshot survey performed on the Second Bologna Catalog (B2CAT) of powerful radio sources. The B2CAT \citep{1970A&AS....1..281C, 1972A&AS....7....1C, 1973A&AS...11..291C, 1974A&AS...18..147F}, listing about 10000 sources detected above \(0.1 \text{ Jy}\) (completeness above \(0.2 \text{ Jy}\)) at \(408 \text{ MHz}\) with Bologna Northern Cross Telescope between \(21\degr\) and \(40\degr\) declination, is well suited to study the properties of extra-galactic radio sources. As a low frequency radio selected sample, its selection criteria are unbiased with respect to X-rays and active galactic nuclei (AGN) viewing angle. From this catalog have been derived well studied samples of radio-loud active galaxies \citep{1987A&AS...69...57F}, as well as radio loud quasars \citep{1986A&AS...64..557R} and spiral galaxies \citep{1980A&AS...41..329G}. The high sensitivity of the B2CAT with respect to other radio samples as e.g. the 3CR allowed to study the properties of low luminosity radio galaxies as Fanaroff-Riley I \citep[FRI,][]{1974MNRAS.167P..31F} radio galaxies, and the non-thermal properties of spiral galaxies and low luminosity quasars.

In addition, the B2CAT spans a wide range in redshift and radio power, and it is augmented by a vast suite of ground and space-based observations at all accessible wavelengths (optical, \citealt{2002A&A...383..104C, 2002A&A...396..857D}; and radio band between \(1.4\) and \(8 \text{ GHz}\), \citealt{1974A&A....32..155F, 1980A&AS...42..319H, 1981A&AS...46..473P, 1995MNRAS.274..939L}). This catalog represents an ideal sample to study the X-ray emission arising from jet knots, hotspots, and nuclei of radio sources, look for new galaxy clusters via the presence of extended X-ray emission unrelated to the radio structures \citep{2007MNRAS.381.1109B, 2013MNRAS.431..858M}, and investigate observational evidence of AGN feedback with the hot gas in galaxies, groups, and clusters of galaxies \citep{2012ARA&A..50..455F, 2012ApJ...749...19K, 2017MNRAS.470.2762M}. The B2CAT therefore represents a powerful tool to optimize the \textit{Chandra} Cool Targets (CCT\footnote{\href{https://cxc.harvard.edu/proposer/CCTs.html}{https://cxc.harvard.edu/proposer/CCTs.html}}) observing strategy, that is, observations acquired while the spacecraft performs pointings to avoid overheating (or excessive cooling) of various observatory sub-systems.

The paper is organized as follows. A brief description of the B2CAT CCT sources observed to date is presented in Sect. \ref{sec:b2cat}. \textit{Chandra} data reduction and analysis are presented in Sect. \ref{sec:analysis}. Results on individual sources imaging and spectral analysis are presented in Sect. \ref{sec:imaging}, Sect. \ref{sec:sources} and Sect. \ref{sec:spectral}, while Sect. \ref{sec:summary} is devoted to our conclusions. Unless otherwise stated we adopt cgs units for numerical results and we also assume a flat cosmology with \(H_0=69.6\text{ km}\text{ s}^{-1}\text{ Mpc}^{-1}\), \(\Omega_{M}=0.286\) and \(\Omega_{\Lambda}=0.714\) \citep{2014ApJ...794..135B}.

\section{Sample Description}\label{sec:b2cat}

In the selection of B2CAT CCT survey targets we started from the B2CAT catalog excluding sources already observed by \textit{Chandra}. Taking into account the ecliptic latitude cut (\(\ell > 55\degree\)), we then selected a sample of 3080 sources. We stress that, due to the serendipitous nature of the CCT program, large samples are required to perform such observations, and that the proposed sources will be observed randomly. Finally, for this survey we applied for snapshot (\(16 \text{ ks}\)) observations, following the same approach used for the \textit{Chandra} 3C survey \citep{2010ApJ...714..589M}.

The sample of radio sources discussed in this work is constituted by the first 33 B2CAT targets observed by \textit{Chandra} during the CCT survey up to June 2023. The main properties of these sources are presented in Table \ref{tab:sources}. Redshift measurements are available for only seven sources.

In addition to the newly obtained \textit{Chandra} data, we collected multi-wavelength radio data for the sources in the sample. In particular, to investigate the correlation between the diffuse X-ray emission and the extended radio structures, we collected \(74 \text{ MHz}\) Karl G. Jansky Very Large Array (VLA) data obtained through the VLA Low-frequency Sky Survey Redux \citep[VLSSr,][]{2014MNRAS.440..327L}\footnote{VLSSr images cover an area of \(\sim 30,530\) square degrees with a resolution of \(75 \arcsec\), and an rms of \(\sim 0.1 \text{ Jy}/\text{beam}\).}, \(145 \text{ MHz}\) Low Frequency Array \citep[LOFAR,][]{2013A&A...556A...2V} observations from the forthcoming Data Release 2 (DR2) of LoTSS\footnote{The DR2 v2.2 was run as part of the ddf pipeline (\href{https://github.com/mhardcastle/ddf-pipeline}{https://github.com/mhardcastle/ddf-pipeline}) and the LoTSS DR1 consists of images at \(6 \arcsec\) resolution and \(\sim 70\, \upmu \text{Jy}/\text{beam}\) sensitivity covering an area of \(\sim 400\) square degrees while the footprint of the DR2 covers an area of approximately \(5700\) square degrees, both performed in the northern hemisphere.} processed by the international LOFAR collaboration as part of the LOFAR Data Release 1 and 2 (\citealt{2017A&A...598A.104S, 2019A&A...622A...1S}, and \citealt{2021A&A...648A...1T, 2022A&A...659A...1S}, respectively), \(150 \text{ MHz}\) Giant Metrewave Radio Telescope (GMRT) data obtained from the TIFR GMRT Sky Survey \citep[TGSS,][]{2017A&A...598A..78I}\footnote{TGSS images cover an area of \(\sim 36,900\) square degrees, and have a resolution of \(25 \arcsec\) for Dec \(> 19\degree\) and of \(25 \arcsec/\cos{\left({\text{Dec}-19\degree}\right)}\) for Dec \(< 19\degree\), with a median rms of \(\sim 3.5 \text{ mJy}/\text{beam}\).}, \(1.4 \text{ GHz}\) VLA data obtained through the NRAO VLA Sky Survey \citep[NVSS,][]{1998AJ....115.1693C}\footnote{NVSS images cover the entire north sky above \(-40\degree\) Dec with a resolution of \(45\arcsec\), and an rms of \(\sim 450\, \upmu \text{Jy}/\text{beam}\).}, and \(3 \text{ GHz}\) VLA data obtained through the VLA Sky Survey \citep[VLASS,][]{2020PASP..132c5001L}\footnote{VLASS images cover an area of \(\sim 33,885\) square degrees with Dec \(>-40\degree\) with a resolution of \(2\farcs 5\), with an rms of \(\sim 120\, \upmu \text{Jy}/\text{beam}\) for the single epoch observations and of \(\sim 70\, \upmu \text{Jy}/\text{beam}\) for the three combined epochs.}.

In Appendix \ref{app:a} we report the complete set of radio images (see Fig. \ref{fig:radiomaps}) available in the aforementioned surveys for the B2CAT sources considered here, together with estimates of the radio flux for different radio structures (see Table \ref{tab:radiotable}).

\section{Data Analysis}\label{sec:analysis}

\textit{Chandra} observations of B2CAT sources were retrieved from \textit{Chandra} Data Archive through ChaSeR service\footnote{\href{http://cda.harvard.edu/chaser}{http://cda.harvard.edu/chaser}} (see Table \ref{tab:sources}). They consist of ACIS-S snapshot observations with nominal exposure time of \(16 \text{ ks}\), performed between April 2019 and June 2023 in VFAINT mode. These data have been analyzed with the \textit{Chandra} Interactive Analysis of Observations \citep[\textsc{CIAO},][]{2006SPIE.6270E..1VF} data analysis system version 4.14 and \textit{Chandra} calibration database CALDB version 4.9.8, adopting standard procedures. The observations were filtered for time intervals of high background flux exceeding \(3\sigma\) above the average level with \textsc{deflare} task, to attain the final exposures listed in Table \ref{tab:sources}. Field point sources in the \(0.3-7\text{ keV}\) energy band were detected with the \textsc{wavdetect} task, adopting a \(\sqrt{2}\) sequence of wavelet scales (i.e., 1, 1.41, 2, 2.83, 4, 5.66, 8, 11.31 and 16 pixels) and a false-positive probability threshold of \({10}^{-6}\). Given the relatively short exposure times and the consequent low statistics, we did not correct the absolute astrometry of the \textit{Chandra} ACIS-S images and did not register them to radio maps, as the typical shift for \textit{Chandra} images found during the 3C \textit{Chandra} Snapshot Survey is \(0\farcs 5\) \citep[see e.g.,][]{2011ApJS..197...24M, 2020ApJS..250....7J}.

We produced broad (\(0.3-7\text{ keV}\)), soft (\(0.3-3\text{ keV}\)), and hard (\(3-7\text{ keV}\)) band \textit{Chandra} images centered on ACIS-S chip 7. We also produced Point Spread Function (PSF) maps (with the \textsc{mkpsfmap} task), effective area corrected exposure maps, and flux maps using the \textsc{flux\_obs} task (see Fig. \ref{fig:maps}). The image pixel sizes and the \(\sigma\) widths of the Gaussian kernel used for smoothing are listed in Table \ref{tab:sources}.

\subsection{Imaging Analysis}\label{sec:imaging}

We first proceeded searching for nuclear detections in the broad \(0.3-7\text{ keV}\) band images. Using the higher resolution VLASS data, we defined \(2''\) circular regions coincident with the core emission as identified in the VLASS images. If source cores were not clearly detected in VLASS images, we tentatively identified as source cores those compact X-ray regions lying at the center of the radio structures (see Sect. \ref{sec:sources}). We evaluated the nuclear X-ray fluxes making use of the \textsc{srcflux} \textsc{CIAO} task that evaluates the PSF corrections and the detector effective area and response function at the source location, assuming a power-law spectrum with \(1.8\) slope - as usually observed in AGN nuclear emission - and taking into account the photo-electric absorption by the Galactic column density along the line of sight  \citep{2016A&A...594A.116H}. The nuclear fluxes are listed in Table \ref{tab:sources}.

With this procedure, we confirmed nuclear detections for 28 out of 33 sources, with 19 being detected at least at \(3 \sigma\) significance. For seven other sources (B2.2 0143+24, B2.1 0241+30, B2.1 0455+32B, B2.1 0455+32C, B2.2 0775+24, B2.4 1112+23, and B2.4 2054+22B) we obtained a \(2 \sigma\) significance detection of nuclear emission, for two sources (B2.3 0516+40 and B2.2 0038+25B) we obtained a marginal \(1 \sigma\) significance detection of the nuclear emission, and for two sources (B2.2 1338+27 and B2.3 2334+39) we were only able to put a \(1\sigma\) upper limit on the nuclear flux. For the remaining three sources (B2.1 0302+31, B2.2 1439+25 and B2.2 2133+27), since we do not have any clear indication - neither in radio nor in X-ray data - of the location of the core, we do not report any nuclear flux estimate. We note that the brightest nucleus in our sample, B2.1 0742+31, is significantly affected by pileup, as shown in the map obtained with the \textsc{CIAO} task \textsc{pileup\_map}, therefore the value of \(17.7\pm 0.4 \times {10}^{-13} \text{ erg}\text{ cm}^{-2}\text{ s}^{-1}\) reported in Table \ref{tab:sources} should be considered as a lower limit to the real flux (see Sect. \ref{sec:spectral}).

Correlation between AGN nuclear radio and X-ray emission from ROSAT All Sky Survey \citep{1999A&A...349..389V} has been observed and discussed in several works \citep[e.g.,][]{1994ApJ...427..134W, 2012A&A...543A..57Z}. To investigate this correlation in our sample, in Fig. \ref{fig:xray-radio-corr} we plot the X-ray nuclear fluxes evaluated above versus the \(3 \text{ GHz}\) radio nuclear specific fluxes, evaluated from VLASS maps that show a discernible nuclear emission (regions N in Table \ref{tab:radiotable}). Despite the paucity of the sample, there appears to be a correlation between the nuclear emission at radio and X-ray frequencies, as evaluated through hierarchical Bayesian linear regression \citep{2007ApJ...665.1489K}. In particular, a linear regression of the logarithmic values of both quantities, including the highly piled-up B2.1 0742+31, yields a slope of \({0.95}_{-0.35}^{+0.36}\) with a correlation coefficient of \(0.75\) (red line in Fig. \ref{fig:xray-radio-corr}), while excluding it yields a slope of \({1.31}_{-0.81}^{+0.83}\) with a correlation coefficient of \(0.65\) (blue line in Fig. \ref{fig:xray-radio-corr}), both consistent with previous results on radio loud AGNs \citep[e.g.,][]{1997A&A...319..413B, 2000A&A...356..445B}. We note that the X-ray fluxes expected from the correlation that excludes B2.1 0742+31 lie at larger values than that evaluated for this source, reinforcing the point that the X-ray flux evaluated for B2.1 0742+31 should be regarded as a lower limit.

As shown in Fig. \ref{fig:maps}, many sources in the present sample show hints of diffuse soft X-ray \(0.3-3 \text{ keV}\) band emission associated with the extended radio structures mapped by the GMRT and LOFAR images. In order to get a preliminary characterization of this diffuse emission we evaluated its flux making use of the \textsc{srcflux} \textsc{CIAO} task, assuming a thermal spectrum with \(2 \text{ keV}\) temperature and abundance \(0.25\) solar - as expected from typical ICM emission - and taking again into account the Galactic photo-electric absorption. The fluxes of the extended emissions are listed in Table \ref{tab:sources}.

Also the correlation between radio and X-ray diffuse emission in clusters has been discussed in several studies \citep[e.g.,][]{2017MNRAS.464.2752P, 2019SSRv..215...16V}. To investigate this in our sample, in Fig. \ref{fig:xray-radio-corr-ext} we plot these X-ray fluxes versus the \(145 \text{ MHz}\) specific fluxes evaluated from LOFAR maps from regions of extended radio emission (regions A and B listed in Table \ref{tab:radiotable}). In this case, however, the correlation between the extended emission at radio and X-ray frequencies appears very low, with a slope \({0.20}_{-0.38}^{+0.36}\) and a correlation coefficient of \(0.26\).

To evaluate the significance of this extended emission in the X-rays, we extracted net surface flux profiles in the \(0.3-3 \text{ keV}\) band from concentric annuli centered on the radio core positions, excluding counts from detected sources, and evaluating the background level from source-free regions of chip 7. The width of the bins was adaptively determined to reach a minimum signal to noise ratio of \(3\). In the outer regions, when this ratio could not be reached, we extended the bin width to the edges of the ACIS Chip.

We then compared these profiles with those extracted from simulated ACIS PSFs, using the same procedure fully described in \citet{2020ApJ...902...49F} and that we briefly summarize here. The \textit{Chandra} PSFs were simulated using rays produced by the \textit{Chandra} Ray Tracer (ChaRT\footnote{\href{http://cxc.harvard.edu/ciao/PSFs/chart2/}{http://cxc.harvard.edu/ciao/PSFs/chart2/}}) projected on the image plane by MARX\footnote{\href{http://space.mit.edu/CXC/MARX/}{http://space.mit.edu/CXC/MARX/}}. For each observation, we generated the average from 1000 PSF simulations centered at the coordinates of radio or X-ray cores. We then produced images of these PSFs in the \(0.3-3 \text{ keV}\) band, and extracted profiles from the same annuli used for the source images. Finally, the PSF surface flux profiles were normalized to match the level obtained in the innermost annulus for the source images. 

We found that the \(0.3-3 \text{ keV}\) soft-band emission is extended at least at \(5 \sigma\) significance beyond \(10\arcsec\) for \(5\) sources in the present sample (B2.4 0004+21, B2.1 0302+31, B2.4 0412+23, B2.1 0742+31 and B2.3 2254+35). Fig. \ref{fig:profiles} shows the comparison of the net surface flux profiles for these sources (black dots) and their corresponding PSFs (red dots), and we see that the former are clearly extended above the latter, especially for B2.1 0302+31 and B2.3 2254+35.

\subsection{Individual sources}\label{sec:sources}

In this section, we report X-ray compact features associated with radio structures the sources in our sample, while properties of the extended X-ray emission will be discussed in Sect. \ref{sec:spectral}. The broad-band \(0.3-7 \text{ keV}\) flux maps of the central region of each source are presented in Fig. \ref{fig:zooms}, with overlaid in black the \(3 \text{ GHz}\) VLASS contours from Fig. \ref{fig:maps}.

\subsubsection*{B2.4 0004+21}

Also known as NVSS J000727+220413, this source shows a typical Fanaroff-Riley II \citep[FRII,][]{1974MNRAS.167P..31F} structure, with indication of extended X-ray emission (see Fig. \ref{fig:profiles}). Apart from the bright X-ray nucleus (region 1 in Fig. \ref{fig:zooms}, with a \(0.3-7 \text{ keV}\) flux of \({20}_{-2}^{+2}\times {10}^{-14}\text{ erg}\text{ cm}^{-2}\text{ s}^{-1}\), see Sect. \ref{sec:imaging}), the regions of increased X-ray flux in correspondence with the radio structures (regions 2 and 3 in Fig. \ref{fig:zooms}) have low (\(<2 \sigma\)) significance with respect to the level of the diffuse emission at the same radial distance from the nucleus.

\subsubsection*{B2.2 0038+25B}

This source, also known as PKS 0038+255, shows a FRII radio morphology as mapped by VLASS image. We do not have a clear detection of the radio core, although between the lobes we see a faint region of X-ray emission (region 1 in Fig. \ref{fig:zooms}) with a \(0.3-7 \text{ keV}\) flux of \({5}_{-2}^{+3}\times {10}^{-15}\text{ erg}\text{ cm}^{-2}\text{ s}^{-1}\) (see Sect. \ref{sec:imaging}), thaw we identify as the X-ray nucleus. There are some hints for this region to extend along the radio axis toward the lobes, but the low statistic prevent us from drawing further conclusions.

\subsubsection*{B2.2 0143+24}

This source, also known as NVSS J014628+250616, shows a complex, extended radio morphology in LOFAR images, while VLASS data indicate a typical FRII structure (see Fig. \ref{fig:maps}). In Fig. \ref{fig:zooms}, region 1 indicates the location of the possible source nucleus. This region yields an upper limit on the flux of \({6}_{-3}^{+4} \times {10}^{-15}\text{ erg}\text{ cm}^{-2}\text{ s}^{-1}\) (see Sect. \ref{sec:imaging}). On both sides of region 1, along the radio axis, there are two regions of increased flux in front of the hot-spots (regions 2 and 3). The brightest region 2 has 12 broad-band counts, and by sampling the emission at the same radial distance from the putative nucleus we conclude that this is marginally significant at \(3\sigma\) level.

\subsubsection*{B2.4 0145+22}

This source, known as NVSS J014750+223852, shows in the VLASS image a FRII structure, without indications of a clearly detected radio core. In Fig. \ref{fig:zooms}, we indicate with region 1 the position of a bright X-ray source between the two radio jets that we identify as the nucleus for which we estimated a broad-band  flux of \({8}_{-1}^{+1} \times {10}^{-14}\text{ erg}\text{ cm}^{-2}\text{ s}^{-1}\) (see Sect. \ref{sec:imaging}). On the east side of the nucleus there is a region of increased X-ray flux co-spatial with the east radio jet (region 2), possibly connected with a jet knot. This region contains 12 broad-band counts, significant at \(4\sigma\) level above the emission at the same radial distance from the nucleus.

\subsubsection*{B2.4 0229+23}

At a \(z=3.420\) \citep{2002MNRAS.329..700S}, this source, known as NVSS J023220+231756, shows in the VLASS image a compact structure, coincident with a bright X-ray source that in Fig. \ref{fig:zooms} we indicate as region 1. We identify this region as the source nucleus, for which we estimated the broad-band  flux of \({58}_{-3}^{+3} \times {10}^{-14}\text{ erg}\text{ cm}^{-2}\text{ s}^{-1}\) (see Sect. \ref{sec:imaging}). No other significant structures are visible in the broad-band  \textit{Chandra} ACIS-S flux map.

\subsubsection*{B2.2 0241+30}

This source, known as NVSS J024443+302117, has a FRII structure, with edge-brightened radio lobes. In Fig. \ref{fig:zooms} region 1 marks the location of the faint X-ray nucleus, coincident with the radio core, for which we estimated the broad-band  flux of \({5}_{-2}^{+3} \times {10}^{-15}\text{ erg}\text{ cm}^{-2}\text{ s}^{-1}\) (see Sect. \ref{sec:imaging}). On the south-west side of the nucleus there is a region of increased X-ray flux coincident with the south-west radio lobe (region 2). This region contains 7 broad-band counts, being marginally significant at \(3\sigma\) level above the emission at the same radial distance from the nucleus.

\subsubsection*{B2.2 0302+31}

This source, also known as NVSS J030524+312928, shows evidence of significant extended X-ray emission (see Fig. \ref{fig:profiles}), with the western radio lobe showing a FRII edge-brightened structure, while the eastern lobe appears edge-darkened as for FRI radio sources. The global radio structure, roughly connected with the X-ray diffuse emission, can be therefore classified as a Hybrid Morphology Radio Source \citep[HyMoR,][]{2000A&A...363..507G}. In addition, the eastern radio jet appears bent in the southeast direction, more evidently in the LOFAR data (see Fig. \ref{fig:maps}), as observed in wide-angle tailed radio galaxies \citep[WATs,][]{1976ApJ...205L...1O, 1993LNP...421....1L} that usually coincide with the brightest galaxy at the center of a cluster \citep{2019A&A...626A...8M} This source shows no identifiable nuclear emission, either in the radio or in the X-ray bands. In correspondence with the western radio lobe, there is a region of increased flux (region 1 in Fig. \ref{fig:zooms}) that, with 29 broad-band counts, significantly rises above the level of the surrounding emission at \(5\sigma\) level.

\subsubsection*{B2.4 0401+23}

This source, known as NVSS J040452+240656, shows an edge-brightened FRII structure. Region 1 in Fig. \ref{fig:zooms} marks the X-ray nucleus with a flux \({4}_{-1}^{+1}\times {10}^{-14}\text{ erg}\text{ cm}^{-2}\text{ s}^{-1}\) (see Sect. \ref{sec:imaging}). Along the radio axis there are two regions of increased X-ray flux (regions 2 and 3 in Fig. \ref{fig:zooms}), however only region 3 with its 6 broad-band counts is marginally significant at \(3\sigma\) level above the emission at the same radial distance from the nucleus.

\subsubsection*{B2.2 0410+26}

This source, also known as NVSS J041323+264916, shows a rather compact radio structure in both the LOFAR and the VLASS radio maps, where only the core is detected. Region 1 marks the faint X-ray nucleus with a \({11}_{-3}^{+4}\times {10}^{-15}\text{ erg}\text{ cm}^{-2}\text{ s}^{-1}\) flux (see Sect. \ref{sec:imaging}) in correspondence with the radio core. No other significant structures are visible in the broad-band  \textit{Chandra} ACIS-S flux map.

\subsubsection*{B2.4 0412+23}

This source, also known as NVSS J041512+234751, shows a radio structure elongated in the north-south direction, as shown in the LOFAR and GMRT data. The VLASS image shows the radio core and the two compact radio lobes. Apart from the bright X-ray nucleus coincident with the radio core (labeled region 1 in Fig. \ref{fig:zooms}) with a broad-band flux of \({27}_{-2}^{+2} \times {10}^{-14}\text{ erg}\text{ cm}^{-2}\text{ s}^{-1}\) (see Sect. \ref{sec:imaging}), B2.4 0412+23 has evidence of extended X-ray emission (see Fig. \ref{fig:profiles}), with several regions of extended flux along the radio axis (region 2 in in Fig. \ref{fig:zooms}), in correspondence of the radio lobes (regions 3 and 4 in in Fig. \ref{fig:zooms}) and on the western side of the nucleus (region 5 in Fig. \ref{fig:zooms}). These regions have however a low \(\lesssim 2 \sigma\) significance with respect to the surrounding emission, with the exception of region 5, that with 16 broad-band counts reaches a significance of almost \(6 \sigma\).

\subsubsection*{B2.3 0454+35}

This source shows an elongated radio structure in the north-south direction. In particular the GMRT and VLASS data indicate a bright hot-spot on the north and a narrow jet in the southern direction - possibly a one sided relativistic jet - , which originates from a bright compact X-ray source (region 1 in Fig. \ref{fig:zooms}) with a broad-band flux of \({80}_{-3}^{+3} \times {10}^{-14}\text{ erg}\text{ cm}^{-2}\text{ s}^{-1}\) (see Sect. \ref{sec:imaging}), that we identify as the X-ray nucleus. No other radio structure (including the northern hot-spot) shows significant X-ray emission.

\subsubsection*{B2.1 0455+32B}

This source, also known as NVSS J045906+323613, shows a rather compact radio structure in LOFAR and GMRT images. Its VLASS data, instead, reveal two compact radio lobes along the east-west direction. Between these two lobes there is a faint X-ray source (region 1 in Fig. \ref{fig:zooms}) with a broad-band flux of \({8}_{-3}^{+4} \times {10}^{-15}\text{ erg}\text{ cm}^{-2}\text{ s}^{-1}\) (see Sect. \ref{sec:imaging}), possibly the nucleus of the source. The two radio lobes do not show any significant X-ray emission.

\subsubsection*{B2.1 0455+32C}

The LOFAR and VLASS images of this source, also known as NVSS J045913+322607, show two radio lobes along the east-west direction, with an hint of edge-brightened FRII structure. Between the radio lobes there is a faint X-ray source (region 1 in Fig. \ref{fig:zooms}) with a broad-band flux of \({7}_{-3}^{+4} \times {10}^{-15}\text{ erg}\text{ cm}^{-2}\text{ s}^{-1}\) (see Sect. \ref{sec:imaging}), that we identify as the source nucleus. No other radio structure shows significant X-ray emission.

\subsubsection*{B2.3 0516+40}

This source (also known as NVSS J051946+401507), shows a compact radio structure. The nuclear region (marked as 1 in Fig. \ref{fig:zooms}) has a faint broad-band X-ray flux of \({4}_{-3}^{+4} \times {10}^{-15}\text{ erg}\text{ cm}^{-2}\text{ s}^{-1}\) (see Sect. \ref{sec:imaging}). There are two compact regions of enhanced X-ray emission south-west of the nucleus (regions 2 and 3 in Fig. \ref{fig:zooms}) that, compared to the level of the diffuse emission at the same radial distance from the nucleus, have a significance of \(> 3 \sigma\) and \(> 4 \sigma\), respectively. However, they appear disconnected from the radio structure.

\subsubsection*{B2.1 0536+33B}

This source, known as NVSS J054003+334200, shows an edge-brightened FRII structure in its VLASS image elongated in the east-west direction. Region 1 in Fig. \ref{fig:zooms} marks the X-ray nucleus with a \({29}_{-6}^{+7} \times {10}^{-15}\text{ erg}\text{ cm}^{-2}\text{ s}^{-1}\) flux (see Sect. \ref{sec:imaging}). No other significant structures are visible in the broad-band  \textit{Chandra} ACIS-S flux map.

\subsubsection*{B2.1 0549+29}

VLASS data show for B2.1 0549+29, also known as NVSS J055255+293203, a rather compact radio structure, with a core and the two lobes along the east-west direction. In Fig. \ref{fig:zooms} region 1 marks the X-ray nucleus with a \({7}_{-1}^{+1} \times {10}^{-14}\text{ erg}\text{ cm}^{-2}\text{ s}^{-1}\) flux (see Sect. \ref{sec:imaging}). Also in this case, no other significant structures are visible in the broad-band  \textit{Chandra} ACIS-S flux map.

\subsubsection*{B2.1 0643+30}

The radio structure of this source (also known as NVSS J064615+304123) as imaged by VLASS data is compact, showing only the core emission coincident with the X-ray nucleus emitting a \({8}_{-1}^{+1} \times {10}^{-14}\text{ erg}\text{ cm}^{-2}\text{ s}^{-1}\) broad-band flux (see Sect. \ref{sec:imaging}), marked in Fig. \ref{fig:zooms} as region 1. Again, the broad-band  \textit{Chandra} ACIS-S flux map shows no other significant structures.

\subsubsection*{B2.1 0742+31}

This source (also known as NVSS J074542+314252) at a redshift \(0.461\) \citep{1986MNRAS.221.1023K}, features a FRII edge-brightened radio structure as shown in LOFAR and VLASS data (see Fig. \ref{fig:maps}). This source has the brightest X-ray nucleus of the present sample (marked as region 1 in Fig. \ref{fig:zooms}), with an estimated flux of \({219}_{-5}^{+5} \times {10}^{-14}\text{ erg}\text{ cm}^{-2}\text{ s}^{-1}\) (see Sect. \ref{sec:imaging}), and is therefore affected by significant pileup (see Sect. \ref{sec:spectral}). In addition, B2.1 0742+31 shows significant diffuse X-ray emission (see Fig. \ref{fig:profiles}), both along the radio axis and across it, as shown in Fig. \ref{fig:zooms}. In particular, there are two regions of increased X-ray flux, north of the nucleus (region 2) and in correspondence of the southeast radio lobe (region 3). Region 2, north of the nucleus and connected with the emission surrounding the latter, contains 17 broad-band counts, while region 3 contains 30 broad-band counts. Both regions are significant at \(4 \sigma\) level with respect to the surrounding emission.

\subsubsection*{B2.2 0755+24}

This source (also known as NVSS J075802+242219), at a redshift \(0.502\) \citep{2019AJ....157..126G}, has a compact radio structure, where only the radio lobes are visible in VLASS data (see Fig. \ref{fig:maps}). In Fig. \ref{fig:zooms} we mark as region 1 the location that we identify as the faint X-ray nucleus, with an estimated broad-band flux of \({8}_{-3}^{+4} \times {10}^{-15}\text{ erg}\text{ cm}^{-2}\text{ s}^{-1}\) (see Sect. \ref{sec:imaging}). No other significant structures are visible in the broad-band  \textit{Chandra} ACIS-S flux map.

\subsubsection*{B2.3 0848+34}

This source (also known as J085108+341925), at a redshift of \(0.697\) \citep{2015ApJS..219...12A}, shows a rather compact radio structure as imaged by LOFAR data, with a slight extension toward the south. The VLASS data reveal a slightly elongated structure, with two small lobes along the east-west direction. Between the lobes we see a compact region of increased X-ray flux (marked as 1 in Fig. \ref{fig:zooms}) with a broad-band X-ray flux of \({9}_{-3}^{+4} \times {10}^{-15}\text{ erg}\text{ cm}^{-2}\text{ s}^{-1}\) (see Sect. \ref{sec:imaging}) that we identify as the source nucleus. Besides this nuclear region, there are no other significant structures in the broad-band \textit{Chandra} ACIS-S flux map.

\subsubsection*{B2.4 0939+22A}

This source (also known as NVSS J094158+214743), with a redshift of \(0.572\) \citep{2018ApJ...852...48S}, has FRII radio structure, as imaged with VLASS data, with the radio axis aligned along the northeast-southwest direction. The southwestern radio lobe is bent in the northwestern direction. The radio image does not show a clear core, but there is a faint point-like region between the two radio lobes (marked as region 1 in Fig. \ref{fig:zooms}) that we identify as the X-ray nucleus, with a broad-band flux of \({9}_{-3}^{+4} \times {10}^{-15}\text{ erg}\text{ cm}^{-2}\text{ s}^{-1}\) (see Sect. \ref{sec:imaging}). In addition, there is a region of increased X-ray flux (region 2 in Fig. \ref{fig:zooms}) coincident with the northeastern radio lobe. This region contains 25 broad-band counts, and it is therefore highly significant with respect to the emission at the same radial distance from the nucleus at \(7 \sigma\) level.

\subsubsection*{B2.4 1112+23}

The VLASS data of this source, also known as NVSS J111505+232503, only show the radio core, coincident with a region (marked as region 1 in Fig. \ref{fig:zooms}) of faint X-ray flux \({6}_{-3}^{+4} \times {10}^{-15}\text{ erg}\text{ cm}^{-2}\text{ s}^{-1}\) (see Sect. \ref{sec:imaging}) that we identify as the source nucleus. Besides the nucleus, there are no other significant structures in the broad-band  \textit{Chandra} ACIS-S flux map.

\subsubsection*{B2.3 1234+37}

This source, also known as NVSS J123649+365518, features two radio lobes along the northeast-southwest direction, as imaged by LOFAR and VLASS data. Between the radio lobes there is a faint X-ray source (region 1 in Fig. \ref{fig:zooms}) with a broad-band flux of \({13}_{-4}^{+5} \times {10}^{-14}\text{ erg}\text{ cm}^{-2}\text{ s}^{-1}\) (see Sect. \ref{sec:imaging}), that we identify as the source nucleus. No other radio structure shows significant X-ray emission.

\subsubsection*{B2.2 1334+27}

At a redshift \(z=3.228\) \citep{2015ApJS..219...12A}, this source, also known as NVSS J133641+270401, shows a faint extended radio structure in its LOFAR image. However, the VLASS data only reveal two compact radio lobes along the northwest-southeast direction. Between the radio lobes there is a X-ray source (region 1 in Fig. \ref{fig:zooms}) with a broad-band flux of \({9}_{-1}^{+1} \times{10}^{-14}\text{ erg}\text{ cm}^{-2}\text{ s}^{-1}\) (see Sect. \ref{sec:imaging}), possibly the X-ray source nucleus. No other radio structure shows significant X-ray emission.

\subsubsection*{B2.2 1338+27}

This source, also known as NVSS J134029+272326, shows two radio hot-spots along the northwest-southeast direction, as imaged by LOFAR, GMRT and VLASS data. Between the radio lobes there is a faint X-ray source (region 1 in Fig. \ref{fig:zooms}) with a broad-band flux of \(<9 \times{10}^{-15}\text{ erg}\text{ cm}^{-2}\text{ s}^{-1}\) (see Sect. \ref{sec:imaging}), possibly the X-ray source nucleus. We detected no significant X-ray emission in correspondence with the radio lobes.

\subsubsection*{B2.2 1439+25}

The VLASS data of this source, also known as NVSS J144204+250335, show a FRII radio structure extending along the north-south direction without revealing the radio core. No significant X-ray structure is revealed in the broad-band  \textit{Chandra} ACIS-S flux map.

\subsubsection*{B2.4 1512+23}

This source (also known as NVSS J151414+232711), at a redshift of \(0.088\) \citep{2012ApJS..203...21A}, shows a FRII radio structure extending along the north-south direction, as imaged by VLASS data. In particular, the southern lobe appears connected to the central region with a jet-shaped structure, originating from a bright X-ray compact region (marked as 1 in Fig. \ref{fig:zooms}) with broad-band X-ray flux of \({16}_{-5}^{+5} \times {10}^{-14}\text{ erg}\text{ cm}^{-2}\text{ s}^{-1}\) (see Sect. \ref{sec:imaging}), that we identify as the source nucleus. There are regions of enhanced X-ray emission in correspondence with the northern lobe (region 2 in Fig. \ref{fig:zooms}) and with the southern hot-spot (region 3 in Fig. \ref{fig:zooms}). Region 3 has a low \(<2 \sigma\) significance, while region 2 is significant at a \(4\sigma\) level when compared with respect to the level of the diffuse emission at the same radial distance from the nucleus.

\subsubsection*{B2.4 2054+22B}

This source, known as NVSS J205658+222954, shows a FRII radio structure in VLASS data, with lobes extending across the east-west direction. A faint region of enhanced X-ray emission (indicated as region 1 in Fig. \ref{fig:zooms}) features a broad-band flux of \({6}_{-3}^{+4} \times {10}^{-15}\text{ erg}\text{ cm}^{-2}\text{ s}^{-1}\) (see Sect. \ref{sec:imaging}), that we identify as the source nucleus. North of the nucleus there is another compact region of increased X-ray flux (region 2 in Fig. \ref{fig:zooms}) that is significant at \(3\sigma\) level compared to the level of the diffuse emission at the same radial distance from the nucleus. This region, however, does not appear to be clearly connected with the radio structure.

\subsubsection*{B2.2 2104+24}

This source, known as NVSS J210621+243324, shows a FRII VLASS radio structure, with a bright X-ray nucleus (region 1 in Fig. \ref{fig:zooms}) emitting a broad-band flux of \({30}_{-2}^{+2} \times {10}^{-14}\text{ erg}\text{ cm}^{-2}\text{ s}^{-1}\) (see Sect. \ref{sec:imaging}). Besides the nucleus, there are no other significant X-ray structures in the broad-band  \textit{Chandra} ACIS-S flux map.

\subsubsection*{B2.2 2133+27}

The radio structure of this source (known as NVSS J213516+271626), as imaged by VLASS data, does not have a clear shape, with the radio axis lying along the northwest-southeast direction, and without a clear radio core detection. There seems to be a region of increased X-ray flux in correspondence with the northwestern radio lobe (indicated with region 1 in Fig. \ref{fig:zooms}), but its significance with respect to the surrounding emission is only at \(2 \sigma\) level.

\subsubsection*{B2.3 2254+35}

This source (also known as NVSS J225645+354127), at a redshift of \(0.114\) \citep{2014ApJS..210....9B}, has a complex radio and X-ray morphology. The VLASS data reveal the location of the radio core and an edge-darkened FRI structure, with a jet extending toward the northern direction, and the other extending toward the eastern direction. The latter jet, in particular, appears bent in the south-east direction at larger radii, as even more evident in the large scale LOFAR data (see Fig. \ref{fig:maps}), revealing a WAT morphology. The region marked as region 1 in Fig. \ref{fig:zooms} is coincident with the radio core, and emits a broad-band X-ray flux of \({10}_{-3}^{+4} \times {10}^{-15}\text{ erg}\text{ cm}^{-2}\text{ s}^{-1}\) (see Sect. \ref{sec:imaging}). The X-ray emission is clearly extended (see Fig. \ref{fig:profiles}) and shows a complex morphology with two regions of decreased flux (marked with 2 and 3 in Fig. \ref{fig:zooms}) that look like X-ray cavities. Region 2, however, is less luminous than the emission at the same radial distance from the nucleus only at \(2 \sigma\) level, while for region 3 this significance increases to \(\sim 4 \sigma\) level.

\subsubsection*{B2.2 2328+26}

The radio structure of this source (known as NVSS J233032+270614) as mapped by VLASS data appears slightly elongated in the northeast-southwest direction. The region marked with 1 in Fig. \ref{fig:zooms} indicates the point-like source that we identify as the X-ray nucleus, with a broad-band flux of \({12}_{-0.4}^{+0.4} \times {10}^{-15}\text{ erg}\text{ cm}^{-2}\text{ s}^{-1}\) (see Sect. \ref{sec:imaging}). This is the only significant X-ray feature revealed in the broad-band  \textit{Chandra} ACIS-S flux map.

\subsubsection*{B2.3 2334+39}

The VLASS data of this source, also known as NVSS J233655+400546, only reveal the location of the radio core and that of the lobes, aligned along the northwest-southeast direction. The LOFAR data (see Fig. \ref{fig:maps}), on the other hand, indicate a FRI edge-darkened radio morphology, with the southeastern lobe bending toward the northeast direction, and the northwestern lobe bending toward the southeast direction. The region marked with 1 in Fig. \ref{fig:zooms} indicates the faint point-like source that we identify with the X-ray nucleus, with a broad-band flux \(< 5 \times {10}^{-15}\text{ erg}\text{ cm}^{-2}\text{ s}^{-1}\) (see Sect. \ref{sec:imaging}). There are no other significant X-ray features in the broad-band  \textit{Chandra} ACIS-S flux map.

\subsection{Spectral Analysis}\label{sec:spectral}

To characterize in more detail the sources in the present sample, we performed a spectral analysis of their nuclear emission. We extracted nuclear spectra in a \(2\arcsec\) circular region centrered at the coordinates of the radio or X-ray core, while background spectra were extracted in source-free regions as close as possible to the nuclear extraction region to avoid vignetting effects at the CCD edge, but far enough to exclude contamination from eventual diffuse emission. We produced auxiliary response files and spectral response matrices {both for the nuclear and background spectra,} applying for the former point-source aperture corrections (as appropriate for point-like sources). Spectral fitting was performed in the \(0.3-7\text{ keV}\) energy range with \textsc{Sherpa} application \citep{2001SPIE.4477...76F}.

Due to the low counts, we performed the spectral fits by modeling the background spectra using the prescription given by \citet{2003ApJ...583...70M}, that is, a model comprising a thermal plasma component \citep[MEKAL;][]{kasstra1992} with solar abundances and a power-law. We instead modeled the nuclear spectra with a power-law (\textsc{powerlaw}) model, including photo-electric absorption (\textsc{xstbabs}) by the Galactic column density along the line of sight \citep{2016A&A...594A.116H}. In addition, for the source B2.1 0742+31 we included the \textsc{jdpileup} model \citep{2001ApJ...562..575D} to account for the ACIS-S detector pileup. Spectra were binned to obtain a minimum of \(1\) count per bin, making use of the cash statistic \citep{2021A&A...647A..79P}.

Following this procedure we were able to extract and fit nuclear spectra for 15 sources. The results of these fits are presented in Table \ref{tab:nucl_spectra_noabs} and in Fig. \ref{fig:nucl_spectra_noabs}. Uncertainties correspond to the \(1\)-\(\sigma\) confidence level for one interesting parameter. We note that in many spectra we detect very few counts below \(1 \text{ keV}\), as a result of the degrading \textit{Chandra} effective area at low energies. We see that the intrinsic fluxes estimated from these spectral fits are compatible with those evaluated with \textsc{srcflux} (see Sect. \ref{sec:imaging}), with the exceptions of B2.1 0742+31 - for which the \textsc{jdpileup} model estimates a pileup fraction of \(\sim 20\%\), compatible with the value obtained from the pileup map. In addition, we notice that, while in a number of sources we find slopes \(\Gamma \sim 1.5 - 2.0\) - compatible with what is observed in similar sources \citep{2006MNRAS.370.1893H, 2009MNRAS.396.1929H, 2014MNRAS.440..269M} - in others the spectral fit yields particularly flat slopes, indicating the possible presence of significant intrinsic absorption.

To investigate the presence of intrinsic absorption, we repeated the spectral fitting of the nuclear spectra freezing the power-law slope to \(1.8\) and considering an additional absorption component (\textsc{xsztbabs}) at the source redshift or, if this measurement was not available, at redshift zero. The results of these fits are presented in Table \ref{tab:nucl_spectra_abs} and in Fig. \ref{fig:nucl_spectra_abs}. 
For most of the sources we are only able to put upper limits on the intrinsic absorption column, or find values compatible with the Galactic ones. For sources B2.4 0004+21, B2.4 0145+212 and B2.4 0401+23, instead, we find intrinsic absorbing columns \(\sim {10}^{22} \text{ cm}^{-2}\), while for the source B2.4 0229+23 we find an additional absorbing column of \(\sim \times {10}^{23} \text{ cm}^{-2}\), all significantly larger than the Galactic values.

\begin{figure}
	\includegraphics[scale=0.9]{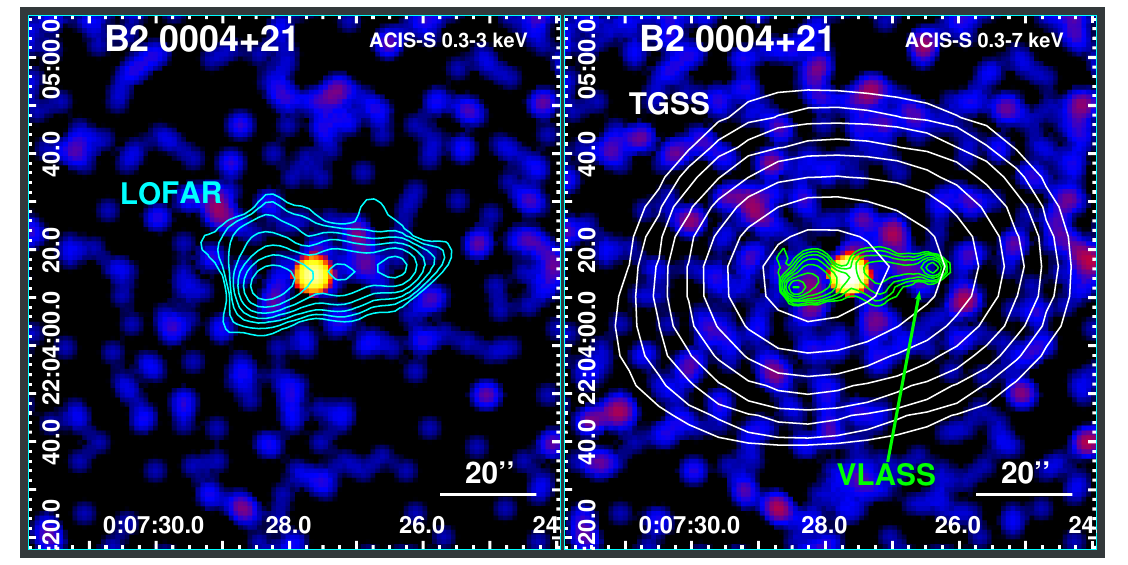}
	\caption{\textit{Chandra} ACIS-S flux maps (in each panel: \(0.3-3 \text{ keV}\) and \(0.3-7 \text{ keV}\) band images on the left and on the right, respectively). Radio map contours are overlaid in cyan (LOFAR \(145 \text{ MHz}\)), white (TGSS \(150 \text{ MHz}\)) and green (VLASS \(3 \text{ GHz}\)). The radio contours start at \(10\) times the rms level, increasing by factors of two. The complete figure set (33 images) is available in the online journal.}\label{fig:maps}
\end{figure}

\begin{figure}
	\centering
	\includegraphics[scale=0.7]{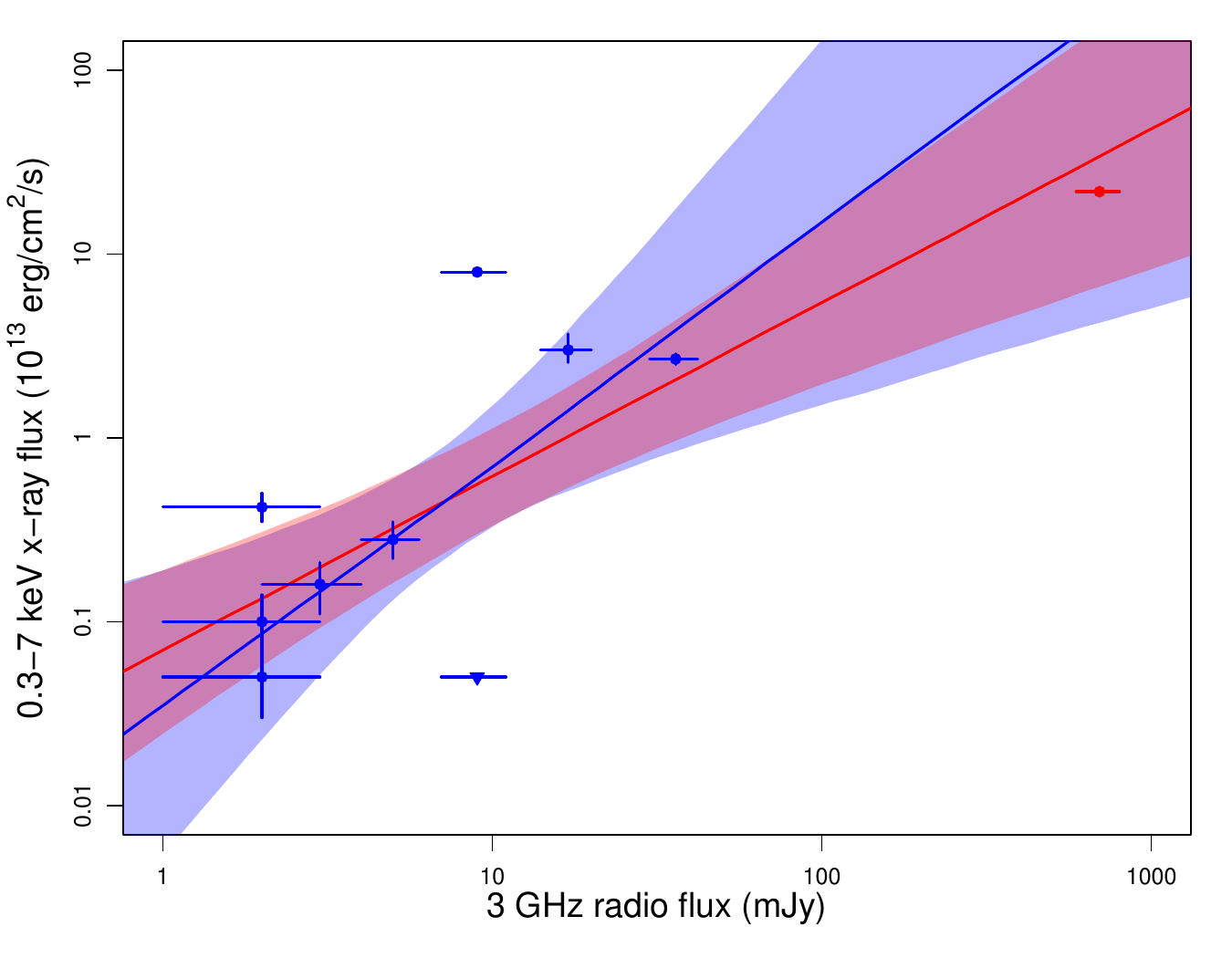}
	\caption{X-ray \(0.3-7 \text{ keV}\) nuclear fluxes versus \(3 \text{ GHz}\) radio nuclear specific fluxes for the sources for both estimates are available (see Sect. \ref{sec:imaging}). Upper limits on X-ray fluxes are represented with triangles pointing down. The red line indicates the linear regression to the logarithmic data that include the highly piled-up source B2.1 0742+31 (indicated with a red circle), while the light red shaded area represents the \(1-\sigma\) uncertainty around the best fit relation. The blue line and light blue shaded area represent the linear regression and uncertainty for the logarithmic data that exclude B2.1 0742+31.}\label{fig:xray-radio-corr}
\end{figure}

\begin{figure}
	\centering
	\includegraphics[scale=0.7]{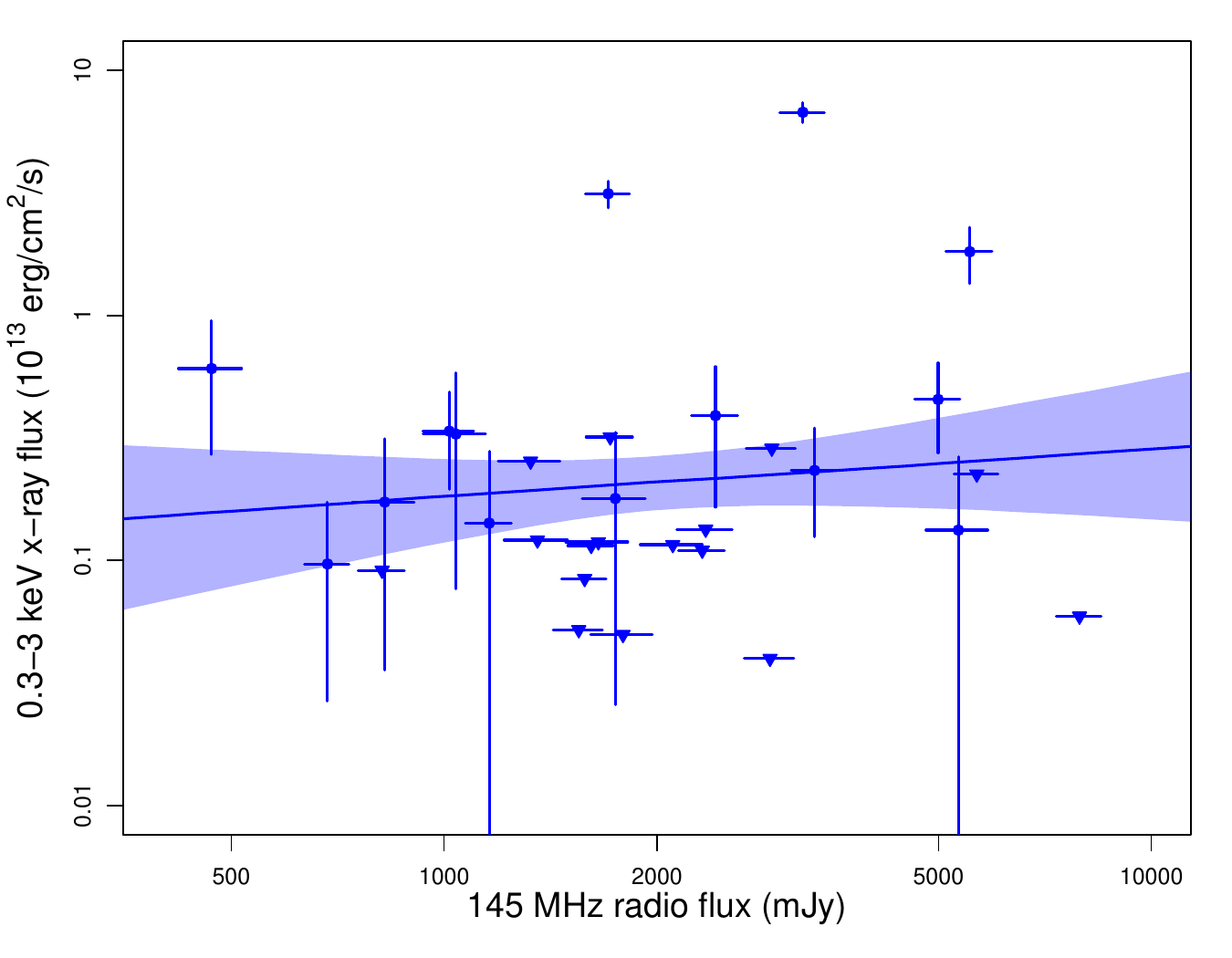}
	\caption{X-ray \(0.3-3 \text{ keV}\) extended emission fluxes versus \(145 \text{ MHz}\) specific fluxes of the extended radio structures (see Sect. \ref{sec:imaging}). Upper limits on X-ray fluxes are represented with triangles pointing down. The blue line indicates the linear regression to the logarithmic data, while the light blue shaded area represents the \(1-\sigma\) uncertainty around the best fit relation.}\label{fig:xray-radio-corr-ext}
\end{figure}

\begin{figure}
	\centering
	\includegraphics[scale=0.29]{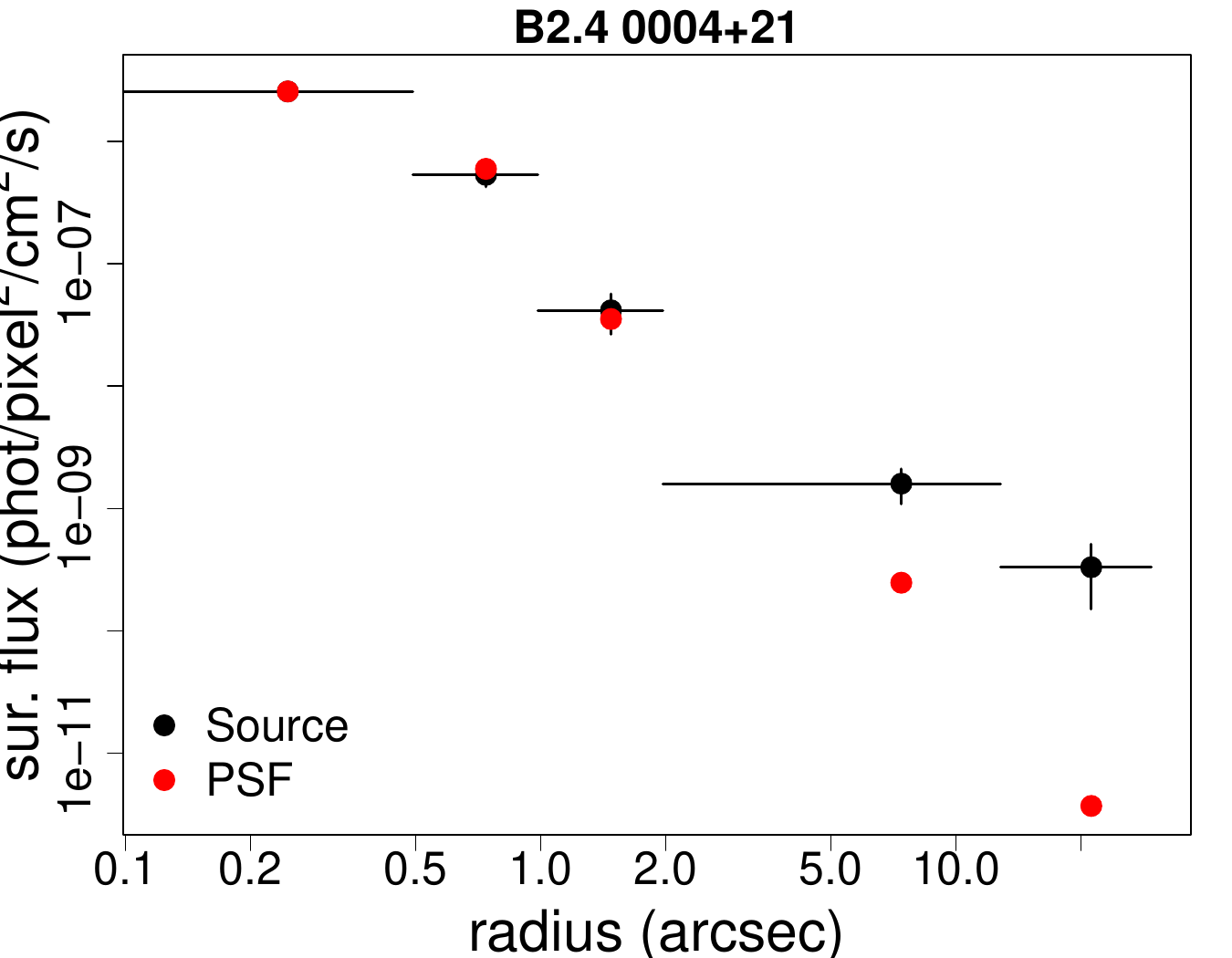}
	\includegraphics[scale=0.29]{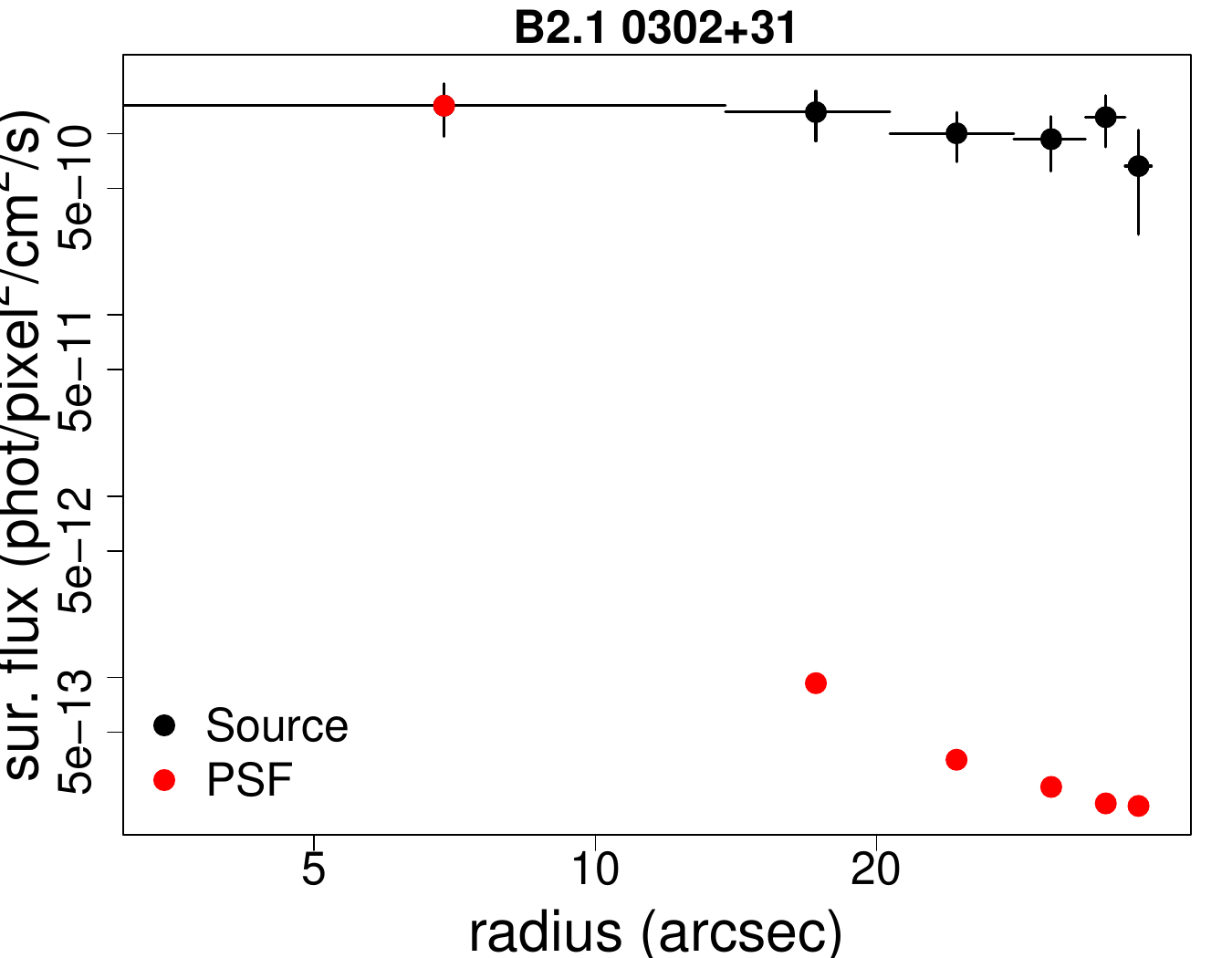}
	\includegraphics[scale=0.29]{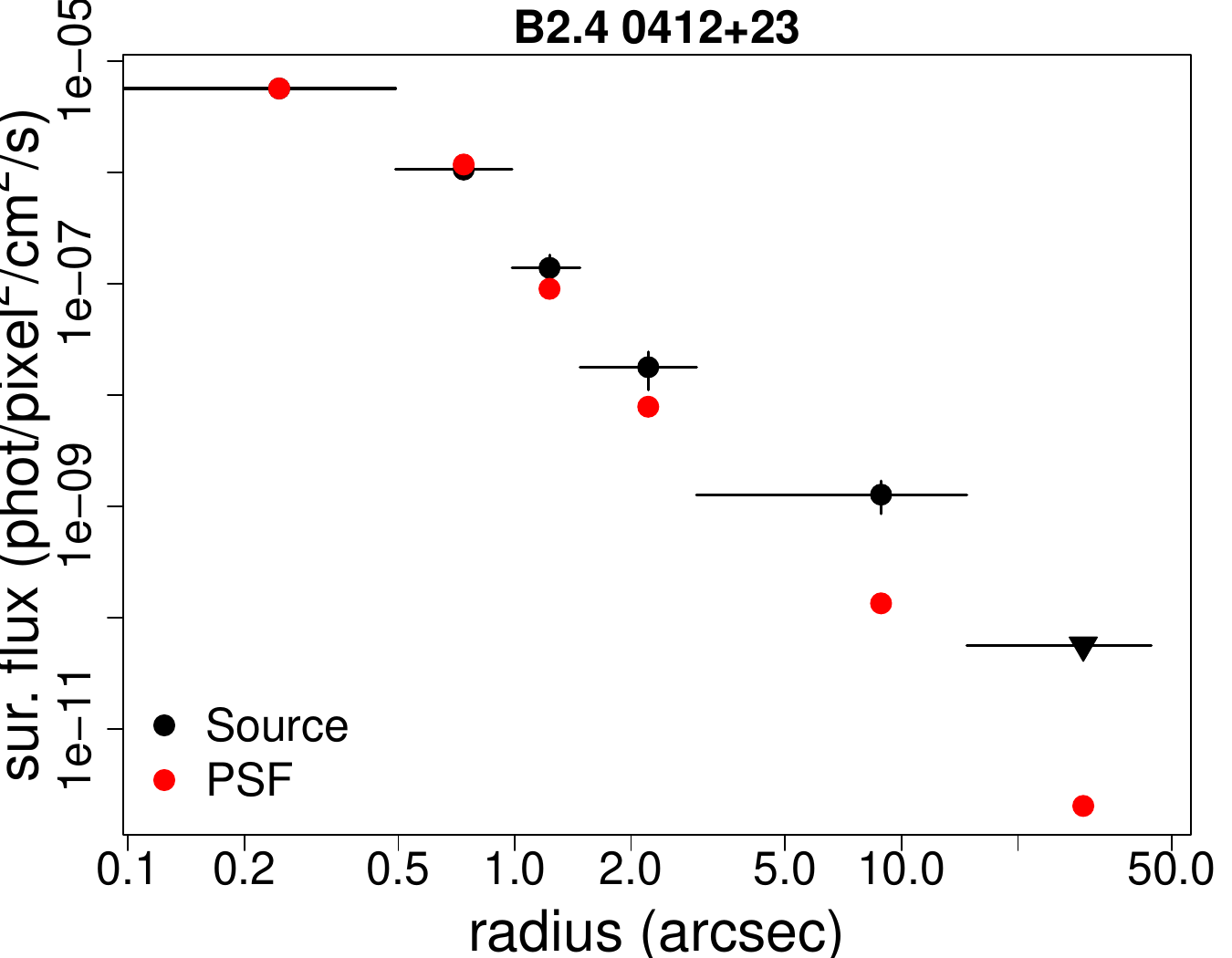}
	\includegraphics[scale=0.29]{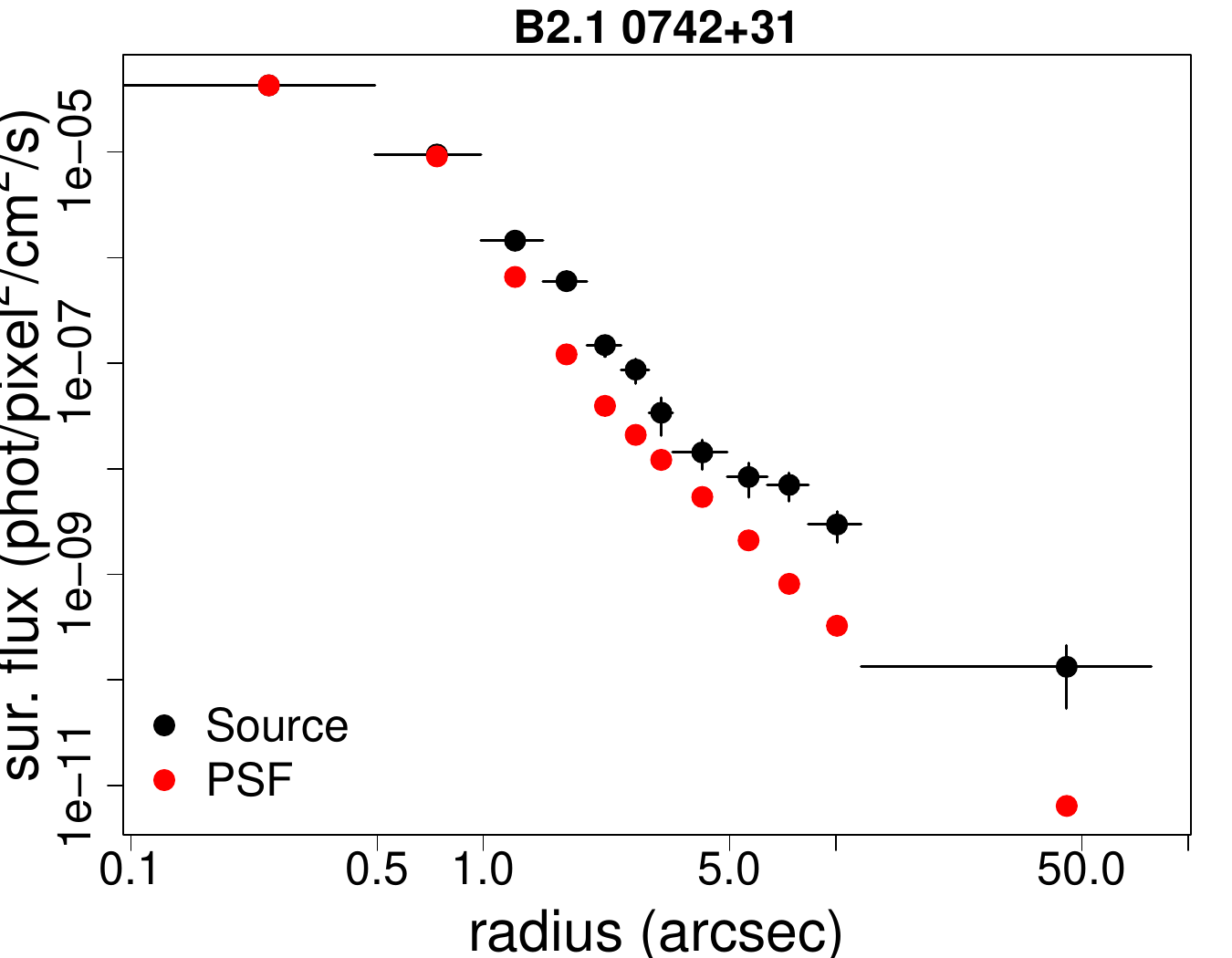}
	\includegraphics[scale=0.29]{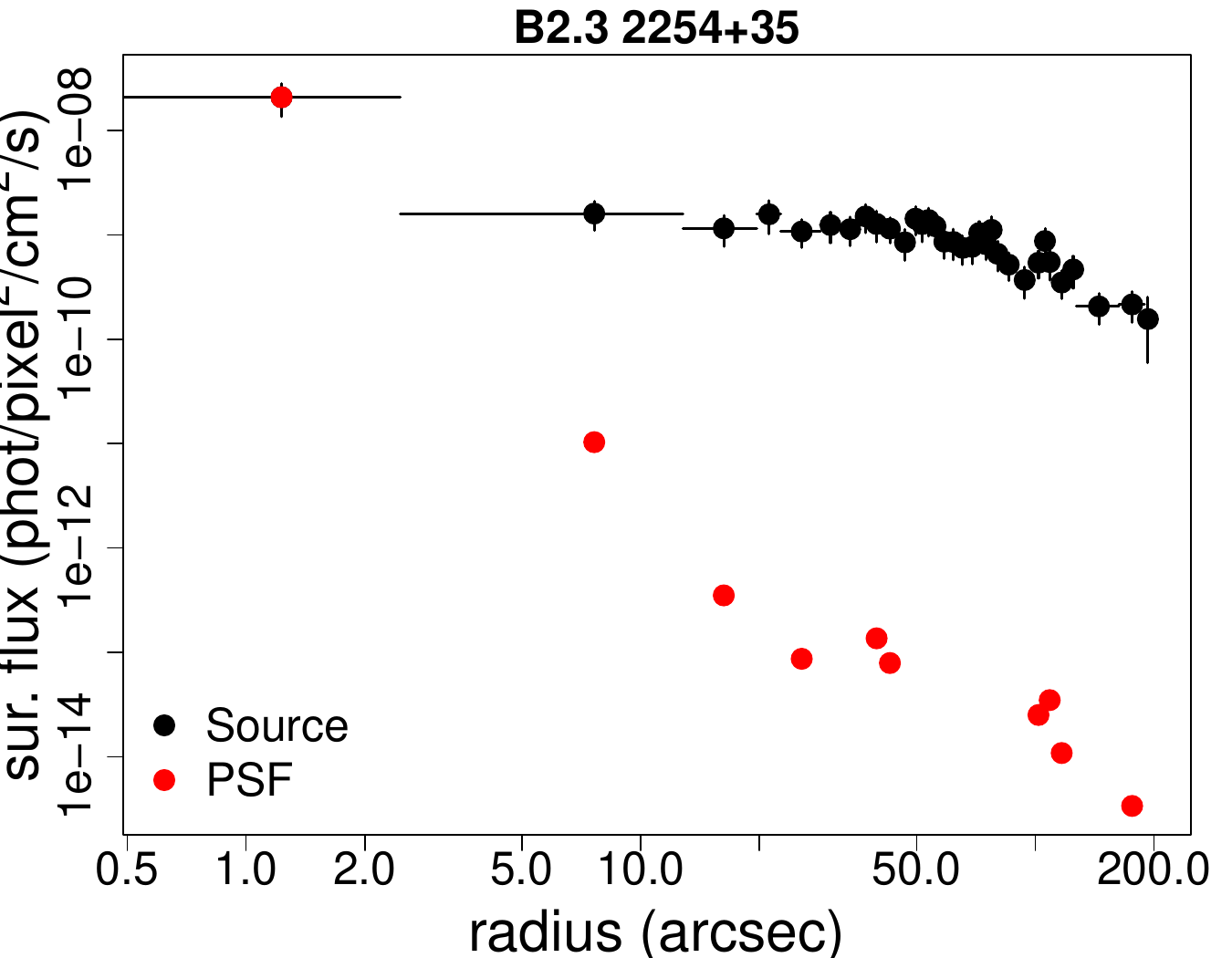}
	\caption{Surface flux profiles for the sources that show evidence of extended X-ray emission in the soft \(0.3-3 \text{ keV}\) band (see Sect. \ref{sec:imaging}). For each source we present with black dots the surface flux profiles from the source itself, and with red dots the profiles from the simulated PSF.}
	\label{fig:profiles}
\end{figure}

\begin{figure}
	\centering
	\includegraphics[scale=0.9]{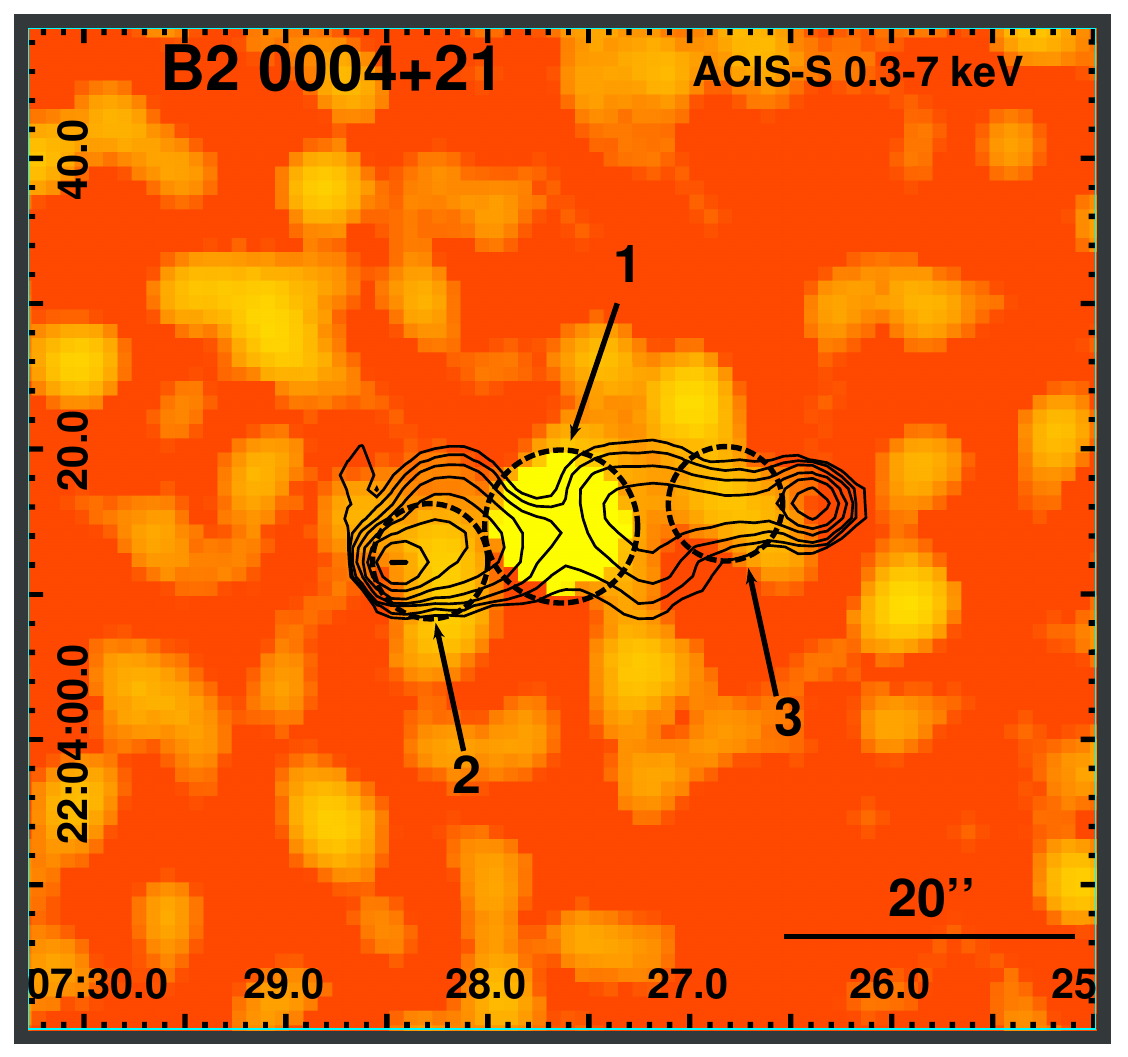}
	\caption{\textit{Chandra} ACIS-S \(0.3-7 \text{ keV}\) flux maps of the central region of the sources in the present sample. In each panel we overlay in black the \(3 \text{ GHz}\) VLASS contours from Fig. \ref{fig:maps}. The black dashed circles marks the regions discussed in detail in Sect. \ref{sec:sources}. For sources B2.1 0302+31, B2.1 0742+31 and B2.3 2254+35 the white ellipse shows the extraction region used for the spectral fittings presented in Table presented in Table \ref{tab:thermal_spectra_radio}, while the blue ellipse shows the region of radio emission excluded in the spectral fittings presented in Table \ref{tab:thermal_spectra_noradio}. The complete figure set (33 images) is available in the online journal.}\label{fig:zooms}
\end{figure}

As discussed in Sect. \ref{sec:imaging}, we have 5 sources that show evidence of significant extended emission in the soft \(0.3-3 \text{ keV}\) band (see Fig. \ref{fig:profiles}). We extracted source spectra in large elliptical regions that encompass the whole extended emission visible in the flux maps (see Fig. \ref{fig:maps}), excluding detected point sources as well as the \(2\arcsec\) nuclear regions. Background spectra were extracted in the same source-free regions used for the nuclear spectral fitting. In this case, we produced spectral response matrices weighted by the count distribution within the aperture (as appropriate for extended sources).

We used the same procedure adopted for the nuclear spectral fitting, that is, modeling the background spectra and using the cash statistics, with spectra binned to obtain a minimum of \(1\) count per bin. The sources B2.4 0004+21 and B2.4 0412+23 did not yield enough counts to allow a reasonable fit, and were therefore excluded from the following analysis. To fit the spectra of the extended emission we used a model comprising the Galactic absorption and a thermal plasma (\textsc{xsapec}\footnote{\href{https://heasarc.gsfc.nasa.gov/xanadu/xspec/manual/XSmodelApec.html}{https://heasarc.gsfc.nasa.gov/xanadu/xspec/manual/XSmodelApec.html}}) with abundance \(0.25\) solar (as expected from typical ICM emission). The redshift of the thermal plasma was set at the value of the source redshift or, if this measurement was not available, at redshift zero.

The results of these fits are presented in Table \ref{tab:thermal_spectra_radio} and in Fig. \ref{fig:thermal_spectra_radio}. Again, uncertainties correspond to the \(1\)-\(\sigma\) confidence level for one interesting parameter. We obtain reasonable best fit temperatures between \(1.9\) and \(2.5 \text{ keV}\). The diffuse emission surrounding these sources, however, can be a combination of thermal emission from hot gas of the ICM and IC/CMB. To minimize the contamination from such non-thermal emission, we repeated the spectral extraction excluding the regions of extended radio emission shown in Fig. \ref{fig:maps}. The results of these fits are presented in Table \ref{tab:thermal_spectra_noradio} and in Fig. \ref{fig:thermal_spectra_noradio}. The temperature values obtained in this way are similar to those obtained previously (although with larger uncertainties), suggesting that in these sources the contribution from non-thermal IC/CMB emission could be sub-dominant with respect to the thermal radiation arising from the ICM. Since the presence of IC/CMB may be revealed by significant X-ray emission above \(2 \text{ keV}\) \citep{2022arXiv220710092M}, we produced \(3-7 \text{ keV}\) hard-band flux images for the \(5\) sources that show evidence of significant extended emission and present them in Fig. \ref{fig:hard}, with radio contours drawn at \(3 \text{ GHz}\) overlaid in green. We see that the only source showing significant X-ray emission in this band in correspondence with the extended radio structures is B2.1 0302+31. In particular the hard-band emission in the east and west radio lobes are detected at \(2.4\sigma\) and \(3.3\sigma\) significance. Although this is conducive of the presence of non-thermal IC/CMB emission in this source, the low statistics do not allow us to draw firm conclusions.

The detection of this extended X-ray emission in any case suggests the presence of ICM, indicating that these sources may belong to groups or clusters of galaxies. In the case of B2.1 0742+31, in particular, this is reinforced by the presence in the source field of 5 additional galaxies at redshift close to that of the galaxy hosting B2.1 0742+31 (see Fig. \ref{fig:0742_z}), that is, with a maximum redshift separation \(\Delta z = 0.005\) (i.e., \(\sim 1500 \text{ km/s}\)) corresponding to the maximum velocity dispersion observed in groups and clusters of galaxies \citep[see, e.g.,][]{1993MNRAS.261..827M, 2004MNRAS.348..866E, 2006ApJS..167....1B}. We therefore compared the properties of the hot gas surrounding these radio sources with those observed in groups and clusters of galaxies. In particular, we are interested in the hot gas X-ray luminosity vs. temperature correlation \citep{2000ARA&A..38..289M}. Since we have redshift estimates only for B2.1 0742+31 and B2.3 2254+35, we restrict our analysis to these two sources. In Fig. \ref{fig:lx_vs_T} we compare the temperature (\(kT\)) and X-ray bolometric luminosity \(L_X\) of the thermal gas surrounding B2.1 0742+31 and B2.3 2254+35 with those of groups and clusters of galaxies from Figure 6 of \citet{2000ARA&A..38..289M}, where the X-ray luminosities have been rescaled to the cosmology adopted in the present analysis.

We see that, while the X-ray emission of B2.1 0742+31 is compatible with the ICM emission of low luminosity clusters of galaxies, the X-ray diffuse emission surrounding the highly disturbed WAT B2.3 2254+35 lies somehow at the edge of the \(L_X - kT\) relation, with a luminosity similar to those of bright groups of galaxies, and a temperature similar to those of low luminosity cluster of galaxies, possibly due the disturbed nature of the gas surrounding this WAT.

Finally, we estimated the mass of the X-ray emitting gas in B2.1 0742+31 and B2.3 2254+35 from the spectral fits. From the normalization of the \textsc{xsapec} models (i.e., their emission measures \(EM\)), we can evaluate the gas proton density \(n_{\text{p}}\). Assuming a uniform particle density in the emitting region, we have a proton density
\begin{equation}\label{eq:gas_den}
	n_{\text{p}} = \sqrt{\frac{{10}^{14}\,EM\,\eta\,4 \pi\, {D_A}^2{\left({1+z}\right)}^2}{V}}\,,
\end{equation}
where \(D_A\) is the angular distance of the source, \(V\) is the emitting region volume, and \(\eta\approx 0.82\) is the ratio of proton to electron density in a fully ionized plasma. We can estimate the total gas mass as \(M_{\text{gas}} = \mu\, m_u n_{\text{tot}} V\), where \(m_u\) is the atomic mass, \(n_{\text{tot}} = n_p (1+1/\eta)\) is the total gas density, and \(\mu=0.6\) is the mean molecular weight \citep{2013SSRv..177..119E}.

To estimate the volumes we model the projected emission regions as ellipses with semi-major and semi-minor axes \(R\) and \(r\), respectively, encompassing the diffuse X-ray and radio emission. Then, we model the emitting regions as ellipsoids with volume \(V=4/3 \, \pi \, R\, r^2\) (when excluding the radio emission region, the volume would be the difference between the X-ray and radio emission region volumes, see Fig. \ref{fig:maps}).

Taking into account the uncertainties on the best fit parameters and the different spectral extraction regions (that is, including or excluding the extended radio structures), we estimate \(M_{\text{gas}} = 2.6 - 9.1 \times{10}^{12}\, M_{\sun}\) and \(M_{\text{gas}} = 0.9 - 1.4 \times{10}^{12}\, M_{\sun}\) for B2.1 0742+31 and B2.3 2254+35, respectively, typical of rich groups \citep[see, e.g.,][]{2000ARA&A..38..289M}.

\section{Summary and Conclusions}\label{sec:summary}

In this work we have analyzed the first 33 \textit{Chandra} ACIS observations obtained through the CCT snapshot campaign on the Second Bologna Catalog of radio sources. The X-ray data have been compared with \(145 \text{ MHz}\) LOFAR, \(150 \text{ MHz}\) GMRT, and \(3 \text{ GHz}\) VLASS data, to study the connection between the X-ray and radio emission in radio galaxies.
The main results of this analysis can be summarized as follows:
\begin{enumerate}	
	\item We detected X-ray nuclear emission for 28 of 33 sources. In particular, 19 nuclei were detected at least at \(3 \sigma\) significance, 7 were detected at \(2 \sigma\) significance, and 2 were detected with a \(1 \sigma\) marginal significance. For two other sources we were only able to put an upper limit on the nuclear flux, and for the remaining three sources we do not report any nuclear flux estimate, since we do not have any clear indication of the location of their core.
	\item We found a mild correlation between the X-ray and radio nuclear fluxes, while the flux of diffuse X-ray emission does not appear to correlate with the radio flux of the extended radio structures.
	\item Comparing the X-ray surface flux profiles of the sources with those of simulated PSFs, we detected extended emission with a minimum \(5 \sigma\) significance level beyond \(10\arcsec\) from the nucleus in \(5\) sources.
	\item We detected 8 regions of increased X-ray flux in correspondence with radio hot-spots or jet knots at a minimum significance level of \(3 \sigma\), \(2\) of which above \(5 \sigma\) level of significance. In B2.3 2254+35 we were able to detect a region of decreased flux, possibly associated with an X-ray cavity, at \(4 \sigma\) level of significance.
	\item We performed a X-ray spectral analysis for 15 nuclei with a power-law model, and found for the nuclei of B2.4 0004+21, B2.4 0145+22 and B2.4 0401+23 significant intrinsic absorption \(N_{H,\text{int}} \sim {10}^{22} \text{ cm}^{-2}\), and for B2.4 0229+23 \(N_{H,\text{int}} \sim {10}^{23} \text{ cm}^{-2}\).
	\item We performed a X-ray spectral analysis of the diffuse emission surrounding 3 sources, finding temperatures of the hot plasma \(\sim 2 \text{ keV}\). There is some hint of X-ray emission above \(3 \text{ keV}\) in correspondence with the radio lobes in B2.1 0302+31, which may suggest the presence of IC/CMB in this source. The low statistics however does not allow us to draw firm conclusions.
	\item For two of these sources, B2.1 0742+31 and B2.3 2254+35, we compared the properties of the X-ray emitting gas with those of the ICM surrounding clusters and groups of galaxies. While the hot gas surrounding  B2.1 0742+31 is compatible with the ICM of low luminosity clusters of galaxies, the X-ray diffuse emission surrounding the highly disturbed WAT B2.3 2254+35 features a luminosity similar to those of the ICM of bright groups of galaxies, while having a temperature similar to those of the ICM of low luminosity clusters of galaxies. The mass of these X-ray emitting plasmas is of the order of \(\sim {10}^{12}\, M_{\sun}\), similar to those observed in the ICM of rich groups.
\end{enumerate}

These first results on the B2CAT CCT survey show that the low-frequency radio selection, combined with short X-ray snapshot observations, are a powerful tool to optimize the ``fill-in'' observing strategy of several X-ray telescopes. In particular, this proves to be particularly effective with \textit{Chandra} observatory, since for \textit{XMM-Newton} such short observations tend to be scheduled at the end of its orbits, which are dominated by high particle background.

\begin{acknowledgements}
	We thank the anonymous referee for their useful comments and suggestions.
	This work is supported by the ``Departments of Excellence 2018 - 2022'' Grant awarded by the Italian Ministry of Education, University and Research (MIUR) (L. 232/2016).
	This research has made use of resources provided by the Compagnia di San Paolo for the grant awarded on the BLENV project (S1618\_L1\_MASF\_01) and by the Ministry of Education, Universities and Research for the grant MASF\_FFABR\_17\_01.
	A.P. acknowledges financial support from the Consorzio Interuniversitario per la Fisica Spaziale (CIS) under the agreement related to the grant MASF\_CONTR\_FIN\_18\_02.
	F.M. acknowledges financial contribution from the agreement ASI-INAF n.2017-14-H.0.
	S.E. acknowledges the financial contribution from the contracts ASI-INAF Athena 2019-27-HH.0, ``Attivit\`a di Studio per la comunit\`a scientifica di Astrofisica delle Alte Energie e Fisica Astroparticellare'' (Accordo Attuativo ASI-INAF n. 2017-14-H.0).
	A.P. acknowledges W. R. Forman for useful comments and suggestions.
	This research has made use of data obtained from the \textit{Chandra} Data Archive. This research has made use of software provided by the \textit{Chandra} X-ray Center (CXC) in the application packages CIAO, ChIPS, and Sherpa.
	This research has made use of the NASA/IPAC Extragalactic Database (NED; \href{https://ned.ipac.caltech.edu}{https://ned.ipac.caltech.edu}), which is funded by the National Aeronautics and Space Administration and operated by the California Institute of Technology.
	SAOImageDS9 development has been made possible by funding from the Chandra X-ray Science Center (CXC), the High Energy Astrophysics Science Archive Center (HEASARC) and the JWST Mission office at Space Telescope Science Institute.
	LOFAR data products were provided by the LOFAR Surveys Key Science project (LSKSP; \href{https://lofar-surveys.org/}{https://lofar-surveys.org/}) and were derived from observations with the International LOFAR Telescope (ILT). LOFAR \citep{2013A&A...556A...2V} is the Low Frequency Array designed and constructed by ASTRON. It has observing, data processing, and data storage facilities in several countries, which are owned by various parties (each with their own funding sources), and which are collectively operated by the ILT foundation under a joint scientific policy. The efforts of the LSKSP have benefited from funding from the European Research Council, NOVA, NWO, CNRS-INSU, the SURF Co-operative, the UK Science and Technology Funding Council and the J\"ulich Supercomputing Centre.
	The authors thank the staff of the GMRT that made these observations possible. GMRT is run by the National Centre for Radio Astrophysics of the Tata Institute of Fundamental Research.
	The National Radio Astronomy Observatory is a facility of the National Science Foundation operated under cooperative agreement by Associated Universities, Inc.
	The Pan-STARRS1 Surveys (PS1) have been made possible through contributions of the Institute for Astronomy, the University of Hawaii, the Pan-STARRS Project Office, the Max-Planck Society and its participating institutes, the Max Planck Institute for Astronomy, Heidelberg and the Max Planck Institute for Extraterrestrial Physics, Garching, The Johns Hopkins University, Durham University, the University of Edinburgh, Queen's University Belfast, the Harvard-Smithsonian Center for Astrophysics, the Las Cumbres Observatory Global Telescope Network Incorporated, the National Central University of Taiwan, the Space Telescope Science Institute, the National Aeronautics and Space Administration under Grant No. NNX08AR22G issued through the Planetary Science Division of the NASA Science Mission Directorate, the National Science Foundation under Grant No. AST-1238877, the University of Maryland, and Eotvos Lorand University (ELTE).
\end{acknowledgements}

\software{CIAO \citep{2006SPIE.6270E..1VF}, Sherpa \citep{2001SPIE.4477...76F, 2007ASPC..376..543D, 2020zndo...3944985B}, ChiPS \citep{2006ASPC..351...57G}, SAOImageDS9 \citep{2003ASPC..295..489J}, TOPCAT \citep{2005ASPC..347...29T}.}

\begin{table}
	\caption{Main properties of the radio sources investigated in the present work. In the table we list the source B2CAT name (Source Name), its J2000 coordinates (RA and Dec) and redshift (\(z\)), the \textit{Chandra} OBSID, the net exposure after filtering for time intervals of high background flux (Clean Exp.), the intrinsic \(0.3 - 7 \text{ keV}\) nuclear flux estimated assuming a power-law spectrum with a \(1.8\) slope (Nuclear \(F_{0.3-7 \text{ keV}}\)) and the corresponding nuclear luminosity for the sources with available redshift (Nuclear \(L_{0.3-7 \text{ keV}}\)), the intrinsic \(0.3-3 \text{ keV}\) extended emission flux estimated assuming a thermal spectrum with 2 keV temperature and abundance 0.25 solar (Extended \(F_{0.3-3 \text{ keV}}\)) and the corresponding extended emission luminosity for the sources with available redshift (Extended \(L_{0.3-3 \text{ keV}}\)), the general radio morphology (Radio Morph.), the image pixel size in terms of the native ACIS pixel size (Bin Size) and the \(\sigma\) width expressed in image pixels of the smoothing Gaussian kernel expressed in image pixels (Smoothing) used in Figs \ref{fig:maps}, \ref{fig:zooms} and \ref{fig:hard}.}\label{tab:sources}
	\begin{center}
	\resizebox{\textwidth}{!}{
		\begin{tabular}{|l|cc|c|c|c|c|c|c|c|c|c|c|}
			\hline
			\hline
			Source Name & RA & Dec & \(z\) & \textit{Chandra} OBSID & Clean Exp. & Nuclear \(F_{0.3-7 \text{ keV}}\) & Nuclear \(L_{0.3-7 \text{ keV}}\) & Extended \(F_{0.3-3 \text{ keV}}\) & Extended \(L_{0.3-3 \text{ keV}}\) & Radio Morph. & Bin Size & Smoothing \\
			\hline
			& hh:mm:ss.ss & dd:mm:ss.s & & & ks & \({10}^{-13}\text{ erg}\text{ cm}^{-2}\text{ s}^{-1}\) & \(\text{erg}\text{ s}^{-1}\) & \({10}^{-13}\text{ erg}\text{ cm}^{-2}\text{ s}^{-1}\) & \(\text{erg}\text{ s}^{-1}\) & & & \\
			\hline
			B2.4 0004+21  		   & 00:07:26.69 		  & +22:03:23.8 		 &             	  & 26155 		   & 15.9          & \({1.95}_{-0.15}^{+0.15}\)      & 										    & \({0.45}_{-0.18}^{+0.19}\) & 											& FRII & 2 & 4 \\
			B2.2 0038+25B & 00:41:18.54 & +25:49:50.9 &             	  & 27884 & 14.6 & \({0.05}_{-0.02}^{+0.03}\)      & 										    & \({<0.06}\)                & 											& FRII & 2 & 3 \\
			B2.2 0143+24  		   & 01:46:28.83 		  & +25:06:04.6 		 &             	  & 23074 		   & 15.3          & \({0.06}_{-0.03}^{+0.04}\)      & 										    & \({<0.08}\)                & 											& FRII & 4 & 4 \\
			B2.4 0145+22  		   & 01:47:49.84 		  & +22:38:55.0 		 &             	  & 27519 		   & 15.9          & \({0.79}_{-0.10}^{+0.10}\)      & 										    & \({<0.32}\)                & 											& FRII & 4 & 3 \\
			B2.4 0229+23  		   & 02:32:20.96 		  & +23:17:24.6 		 & 3.420          & 27538 		   & 14.9          & \({5.81}_{-0.29}^{+0.29}\)      & \({6.4}_{-0.3}^{+0.3} \times {10}^{46}\) & \({0.61}_{-0.34}^{+0.35}\) & \({6.7}_{-3.7}^{+3.9} \times {10}^{45}\) & Compact & 2 & 3 \\
			B2.1 0241+30  		   & 02:44:42.80 		  & +30:20:44.5 		 &             	  & 26174 		   & 15.4          & \({0.05}_{-0.02}^{+0.03}\)      & 										    & \({0.23}_{-0.11}^{+0.11}\) &  										& FRII & 2 & 6 \\
			B2.1 0302+31  		   & 03:05:23.16 		  & +31:29:42.5 		 &             	  & 26198 		   & 15.9          &  							     & 										    & \({3.14}_{-0.39}^{+0.39}\) &  										& HyMoR & 2 & 8 \\
			B2.4 0401+23  		   & 04:04:51.68 		  & +24:07:02.3 		 &             	  & 26199 		   & 15.9          & \({0.42}_{-0.07}^{+0.08}\)      & 										    & \({<0.05}\)                &  										& FRII & 2 & 5 \\
			B2.2 0410+26  		   & 04:13:23.64 		  & +26:48:47.5 		 &             	  & 26200 		   & 15.9          & \({0.11}_{-0.03}^{+0.04}\)      & 										    & \({0.13}_{-0.13}^{+0.13}\) &  										& Compact & 2 & 5 \\
			B2.4 0412+23  		   & 04:15:12.82 		  & +23:47:52.3 		 &             	  & 26214 		   & 15.9          & \({2.69}_{-0.18}^{+0.18}\)      & 										    & \({0.39}_{-0.22}^{+0.23}\) &  										& Lobes & 2 & 5 \\
			B2.3 0454+35  		   & 04:58:07.36 		  & +35:45:46.3 		 &             	  & 27589 		   & 15.8          & \({8.00}_{-0.34}^{+0.35}\)      & 										    & \({<0.67}\)                &  										& One-sided jet & 1 & 3 \\
			B2.1 0455+32B  		   & 04:59:05.74 		  & +32:36:30.0 		 &            	  & 27590 		   & 15.9          & \({0.08}_{-0.03}^{+0.04}\)      & 										    & \({<0.12}\)                &  										& Lobes & 2 & 4 \\
			B2.1 0455+32C  		   & 04:59:14.08 		  & +32:26:11.4 		 &            	  & 27591 		   & 15.9          & \({0.07}_{-0.03}^{+0.04}\)      & 										    & \({<0.11}\)                &  										& FRII & 2 & 4 \\
			B2.3 0516+40  & 05:19:45.53 & +40:15:44.5 & 				  & 27750 & 14.9 & \({0.04}_{-0.03}^{+0.04}\)      & 										    & \({<0.25}\)                &  										& Compact & 4 & 3 \\
			B2.1 0536+33B		   & 05:40:03.88 		  & +33:42:04.2 		 &         		  & 22192 		   & 15.9 		   & \({0.28}_{-0.06}^{+0.07}\)      & 										    & \({<0.09}\)                &  										& FRII & 1 & 4\\
			B2.1 0549+29  		   & 05:52:55.28 		  & +29:33:07.9 		 &          	  & 26230 		   & 15.9 		   & \({0.65}_{-0.10}^{+0.11}\)      & 										    & \({0.33}_{-0.25}^{+0.26}\) &  										& Lobes & 1 & 3 \\
			B2.1 0643+30  		   & 06:46:15.48 		  & +30:41:16.0 		 &         		  & 26263 		   & 15.8 		   & \({0.80}_{-0.10}^{+0.10}\)      & 										    & \({0.30}_{-0.16}^{+0.16}\) &  										& Compact & 1 & 3 \\
			B2.1 0742+31  		   & 07:45:41.63 		  & +31:43:10.3 		 & 0.461   		  & 26264 		   & 15.9 		   & \({21.90}_{-0.50}^{+0.50}{}^*\) & \({17.7}_{-0.4}^{+0.4} \times{10}^{44}\) & \({1.82}_{-0.47}^{+0.47}\) & \({1.5}_{-0.4}^{+0.4} \times {10}^{44}\) & FRII & 2 & 8 \\
			B2.2 0755+24  		   & 07:58:02.75 		  & +24:21:58.1 		 & 0.502   		  & 26265 		   & 15.9 		   & \({0.08}_{-0.03}^{+0.04}\)      & \({7.9}_{-3.0}^{+4.0} \times{10}^{42}\)  & \({0.17}_{-0.14}^{+0.14}\) & \({1.7}_{-1.4}^{+1.4} \times {10}^{43}\) & Lobes & 2 & 5 \\
			B2.3 0848+34  & 08:51:08.44 & +34:19:20.4 & 0.697 & 27850 & 14.8 		   & \({0.09}_{-0.03}^{+0.04}\)      & \({1.9}_{-0.7}^{+0.9} \times{10}^{43}\)  & \({<0.12}\)                & \(<2.6 \times {10}^{43}\)                & Lobes & 2 & 4 \\
			B2.4 0939+22A 		   & 09:41:55.62 		  & +21:48:47.8 		 & 0.572   		  & 26291 		   & 15.7 		   & \({0.09}_{-0.03}^{+0.04}\)      & \({1.2}_{-0.4}^{+0.5} \times{10}^{43}\)  & \({0.34}_{-0.14}^{+0.15}\) & \({4.6}_{-1.9}^{+12.0} \times {10}^{43}\) & FRII & 2 & 6 \\
			B2.4 1112+23  		   & 11:15:04.89 		  & +23:25:50.2 		 &         		  & 26333 		   & 15.3 		   & \({0.06}_{-0.03}^{+0.04}\)      & 										    & \({<0.12}\)                & 											& Compact & 2 & 6 \\
			B2.3 1234+37  		   & 12:36:50.64		  & +36:55:30.1 		 &         		  & 27609 		   & 15.9 		   & \({0.13}_{-0.04}^{+0.05}\)      & 										    & \({<0.22}\)                & 											& Lobes & 2 & 3 \\
			B2.2 1334+27  		   & 13:36:41.37		  & +27:03:43.7 		 & 3.228   		  & 27617 		   & 15.9 		   & \({0.92}_{-0.10}^{+0.10}\)      & \({8.8}_{-1.0}^{+1.0} \times{10}^{45}\)  & \({<0.13}\)                & \(<1.2 \times {10}^{45}\)                & Lobes & 2 & 4 \\
			B2.2 1338+27  		   & 13:40:29.96		  & +27:22:14.6 		 &         		  & 27618 		   & 16.9 		   & \(< 0.09\) 					 & 										    & \({<0.04}\)                & 											& Lobes & 4 & 3 \\
			B2.2 1439+25  		   & 14:42:04.09		  & +25:03:30.0 		 &         		  & 27411 		   & 15.9 		   &  								 & 										    & \({0.14}_{-0.14}^{+0.14}\) & 											& Lobes & 4 & 5 \\
			B2.4 1512+23  & 15:14:15.01 & +23:28:53.2 & 0.088 & 27718 & 14.9 & \({0.16}_{-0.05}^{+0.05}\)      & \({3.1}_{-1.0}^{+1.0} \times {10}^{41}\) & \({0.10}_{-0.07}^{+0.08}\) & \({2.0}_{-1.4}^{+1.6} \times {10}^{41}\) & FRII & 2 & 6 \\
			B2.4 2054+22B & 20:56:57.55 & +22:30:11.7 &  			  & 27821 & 15.9 & \({0.06}_{-0.03}^{+0.04}\) 	 &											& \({<0.11}\)                & 											& FRII & 2 & 5 \\
			B2.2 2104+24  		   & 21:06:21.19 		  & +24:33:22.7          &         		  & 22170 		   & 16.4 		   & \({3.01}_{-0.17}^{+0.18}\)      & 										    & \({<0.29}\)                & 											& FRII & 1 & 3 \\
			B2.2 2133+27  		   & 21:35:17.61 		  & +27:16:14.1          &         		  & 26149 		   & 15.9 		   & 								 & 										    & \({0.18}_{-0.15}^{+0.15}\) & 											& Lobes & 1 & 6 \\
			B2.3 2254+35  		   & 22:56:46.03 		  & +35:40:56.8          & 0.114   		  & 22193 		   & 15.9 		   & \({0.10}_{-0.03}^{+0.04}\) 	 & \({3.4}_{-1.0}^{+1.4} \times {10}^{41}\) & \({6.75}_{-0.62}^{+0.65}\) & \({2.3}_{-0.2}^{+0.2} \times {10}^{43}\) & WAT & 4 & 6 \\
			B2.2 2328+26  		   & 23:30:34.44 		  & +27:05:21.2          &         		  & 26122 		   & 15.9 		   & \({0.12}_{-0.04}^{+0.04}\) 	 & 										    & \({<0.05}\)                & 											& Compact & 1 & 4 \\
			B2.3 2334+39  		   & 23:36:55.64 		  & +40:06:06.5          &         		  & 26342 		   & 15.9 		   & \(<0.05\) 					     & 										    & \({<0.11}\)                & 											& FRI & 2 & 6 \\
			\hline
			\hline
		\end{tabular}
	}
	\tablecomments{The flux marked with an asterisk (\({}^*\)) is extracted from a nuclear region affected by significant pileup.}
\end{center}
\end{table}

\begin{table}
	\caption{Results of nuclear spectral fits with a simple power-law model (plus Galactic absorption) for the sources with enough counts in their nuclear region (see Sect. \ref{sec:spectral}). For each source we show the power-law slope (\(\Gamma\)), the power-law normalization (Norm.), the pileup fraction estimated with the \textsc{jdpileup} model (Pileup frac.), the reduced cash statistics and degrees of freedom, and the intrinsic \(0.3-7 \text{ keV}\) flux obtained from spectral fitting (\(F_{0.3-7 \text{ keV}}\)).}\label{tab:nucl_spectra_noabs}
	\begin{center}
		\begin{tabular}{|l|ccccc|}
			\hline
			\hline
			Source Name & \(\Gamma\) & Norm. & Pileup frac. & c (d.o.f.) & \(F_{0.3-7 \text{ keV}}\)\\
			\hline
			& & \({10}^{-5}\text{ cm}^{-2}\text{ s}^{-1}\) & & & \({10}^{-13}\text{ erg}\text{ cm}^{-2}\text{ s}^{-1}\) \\
			\hline
			B2.4 0004+21 & \({0.33}_{-0.16}^{+0.16}\) & \({0.85}_{-0.16}^{+0.18}\) & - & \(0.88(456)\) & \({2.08}_{-0.58}^{+0.59}\) \\
			B2.4 0145+22 & \({0.59}_{-0.47}^{+0.23}\) & \({0.67}_{-0.33}^{+0.61}\) & - & \(0.96(313)\) & \({1.07}_{-0.57}^{+1.07}\) \\
			B2.4 0229+23 & \({1.20}_{-0.10}^{+0.10}\) & \({5.90}_{-0.59}^{+0.65}\) & - & \(1.04(510)\) & \({5.20}_{-0.78}^{+0.71}\) \\
			B2.4 0401+23 & \({0.67}_{-0.42}^{+0.41}\) & \({0.29}_{-0.11}^{+0.17}\) & - & \(1.01(329)\) & \({0.44}_{-0.25}^{+0.64}\) \\
			B2.2 0410+26 & \({1.36}_{-0.79}^{+0.83}\) & \({0.15}_{-0.09}^{+0.19}\) & - & \(1.01(312)\) & \({0.15}_{-0.10}^{+0.18}\) \\
			B2.4 0412+23 & \({1.78}_{-0.14}^{+0.14}\) & \({4.46}_{-0.55}^{+0.62}\) & - & \(1.08(450)\) & \({2.47}_{-0.18}^{+0.26}\) \\
			B2.3 0454+35 & \({1.53}_{-0.09}^{+0.09}\) & \({10.39}_{-0.94}^{+1.01}\) & - & \(1.06(510)\) & \({7.04}_{-0.79}^{+0.70}\) \\
			B2.1 0536+33B & \({1.36}_{-0.48}^{+0.51}\) & \({0.33}_{-0.13}^{+0.19}\) & - & \(0.98(331)\) & \({0.26}_{-0.10}^{+0.13}\) \\
			B2.1 0549+29 & \({2.11}_{-0.38}^{+0.39}\) & \({1.34}_{-0.41}^{+0.55}\) & - & \(0.94(343)\) & \({0.73}_{-0.30}^{+0.20}\) \\
			B2.1 0643+30 & \({1.75}_{-0.27}^{+0.27}\) & \({1.29}_{-0.29}^{+0.35}\) & - & \(1.09(366)\) & \({0.81}_{-0.25}^{+0.25}\) \\
			B2.1 0742+31 & \({2.04}_{-0.15}^{+0.07}\) & \({55.39}_{-2.92}^{+3.06}\) & \(0.21\) & \(0.95(675)\) & \({27.75}_{-1.39}^{+1.34}\) \\
			B2.2 1334+27 & \({1.84}_{-0.28}^{+0.28}\) & \({1.55}_{-0.33}^{+0.40}\) & - & \(1.02(310)\) & \({0.83}_{-0.21}^{+0.29}\) \\
			B2.2 2104+24 & \({1.53}_{-0.11}^{+0.11}\) & \({4.02}_{-0.39}^{+0.42}\) & - & \(1.06(470)\) & \({2.63}_{-0.23}^{+0.37}\) \\
			B2.3 2254+35 & \({2.73}_{-0.92}^{+1.18}\) & \({0.40}_{-0.24}^{+0.42}\) & - & \(1.01(327)\) & \({0.31}_{-0.17}^{+0.21}\) \\
			B2.2 2328+26 & \({0.97}_{-1.07}^{+1.07}\) & \({0.16}_{-0.09}^{+0.18}\) & - & \(0.93(310)\) & \({0.17}_{-0.05}^{+0.10}\) \\
			\hline
			\hline
		\end{tabular}
	\end{center}
\end{table}

\begin{table}
	\caption{Same as Table \ref{tab:nucl_spectra_noabs}, but for a fitting model that includes an additional absorbing component, and with a power-law slope fixed to \(1.8\). For each source we show the additional absorbing column density (\(N_{H,\text{int}}\)), the power-law normalization (Norm.), the pileup fraction estimated with the \textsc{jdpileup} model (Pileup frac.), the reduced cash statistics and degrees of freedom, and the intrinsic \(0.3-7 \text{ keV}\) flux obtained from spectral fitting (\(F_{0.3-7 \text{ keV}}\)).}\label{tab:nucl_spectra_abs}
	\begin{center}
		\begin{tabular}{|l|ccccc|}
			\hline
			\hline
			Source Name & \(N_{H,\text{int}}\) & Norm. & Pileup frac. & c (d.o.f.) & \(F_{0.3-7 \text{ keV}}\)\\
			\hline
			& \({10}^{22} \text{ cm}^{-2}\) & \({10}^{-5}\text{ cm}^{-2}\text{ s}^{-1}\) & & & \({10}^{-13}\text{ erg}\text{ cm}^{-2}\text{ s}^{-1}\) \\
			\hline
			B2.4 0004+21 & \({2.14}_{-0.31}^{+0.33}\) & \({7.95}_{-0.83}^{+0.91}\) & - & \(0.87(456)\) & \({4.49}_{-0.58}^{+0.39}\) \\
			B2.4 0145+22 & \({5.07}_{-2.04}^{+1.88}\) & \({5.70}_{-1.26}^{+1.31}\) & - & \(0.96(313)\) & \({3.15}_{-0.10}^{+0.13}\) \\
			B2.4 0229+23 & \({24.02}_{-4.80}^{+4.96}\) & \({13.84}_{-1.05}^{+1.12}\) & - & \(1.06(510)\) & \({7.65}_{-0.60}^{+0.60}\) \\
			B2.4 0401+23 & \({2.10}_{-0.70}^{+0.78}\) & \({1.70}_{-0.38}^{+0.46}\) & - & \(1.00(329)\) & \({0.95}_{-0.27}^{+0.20}\) \\
			B2.2 0410+26 & \(<2.29\) & \({0.31}_{-0.13}^{+0.20}\) & - & \(1.01(312)\) & \({0.17}_{-0.06}^{+0.08}\) \\
			B2.4 0412+23 & \(<0.05\) & \({4.53}_{-0.30}^{+0.32}\) & - &  \(1.08(450)\) & \({2.52}_{-0.10}^{+0.12}\) \\
			B2.3 0454+35 & \({0.21}_{-0.08}^{+0.09}\) & \({14.71}_{-0.86}^{+0.91}\) & - & \(1.07(510)\) & \({8.12}_{-0.42}^{+0.50}\) \\
			B2.1 0536+33B & \({0.63}_{-0.43}^{+0.55}\) & \({0.66}_{-0.18}^{+0.54}\) & - & \(0.98(331)\) & \({0.37}_{-0.09}^{+0.11}\) \\
			B2.1 0549+29 & \(<0.16\) & \({1.03}_{-0.16}^{+0.18}\) & - & \(0.94(343)\) & \({0.59}_{-0.09}^{+0.09}\) \\
			B2.1 0643+30 & \(<0.23\) & \({1.37}_{-0.19}^{+0.26}\) & - & \(1.09(366)\) & \({0.79}_{-0.20}^{+0.12}\) \\
			B2.1 0742+31 & \(<0.02\) & \({53.83}_{-2.84}^{+3.12}\) & 0.14 & \(0.95(675)\) & \({29.76}_{-1.35}^{+1.13}\) \\
			B2.2 1334+27 & \(<13.99\) & \({1.62}_{-0.27}^{+0.36}\) & - & \(1.02(310)\) & \({0.88}_{-0.13}^{+0.12}\) \\
			B2.2 2104+24 & \({0.21}_{-0.07}^{+0.07}\) & \({5.80}_{-0.44}^{+0.47}\) & - & \(1.04(470)\) & \({3.17}_{-0.24}^{+0.20}\) \\
			B2.3 2254+35 & \(<0.49\) & \({0.20}_{-0.06}^{+0.08}\) & - & \(1.02(327)\) & \({0.11}_{-0.03}^{+0.03}\) \\
			B2.2 2328+26 & \(<1.56\) & \({0.41}_{-0.18}^{+0.30}\) & - & \(0.93(310)\) & \({0.22}_{-0.03}^{+0.04}\) \\	
			\hline
			\hline
		\end{tabular}
	\end{center}
\end{table}

\begin{table}
	\caption{Results of spectral fits of the diffuse X-ray emission with a thermal plasma model plus Galactic absorption (see Sect. \ref{sec:spectral}). For each source we show the thermal plasma temperature (\(k T\)), the thermal plasma normalization (Norm.), the reduced cash statistics and degrees of freedom, and the intrinsic \(0.3-7 \text{ keV}\) flux obtained from spectral fitting (\(F_{0.3-7 \text{ keV}}\)).}\label{tab:thermal_spectra_radio}
	\begin{center}
		\begin{tabular}{|l|cccc|}
			\hline
			\hline
			Source Name & \(k T\) & Norm. & c (d.o.f.) & \(F_{0.3-7 \text{ keV}}\)\\
			\hline
			& keV & \({10}^{-5}\text{ cm}^{-5}\) & & \({10}^{-13}\text{ erg}\text{ cm}^{-2}\text{ s}^{-1}\) \\
			\hline
			B2.1 0302+31 & \({2.45}_{-0.47}^{+0.97}\) & \({21.78}_{-2.62}^{+2.58}\) & \(1.02(640)\) & \({2.53}_{-0.55}^{+0.49}\) \\
			B2.1 0742+31  & \({1.93}_{-0.34}^{+0.56}\) & \({40.55}_{-7.33}^{+7.82}\) & \(0.96(607)\) & \({1.81}_{-0.15}^{+0.23}\) \\
			B2.3 2254+35  & \({2.47}_{-0.32}^{+0.39}\) & \({58.38}_{-4.19}^{+4.26}\) & \(1.04(712)\) & \({5.34}_{-0.38}^{+0.36}\) \\
			\hline
			\hline
		\end{tabular}
	\end{center}
\end{table}

\begin{table}
	\caption{Results of spectral fits of the diffuse X-ray emission with a thermal plasma model (plus Galactic absorption), excluding the regions spatially associated with extended radio structures (see Sect. \ref{sec:spectral}). For each source we show the thermal plasma temperature (\(k T\)), the thermal plasma normalization (Norm.), the reduced cash statistics and degrees of freedom, and the intrinsic \(0.3-7 \text{ keV}\) flux obtained from spectral fitting (\(F_{0.3-7 \text{ keV}}\)).}\label{tab:thermal_spectra_noradio}
	\begin{center}
		\begin{tabular}{|l|cccc|}
			\hline
			\hline
			Source Name & \(k T\) & Norm. & c (d.o.f.) & \(F_{0.3-7 \text{ keV}}\)\\
			\hline
			& keV & \({10}^{-5}\text{ cm}^{-5}\) & & \({10}^{-13}\text{ erg}\text{ cm}^{-2}\text{ s}^{-1}\) \\
			\hline
			B2.1 0302+31 & \({2.46}_{-0.54}^{+1.30}\) & \({12.81}_{-2.03}^{+1.98}\) & \(0.94(587)\) & \({1.45}_{-0.27}^{+0.26}\) \\
			B2.1 0742+31  & \({1.93}_{-0.67}^{+5.58}\) & \({10.24}_{-6.10}^{+5.14}\) & \(0.99(528)\) & \({0.48}_{-0.15}^{+0.12}\) \\
			B2.3 2254+35  & \({2.45}_{-0.39}^{+0.54}\) & \({32.21}_{-2.99}^{+3.09}\) & \(1.02(648)\) & \({2.91}_{-0.19}^{+0.20}\) \\
			\hline
			\hline
		\end{tabular}
	\end{center}
\end{table}

\begin{figure}
	\centering
\includegraphics[scale=0.29]{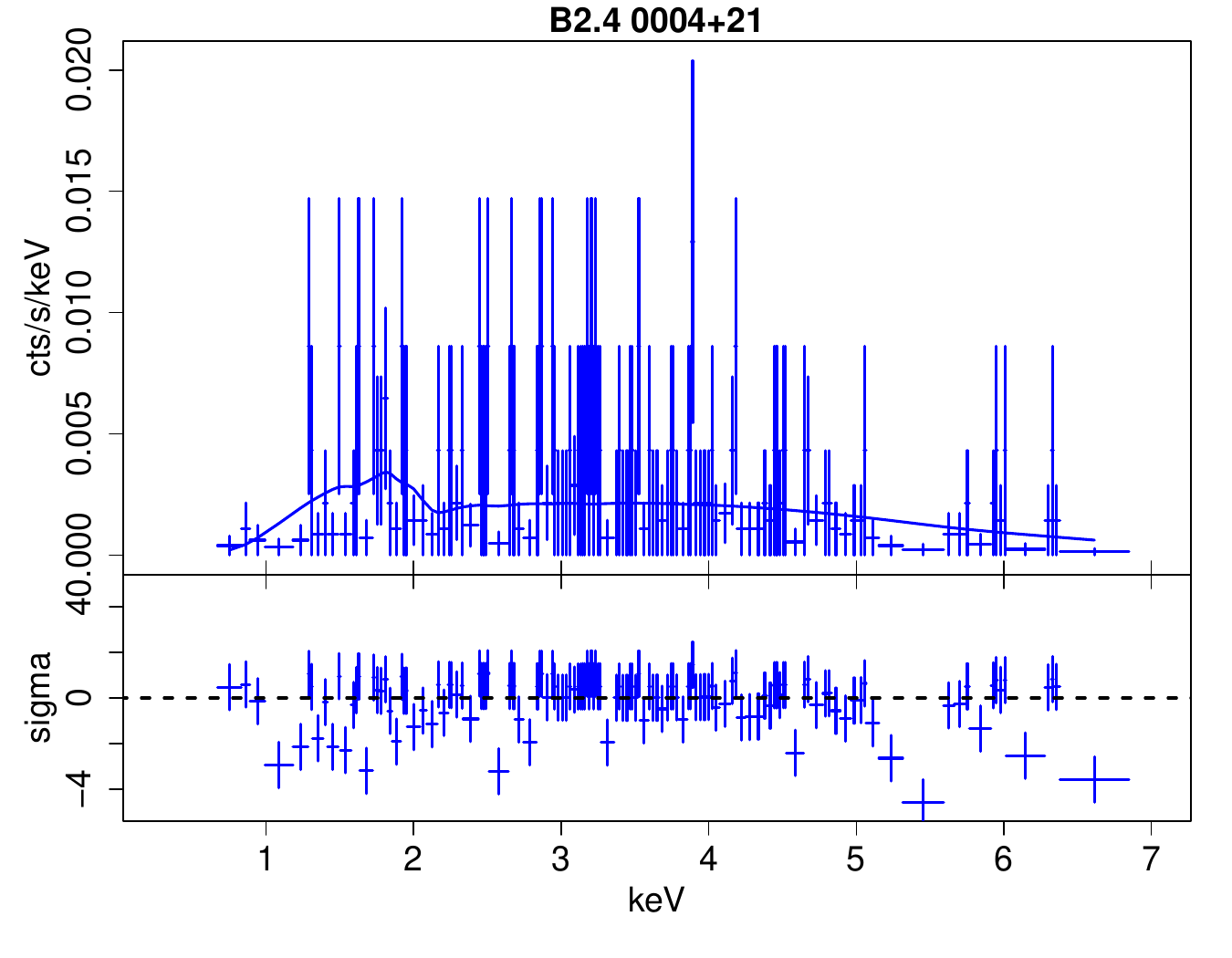}
\includegraphics[scale=0.29]{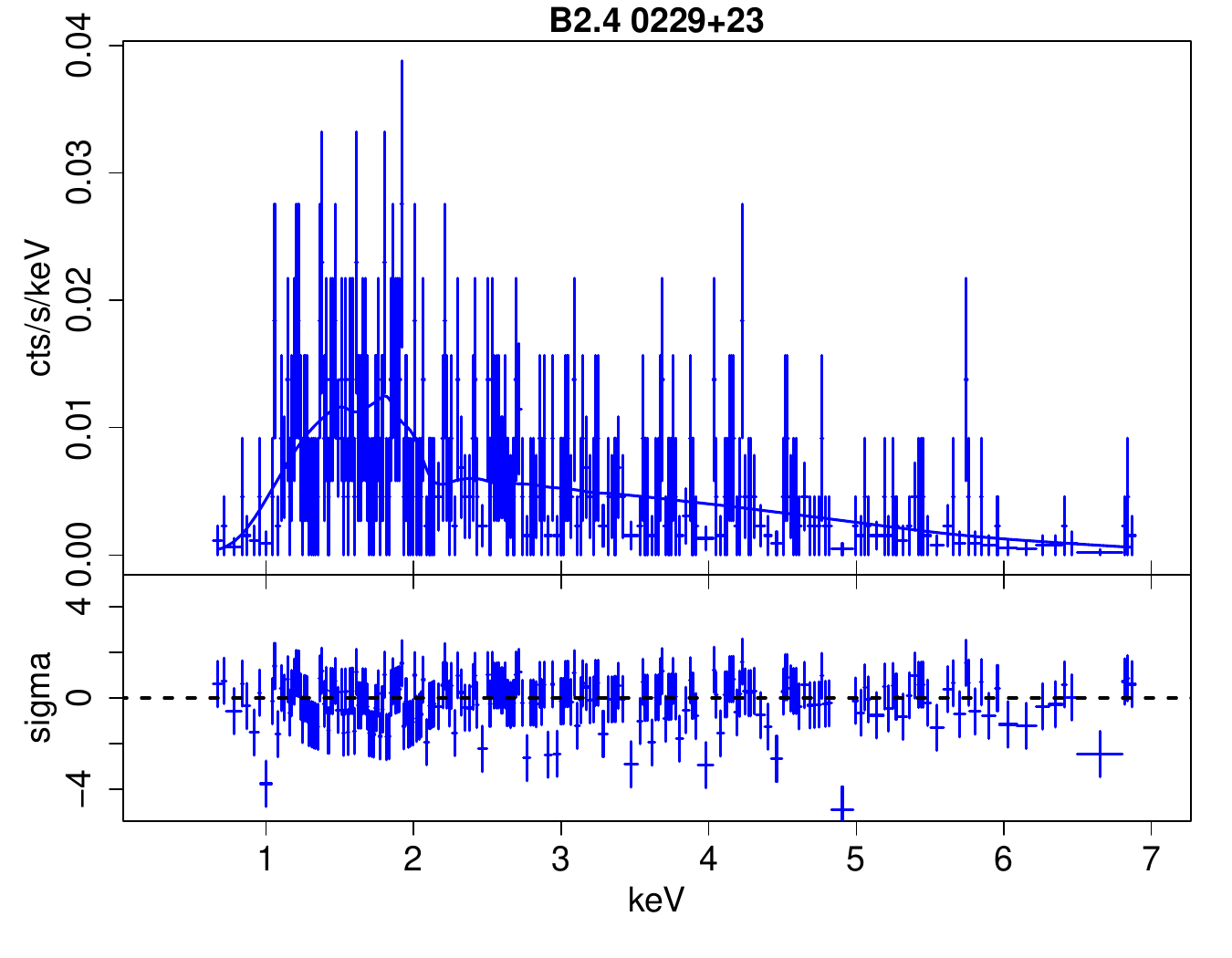}
\includegraphics[scale=0.29]{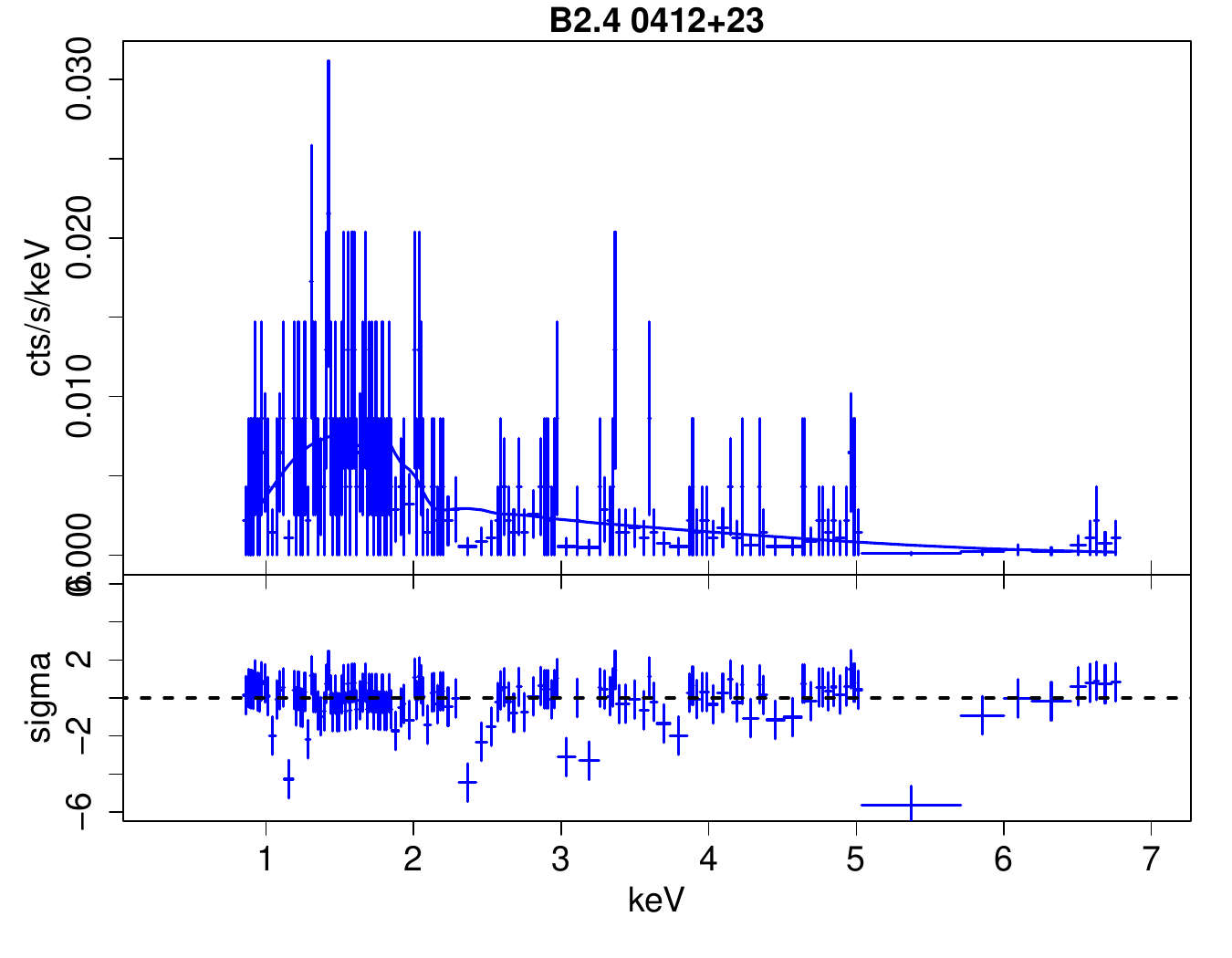}
\includegraphics[scale=0.29]{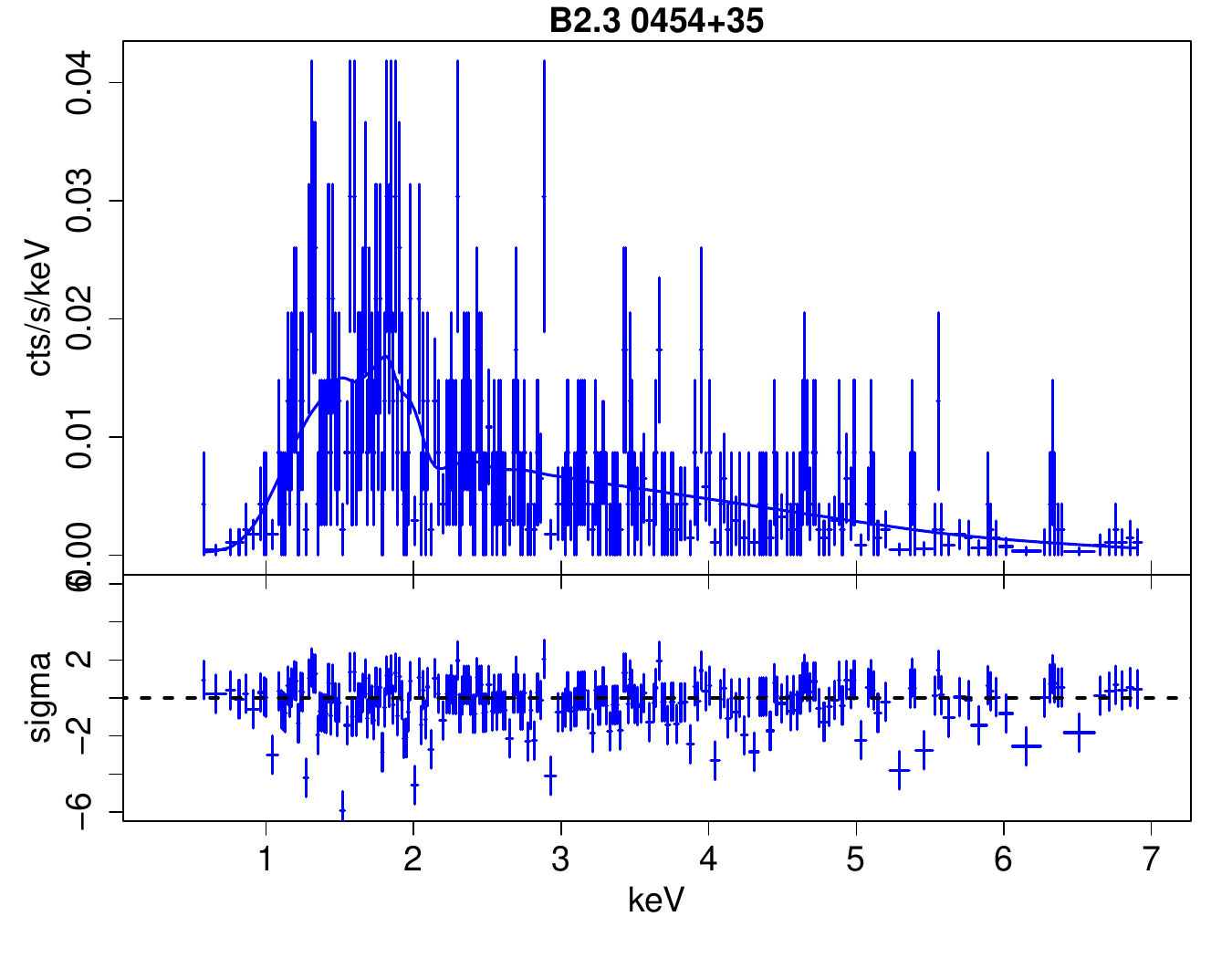}
\includegraphics[scale=0.29]{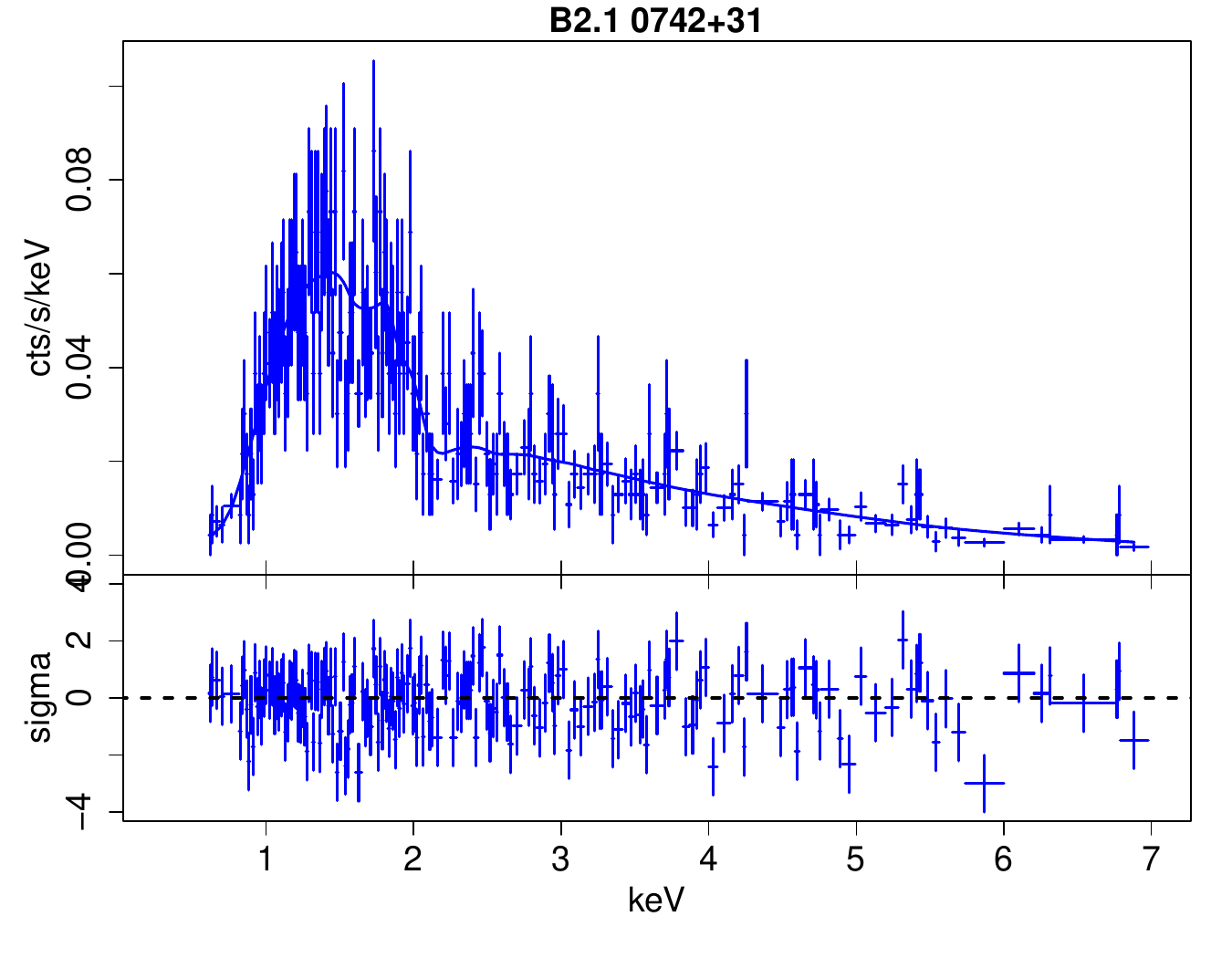}
\includegraphics[scale=0.29]{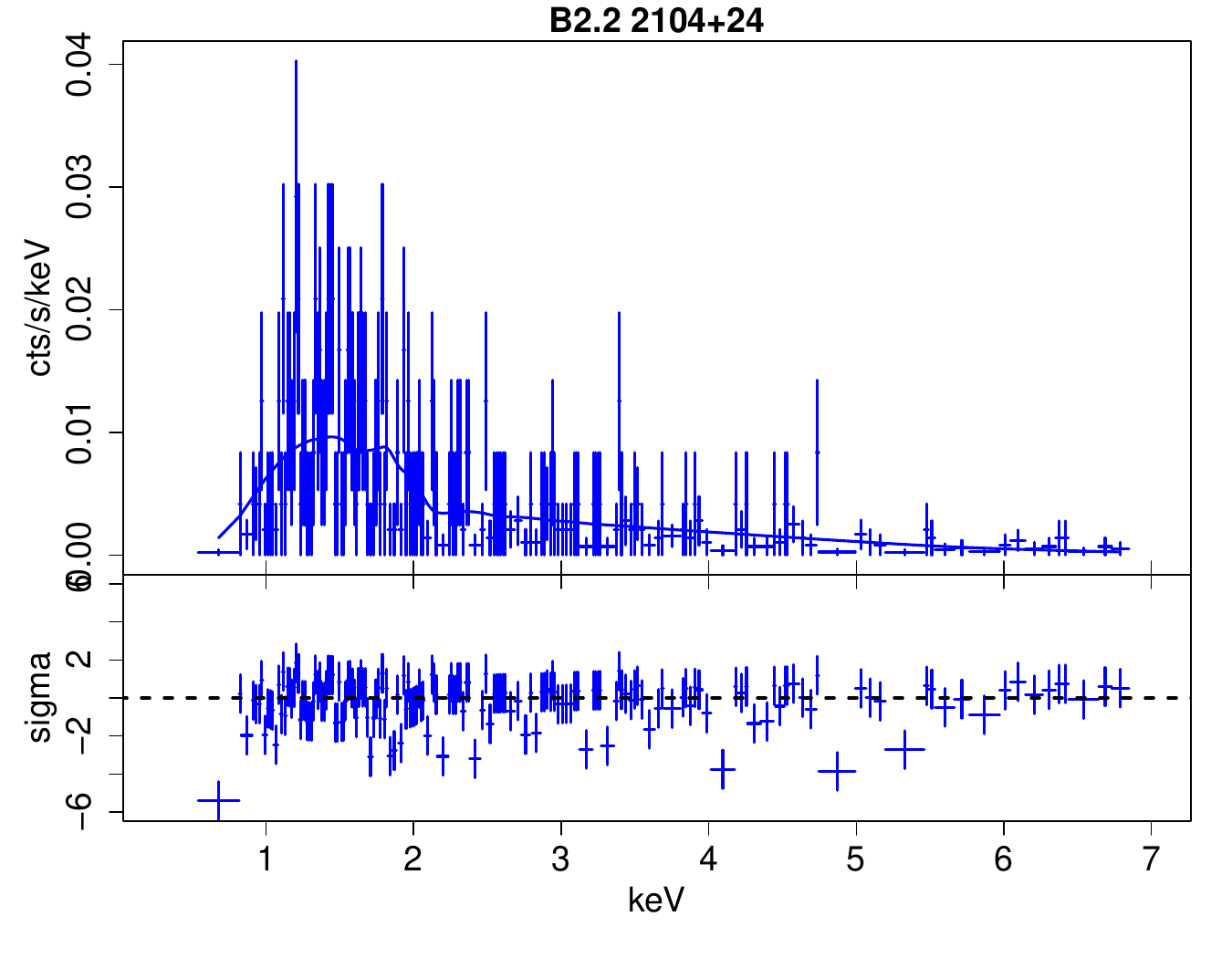}
\includegraphics[scale=0.29]{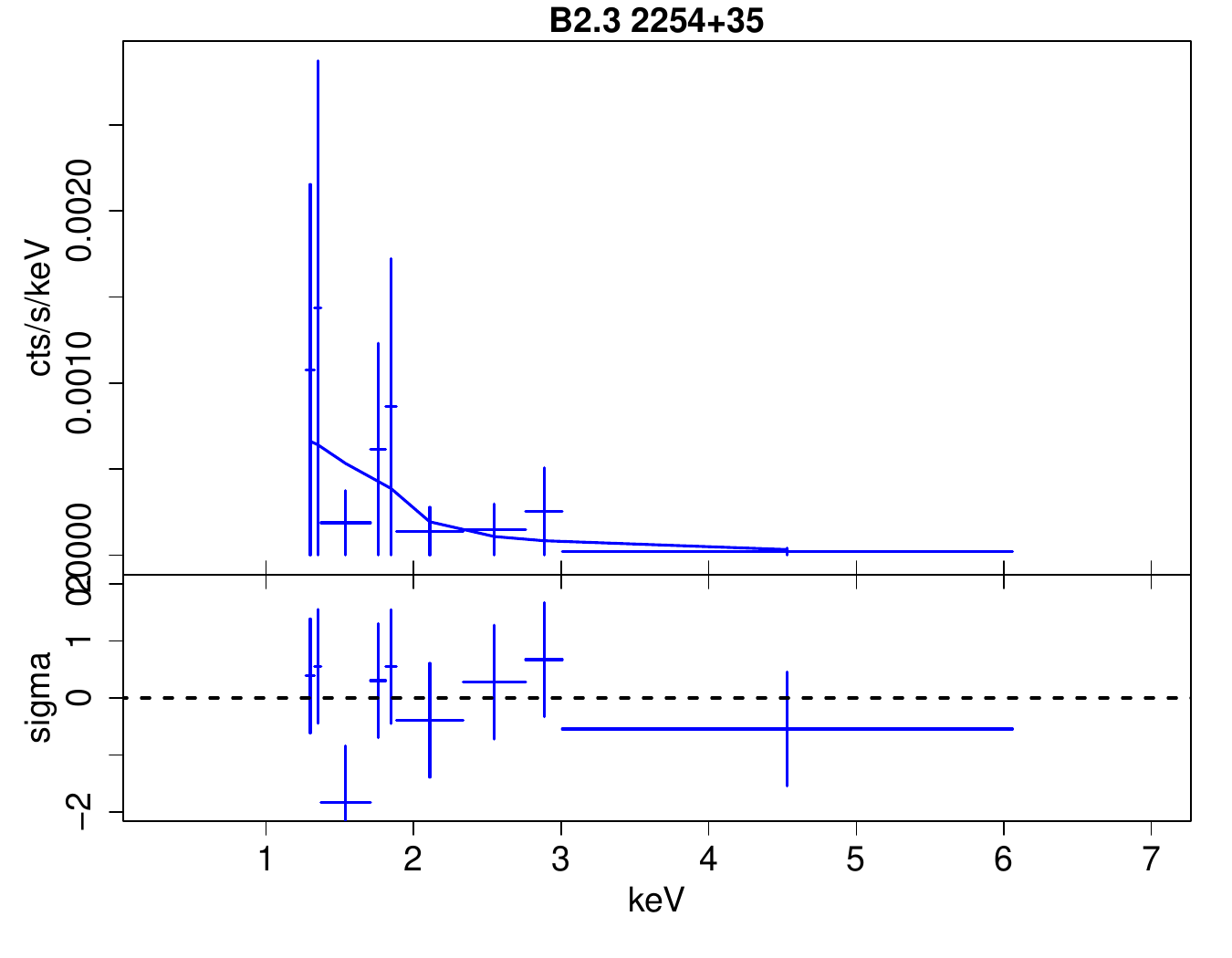}
\includegraphics[scale=0.29]{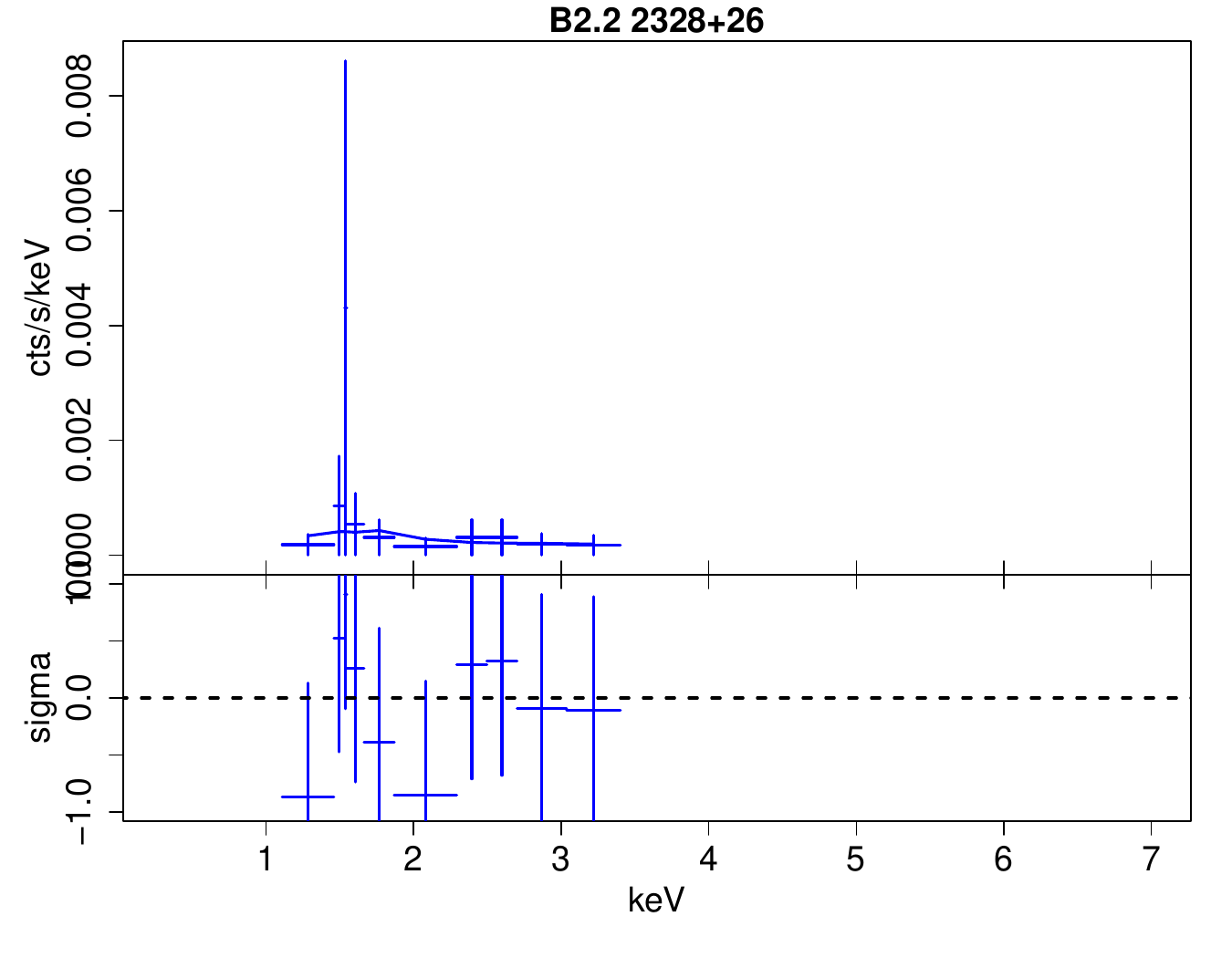}
	\caption{Nuclear spectral fits presented in Table \ref{tab:nucl_spectra_noabs} with a power-law plus Galactic absorption model. For each source, in the upper part of the panel, we show the extracted spectrum with blue crosses and the best fit model with a blue line, while in lower part of the panel we show the residuals.}
\label{fig:nucl_spectra_noabs}
\end{figure}

\begin{figure*}
	\figurenum{6}
	\centering
\includegraphics[scale=0.29]{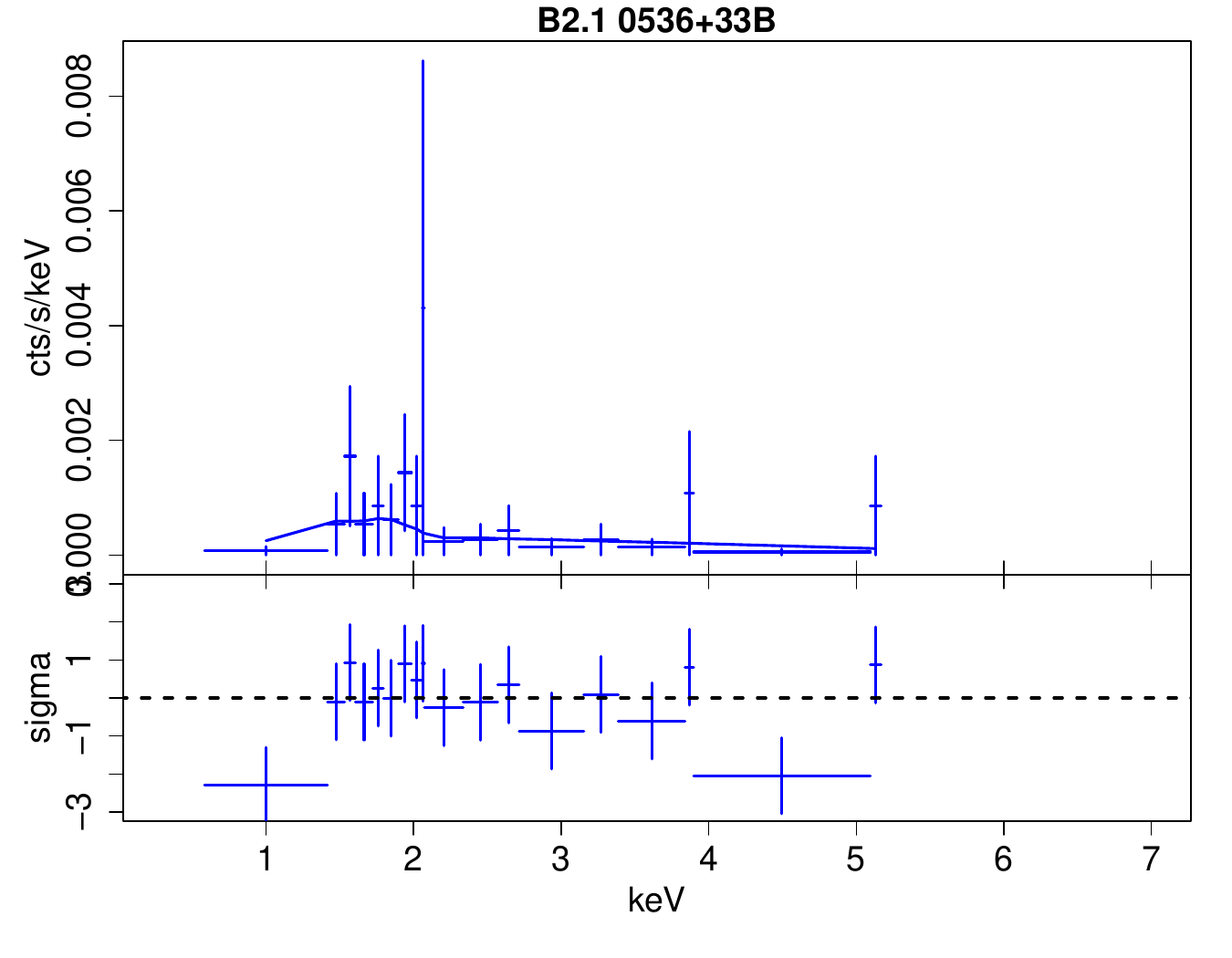}
\includegraphics[scale=0.29]{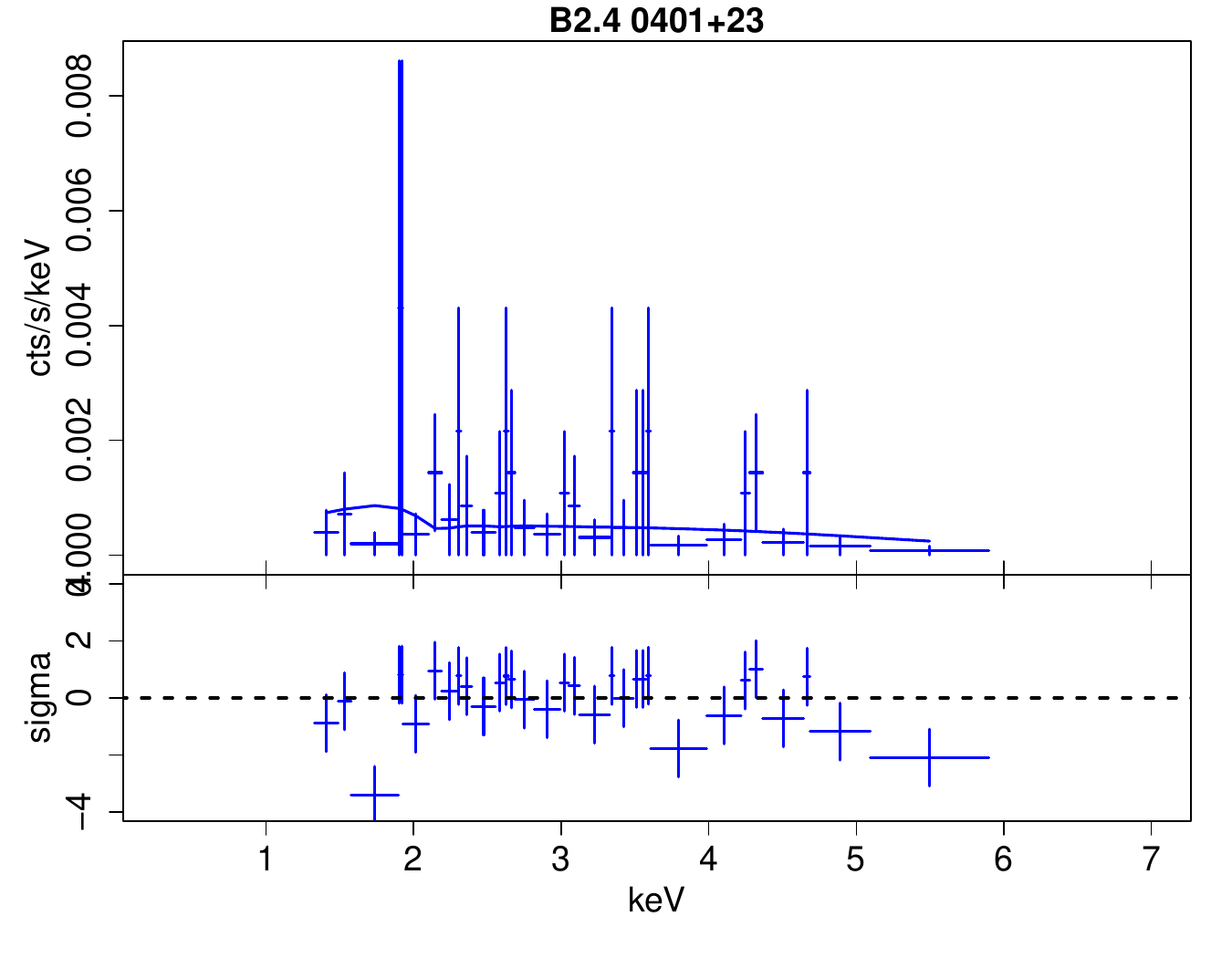}
\includegraphics[scale=0.29]{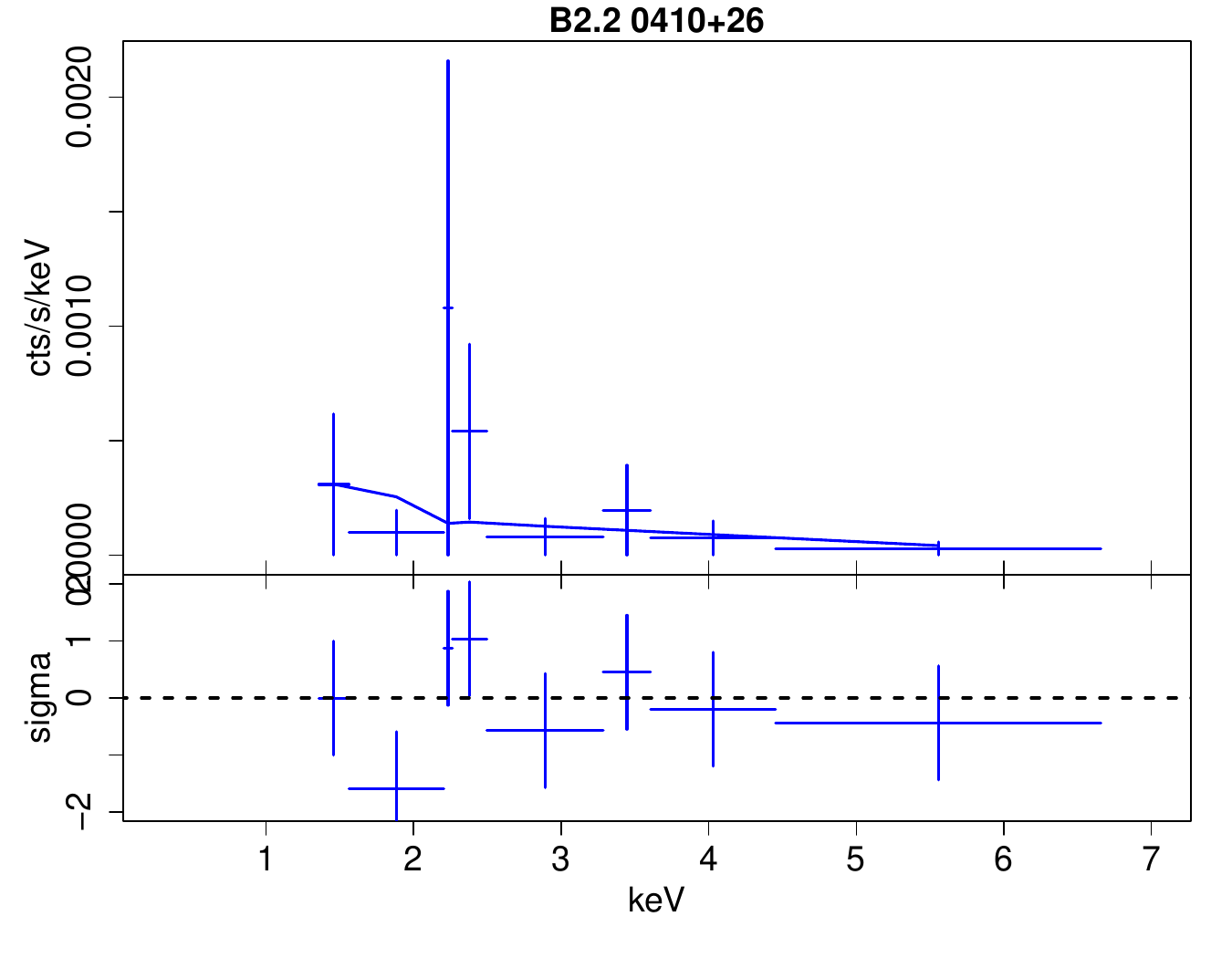}
\includegraphics[scale=0.29]{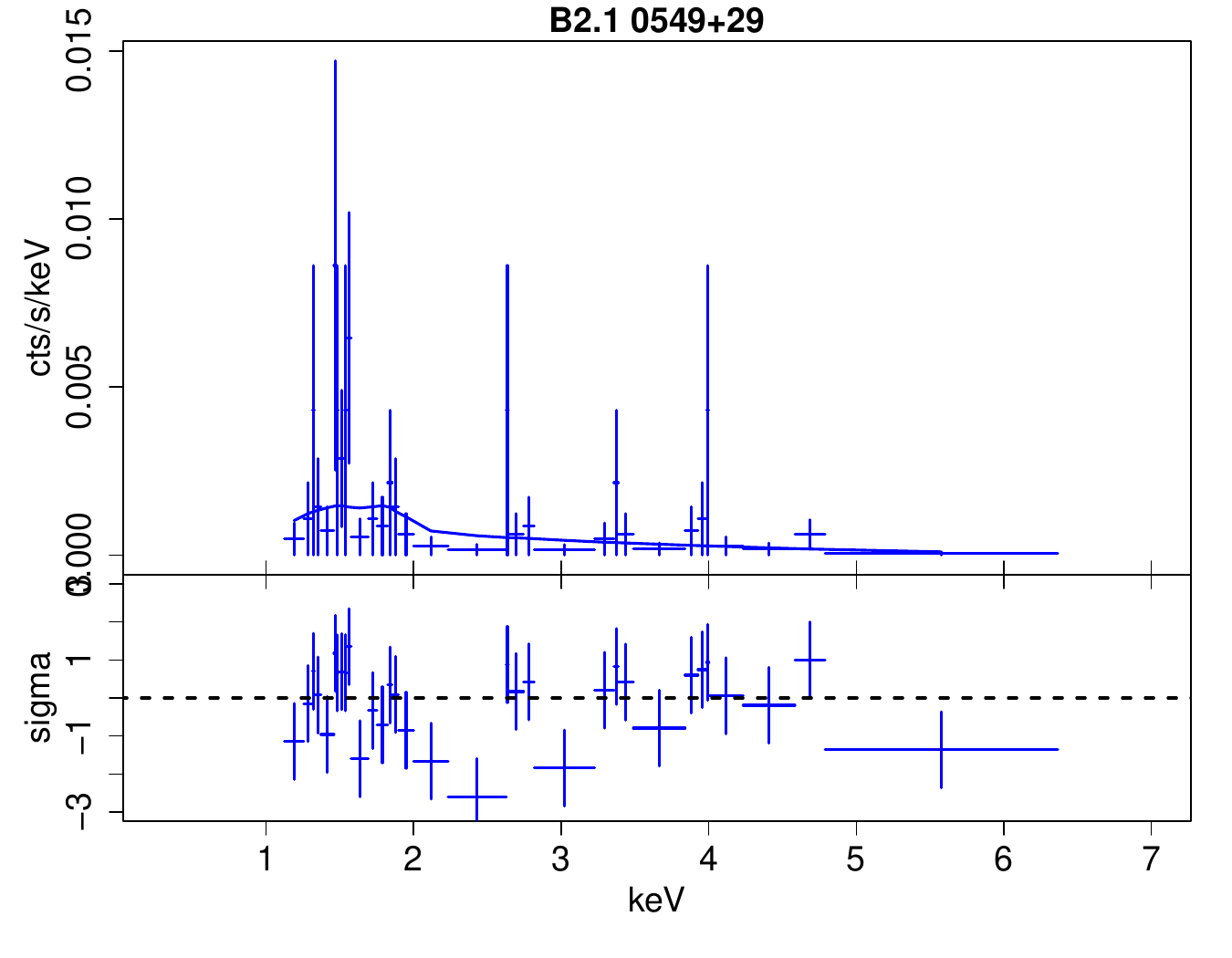}
\includegraphics[scale=0.29]{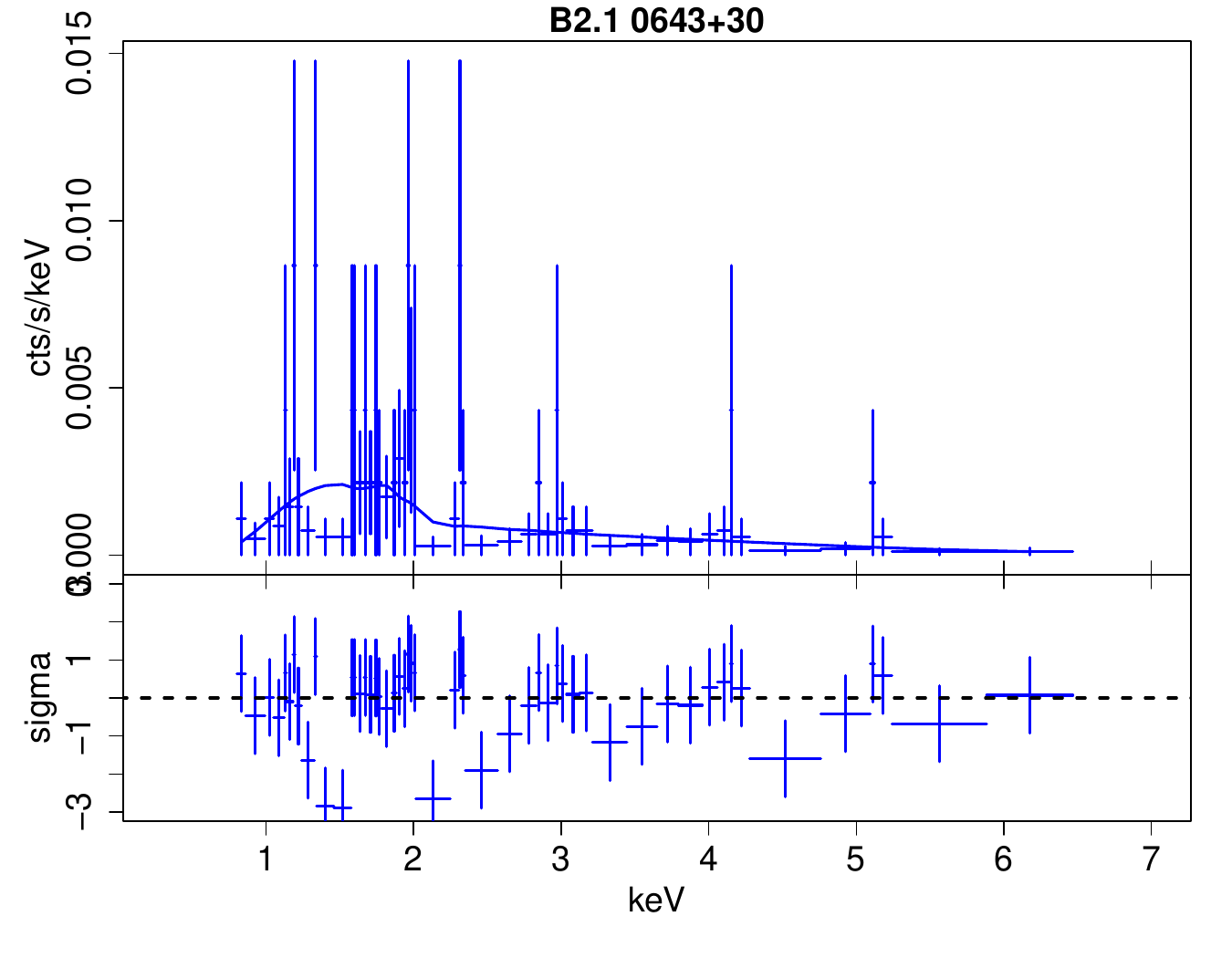}
\includegraphics[scale=0.29]{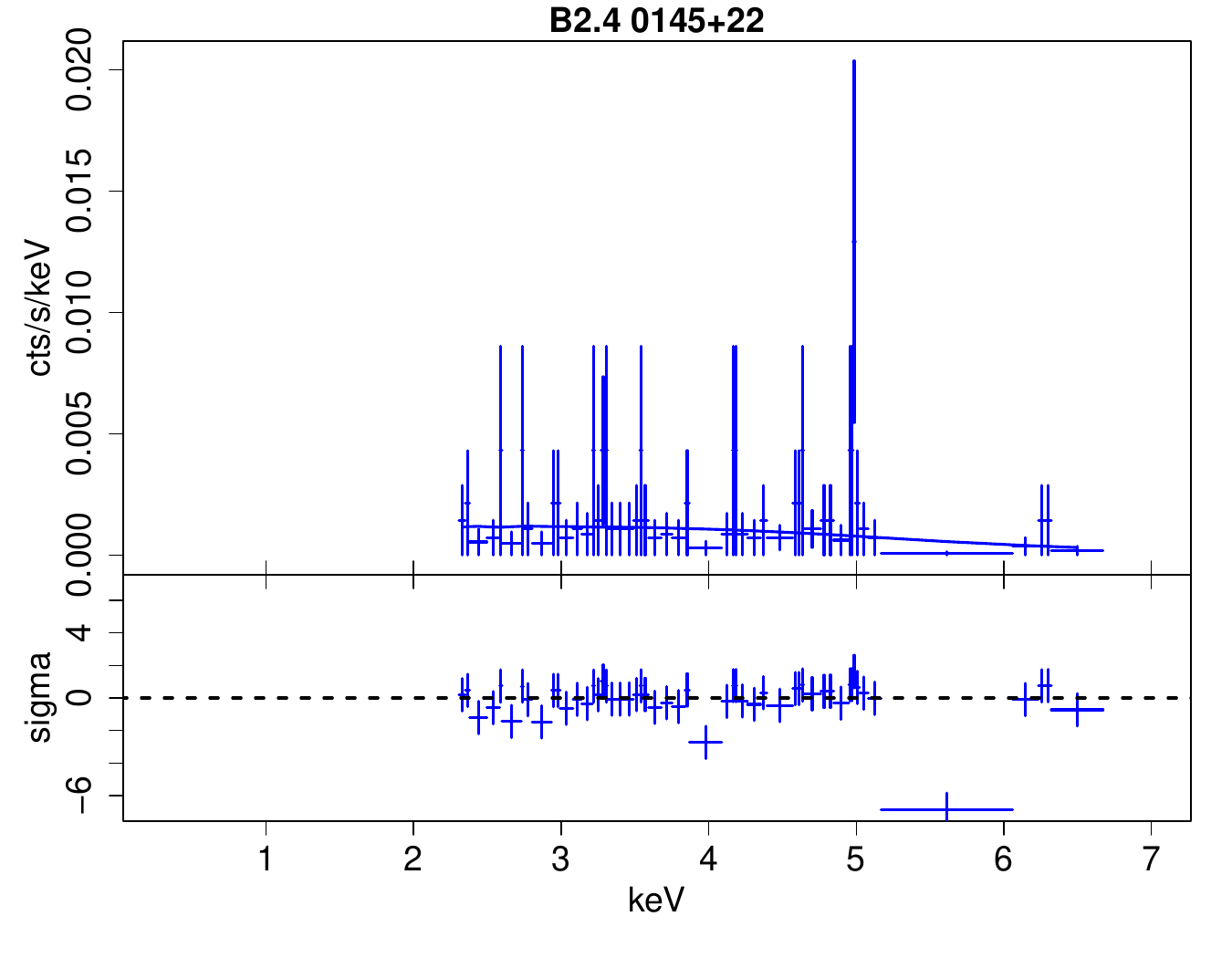}
\includegraphics[scale=0.29]{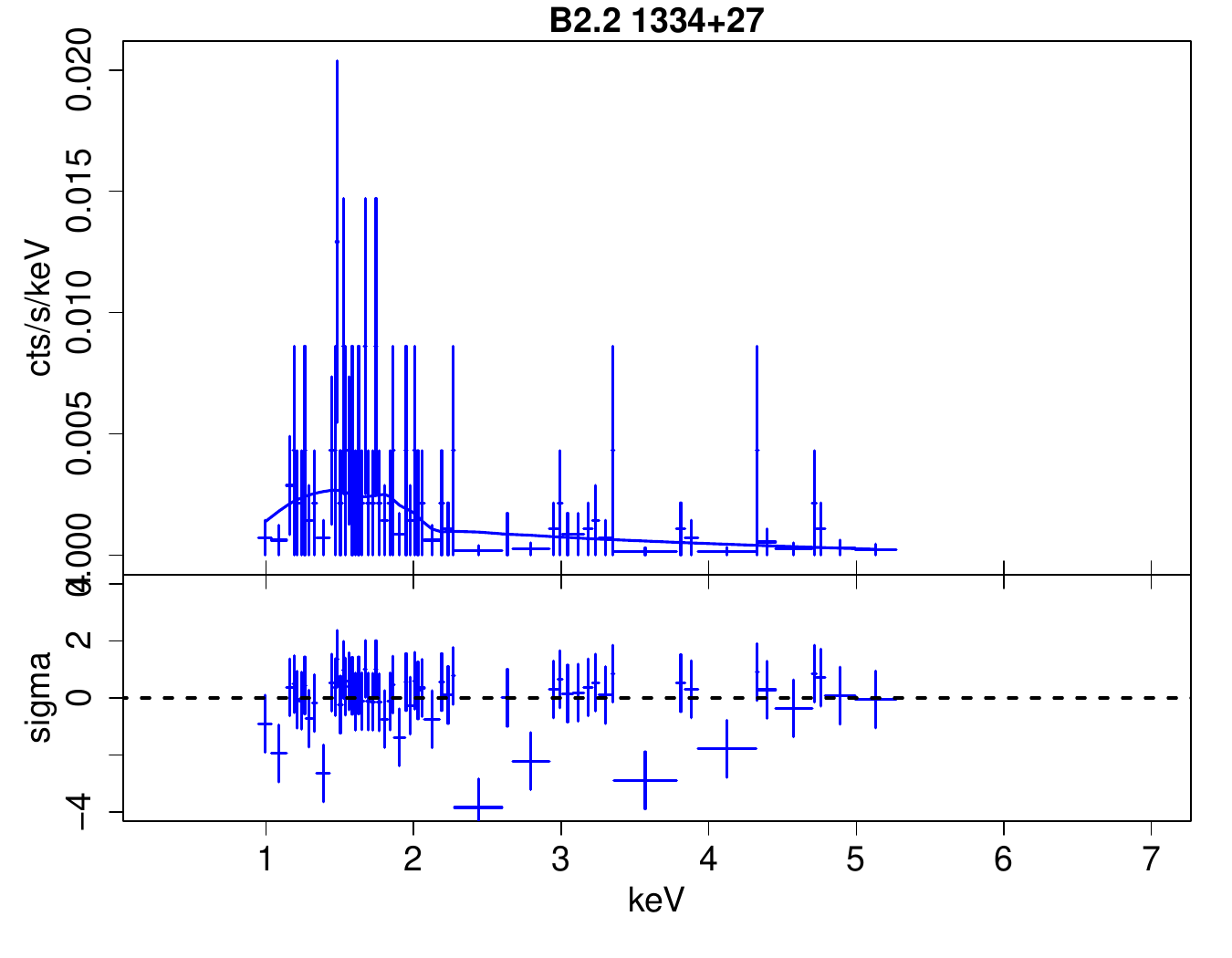}
	\caption{Continued.}
\end{figure*}

\begin{figure}
	\centering
\includegraphics[scale=0.29]{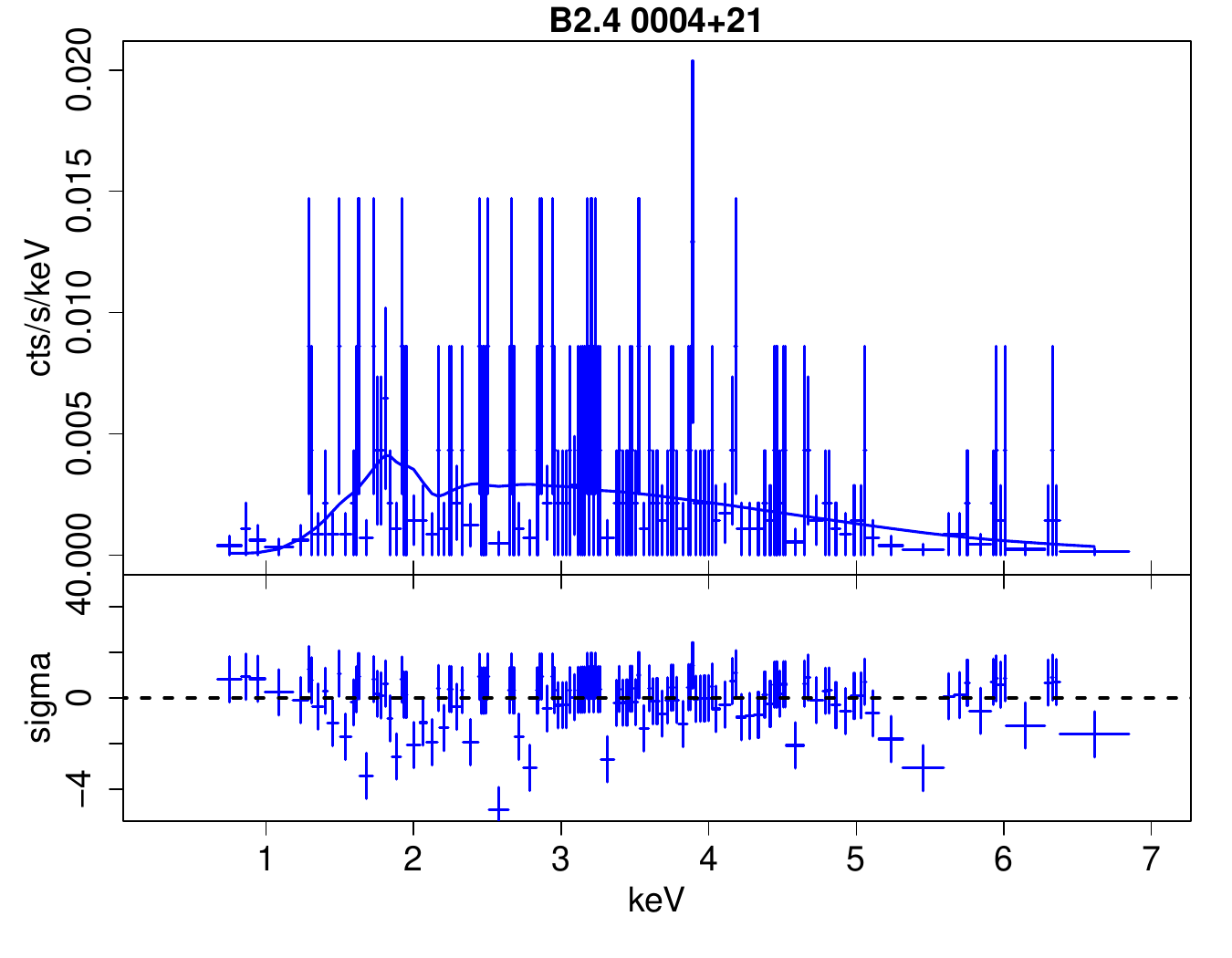}
\includegraphics[scale=0.29]{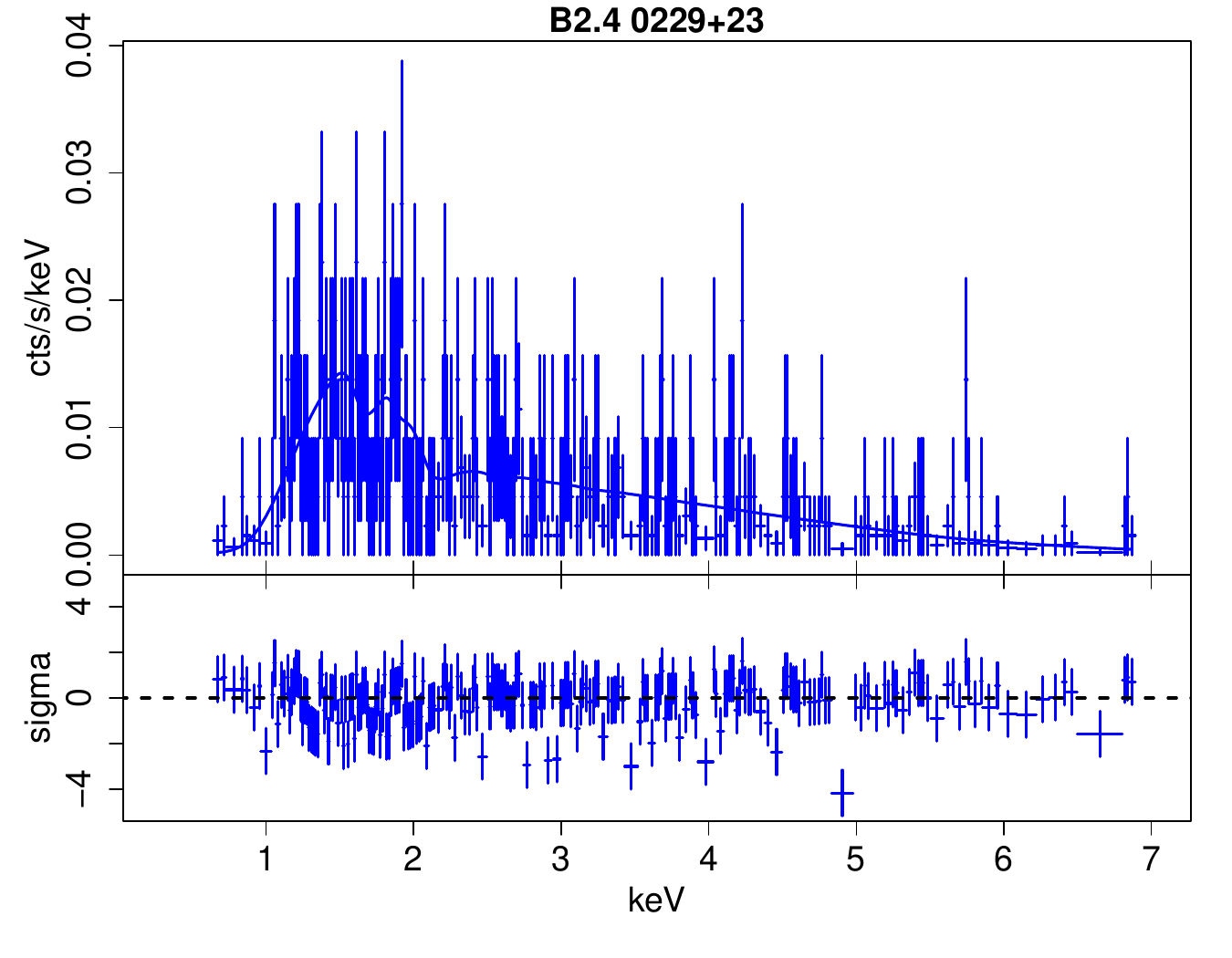}
\includegraphics[scale=0.29]{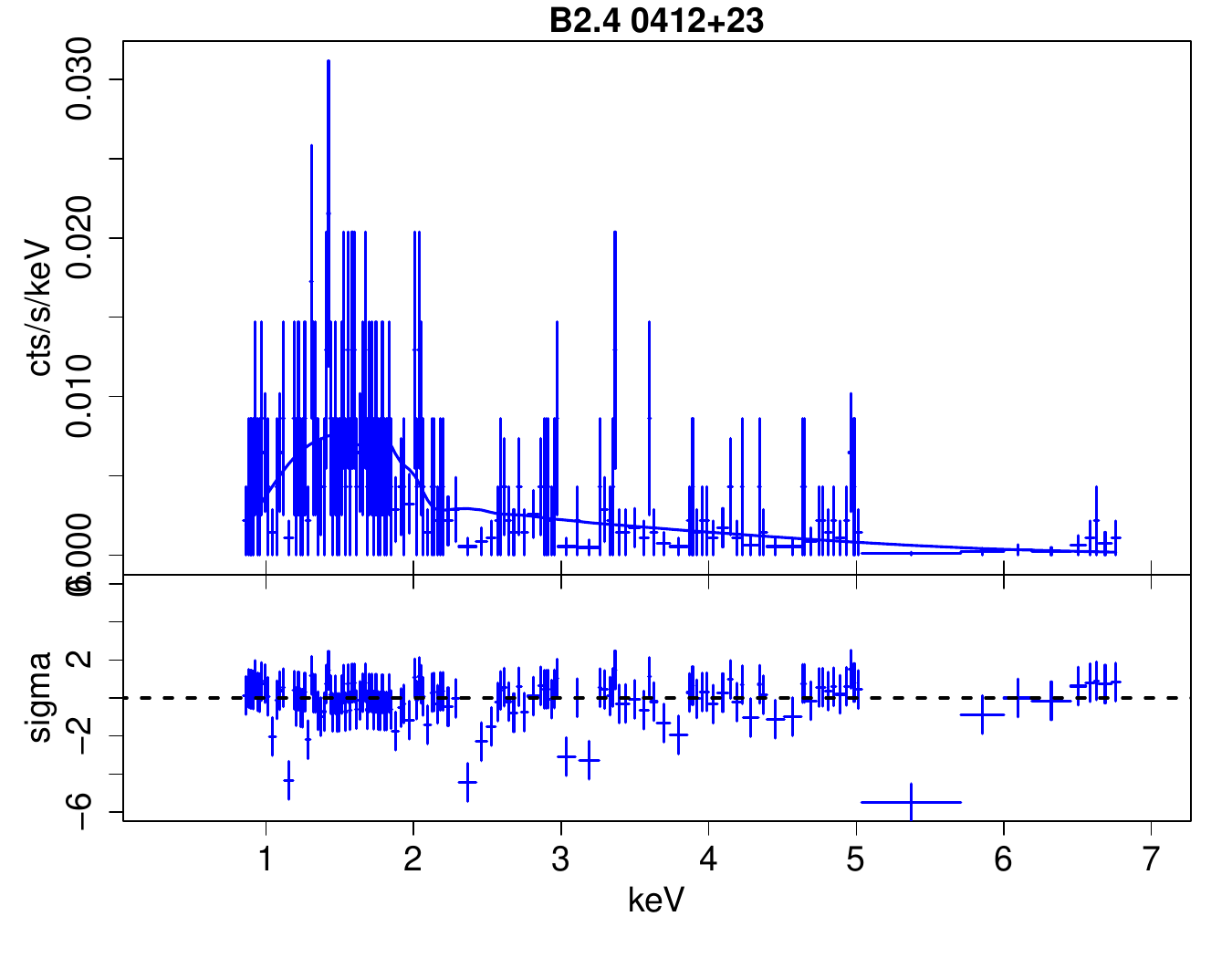}
\includegraphics[scale=0.29]{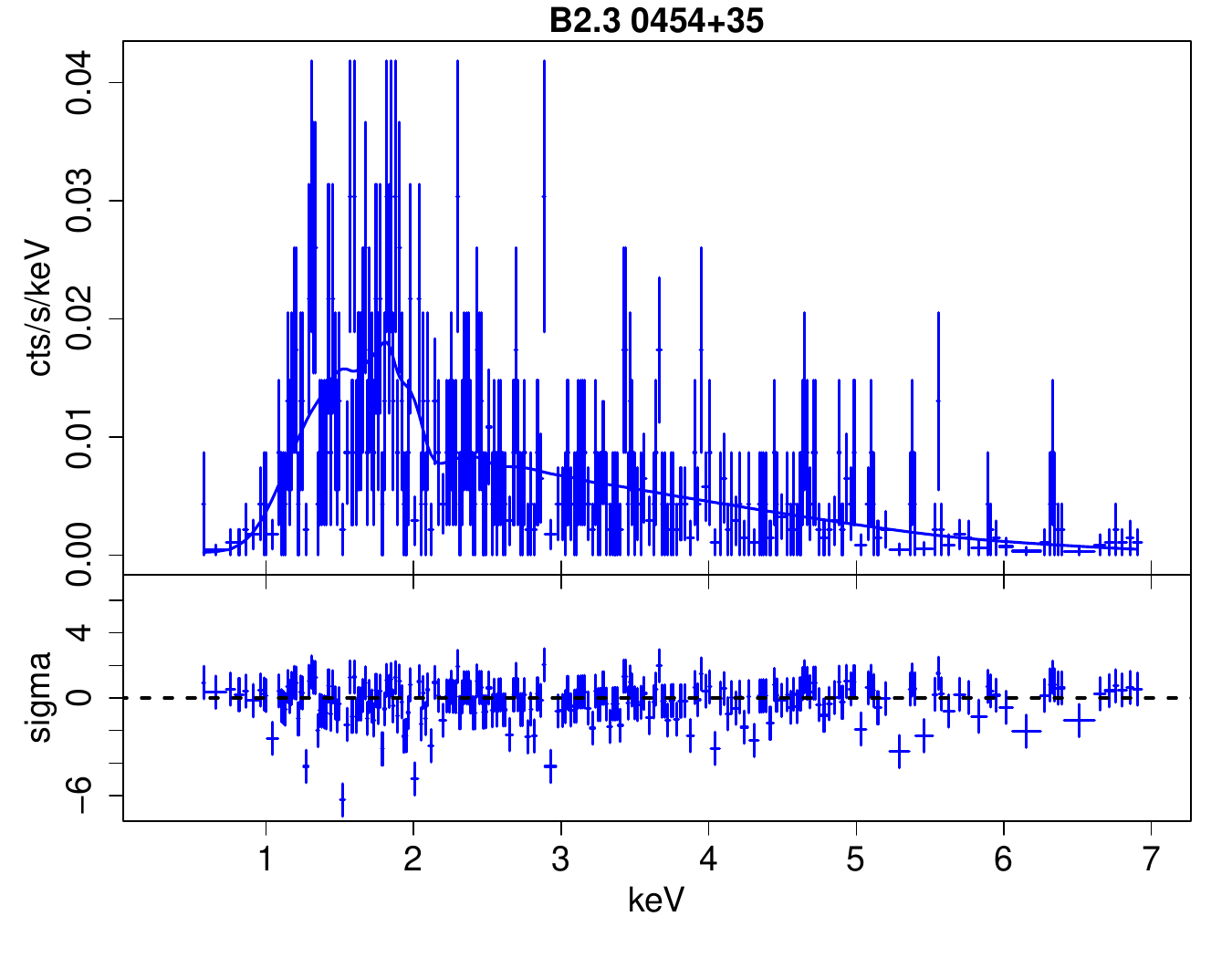}
\includegraphics[scale=0.29]{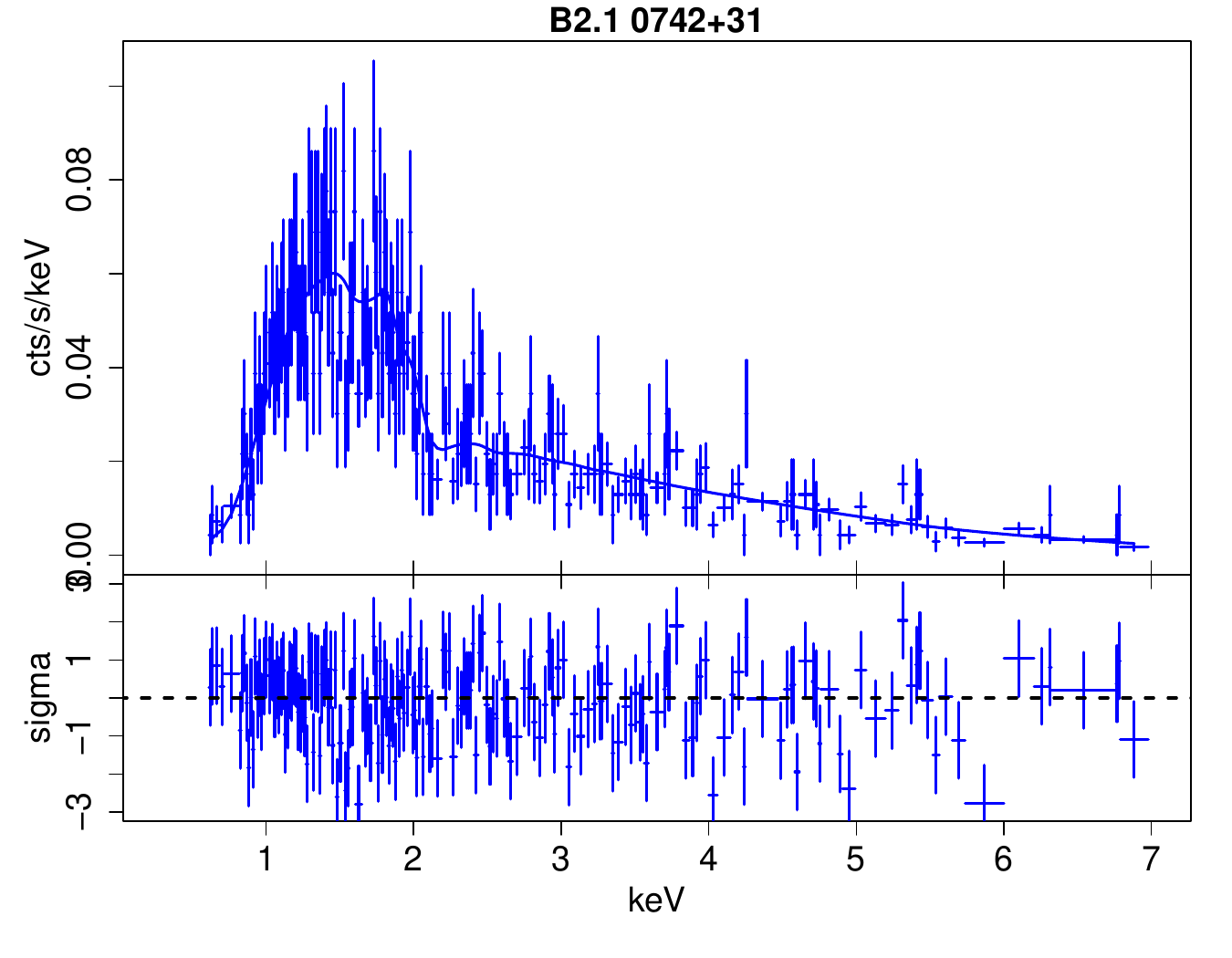}
\includegraphics[scale=0.29]{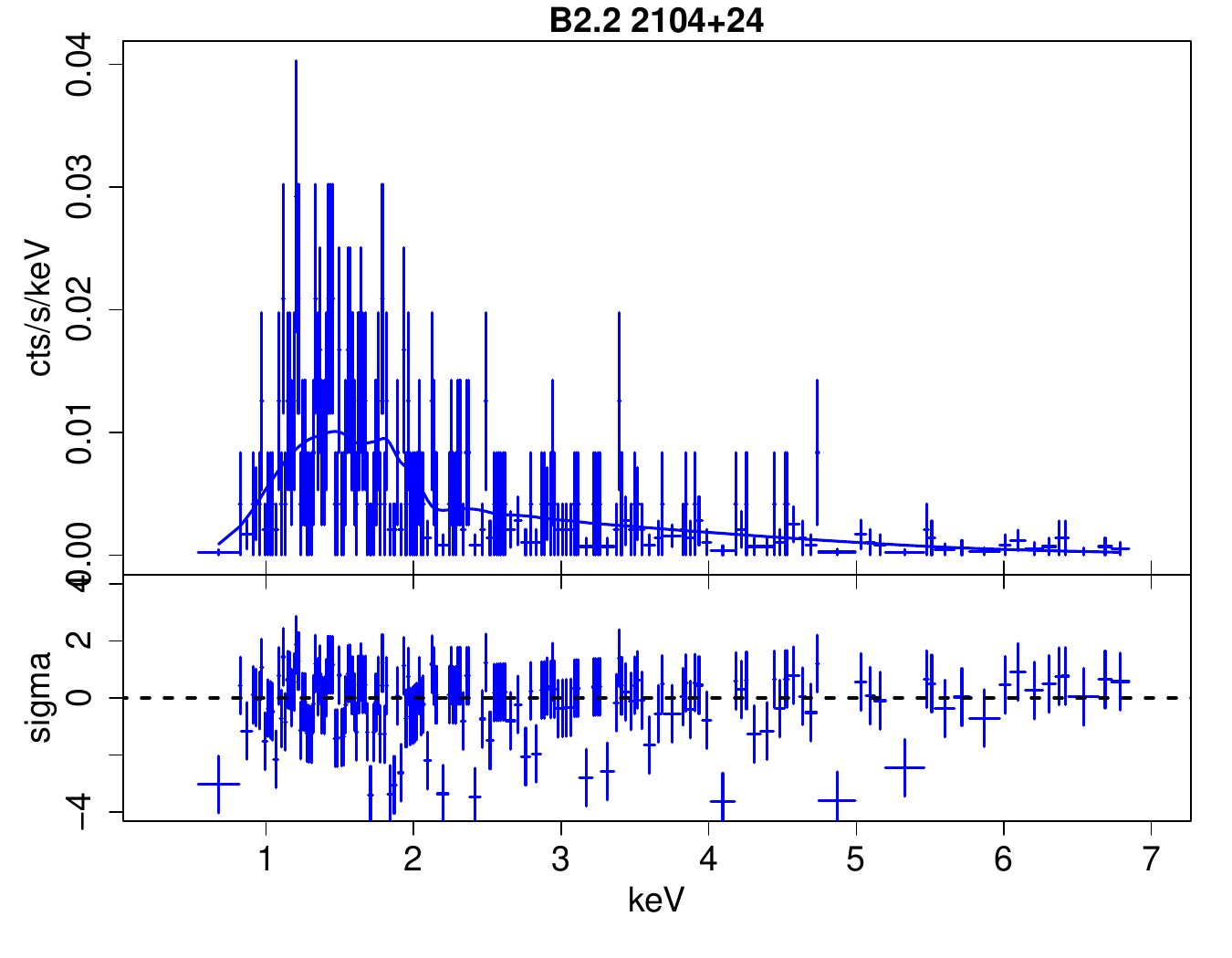}
\includegraphics[scale=0.29]{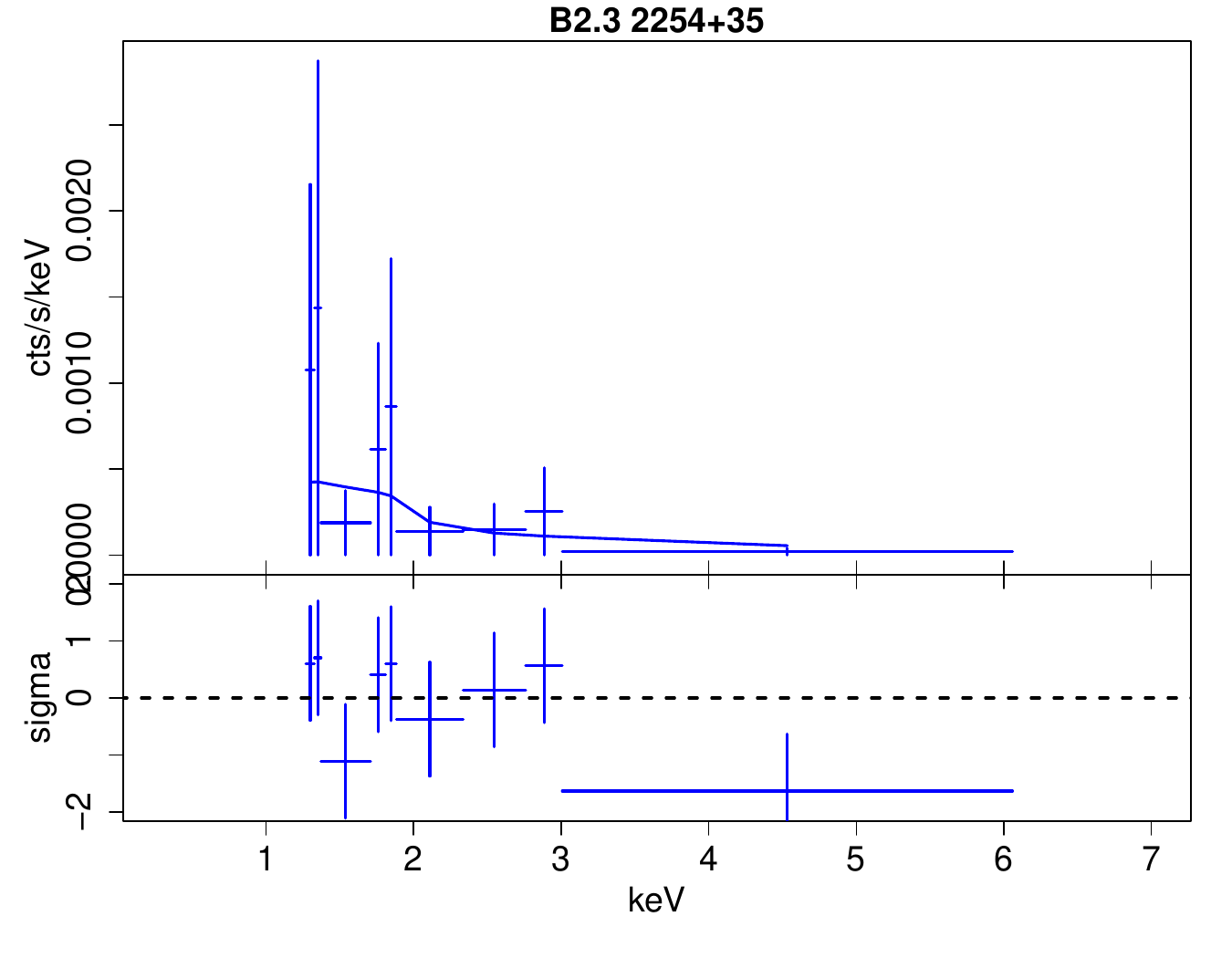}
\includegraphics[scale=0.29]{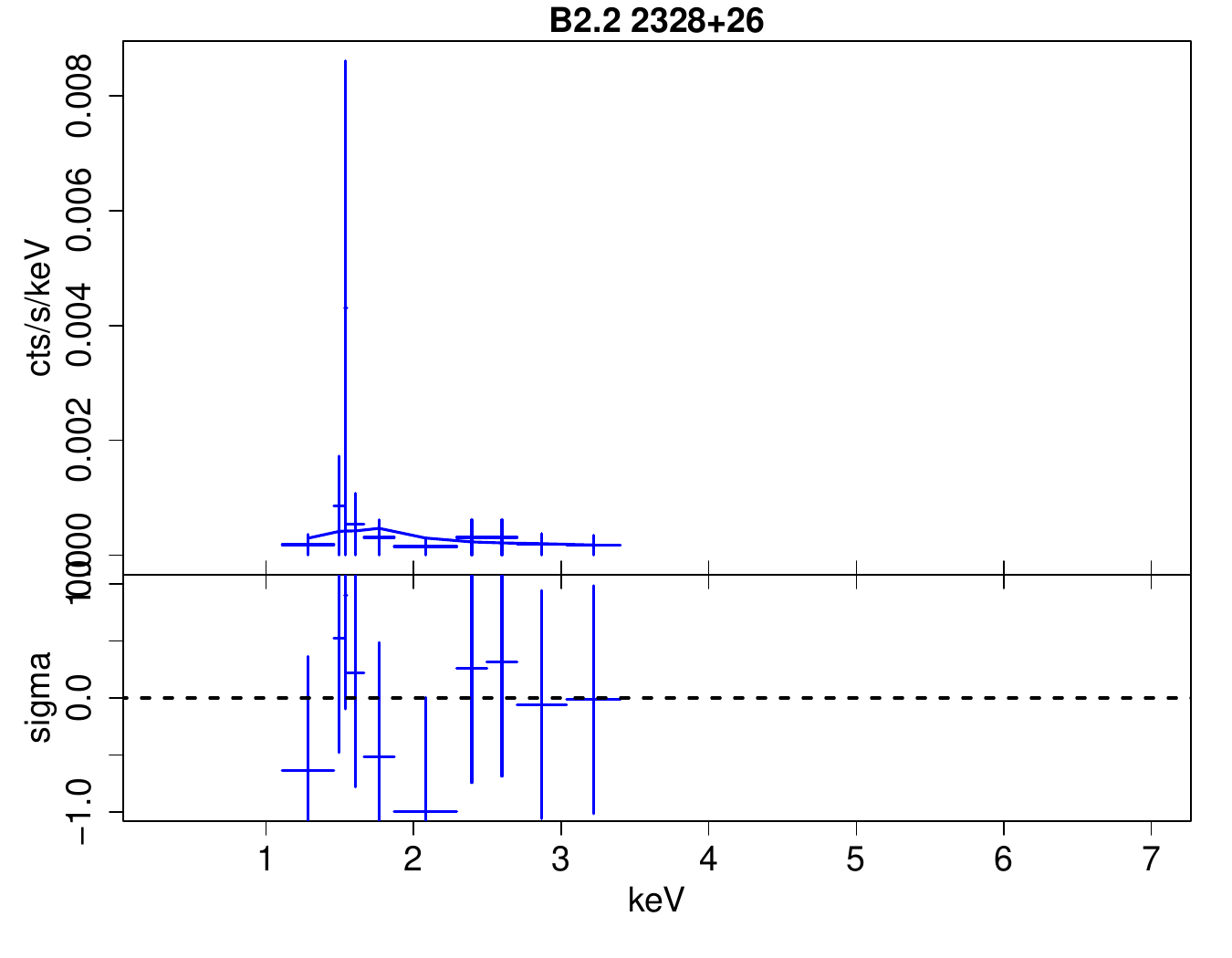}
	\caption{Nuclear spectral fits presented in Table \ref{tab:nucl_spectra_abs} with a power-law model with slope fixed to \(1.8\) plus Galactic absorption, and an additional absorption component. For each source, in the upper part of the panel we show the extracted spectrum with blue crosses and the best fit model with a blue line, while in lower part of the panel we show the residuals.}
\label{fig:nucl_spectra_abs}
\end{figure}

\begin{figure*}
	\figurenum{7}
	\centering
\includegraphics[scale=0.29]{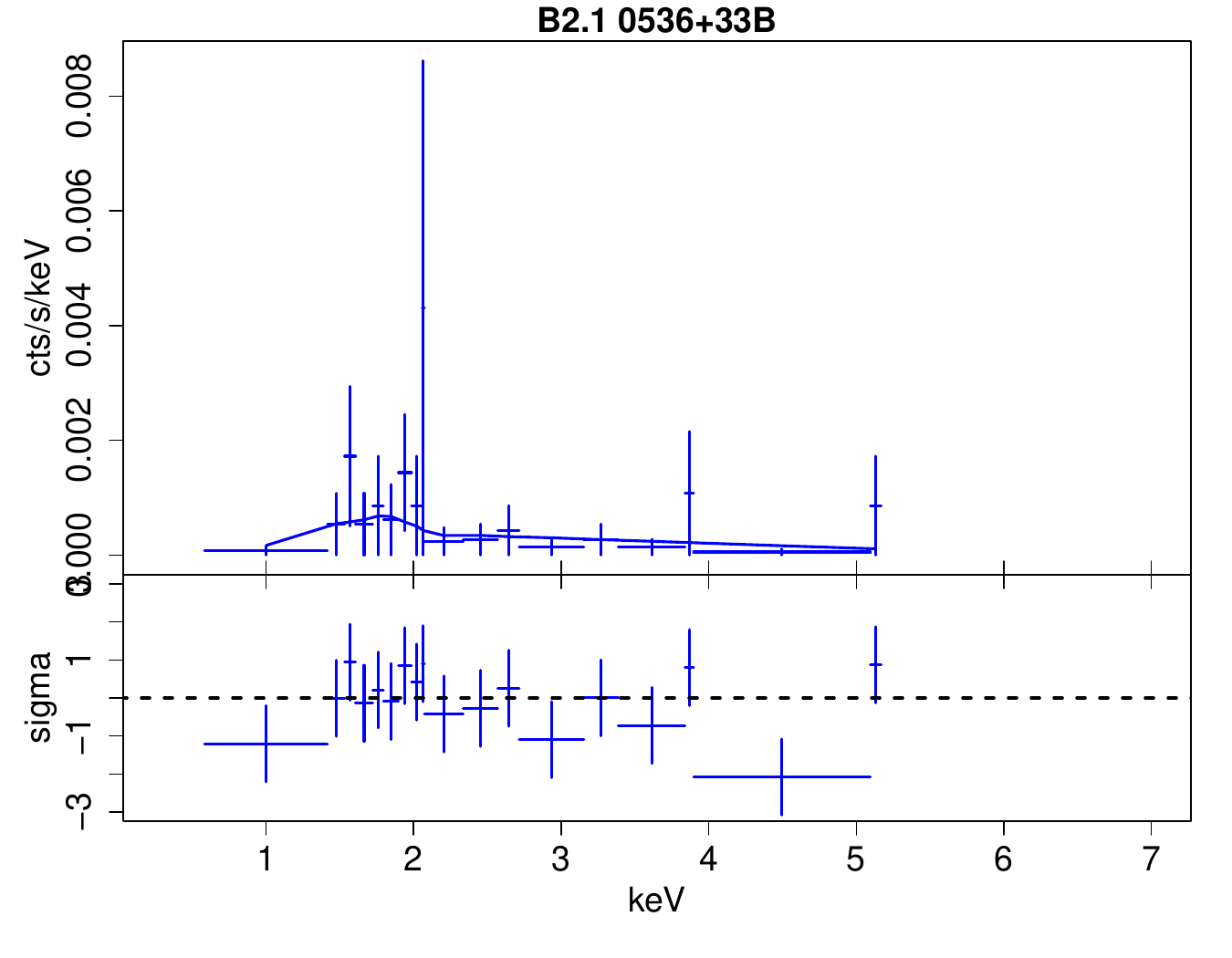}
\includegraphics[scale=0.29]{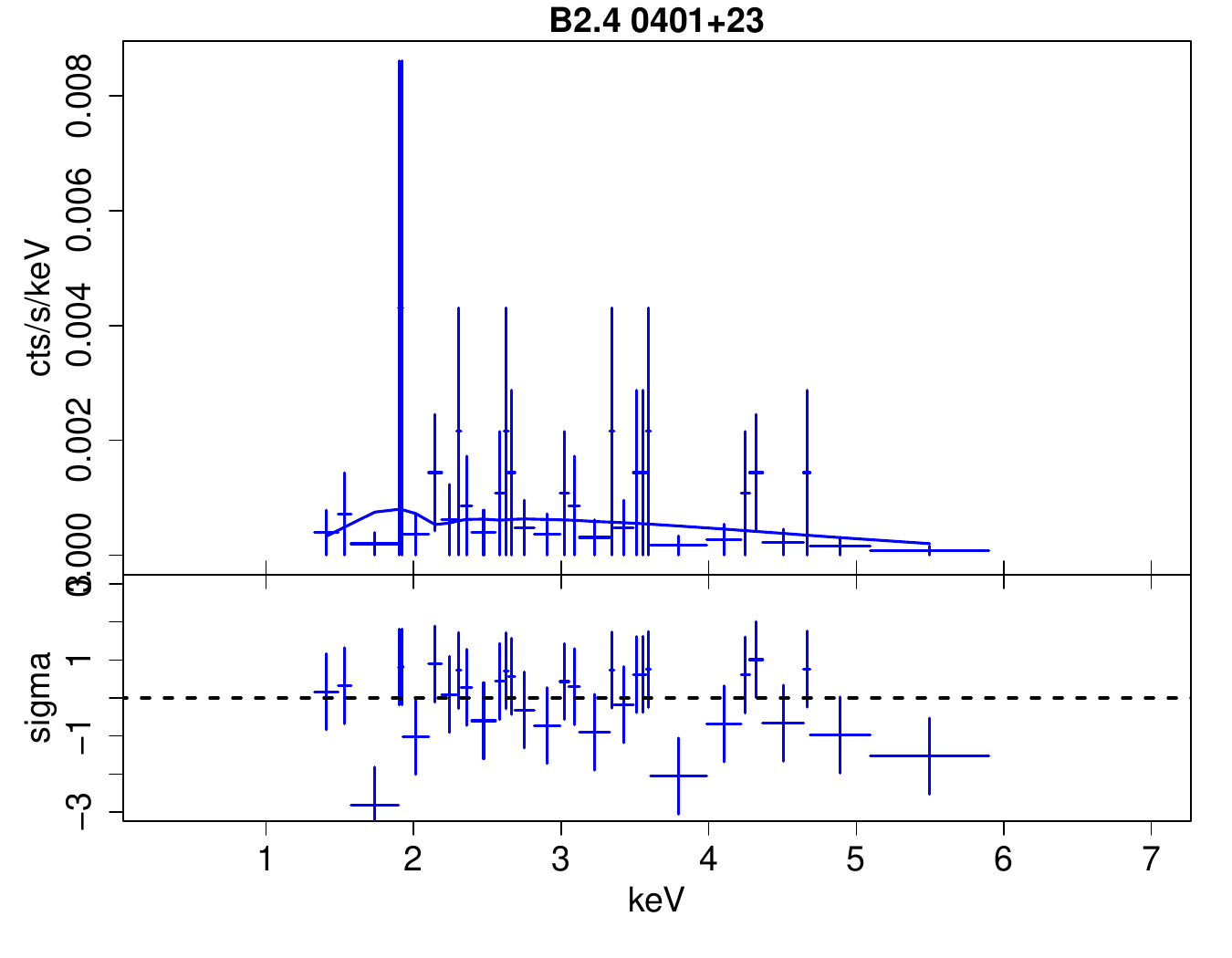}
\includegraphics[scale=0.29]{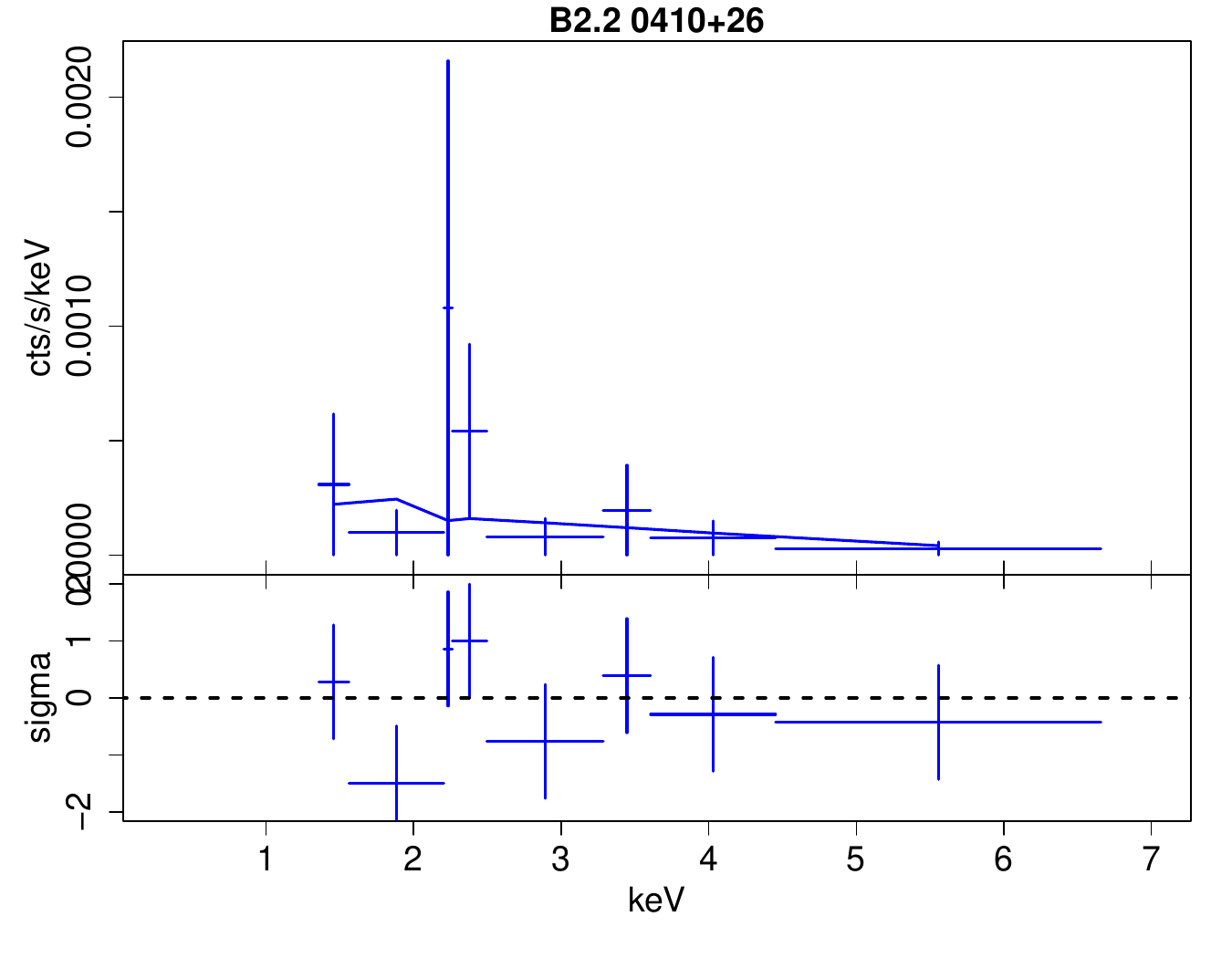}
\includegraphics[scale=0.29]{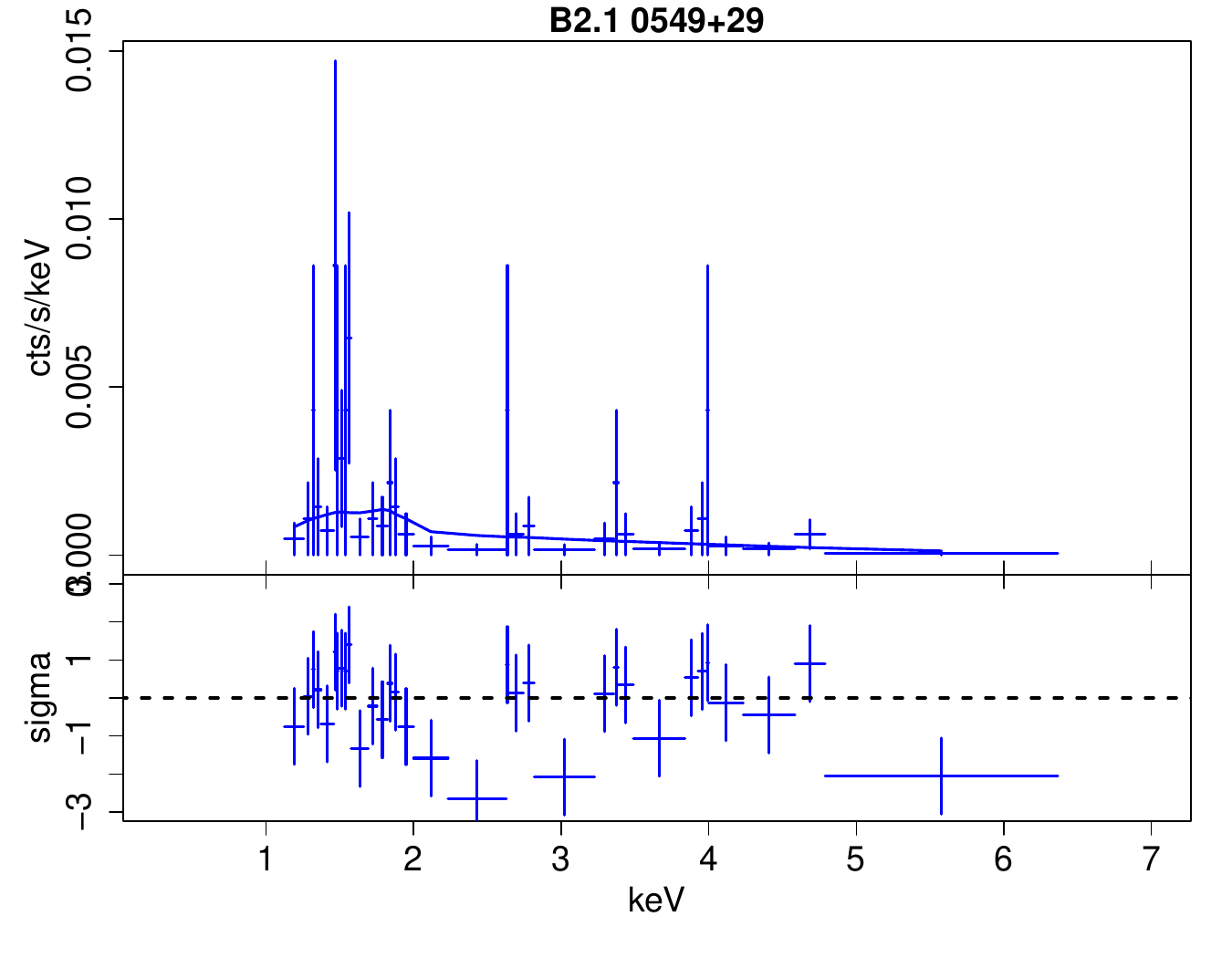}
\includegraphics[scale=0.29]{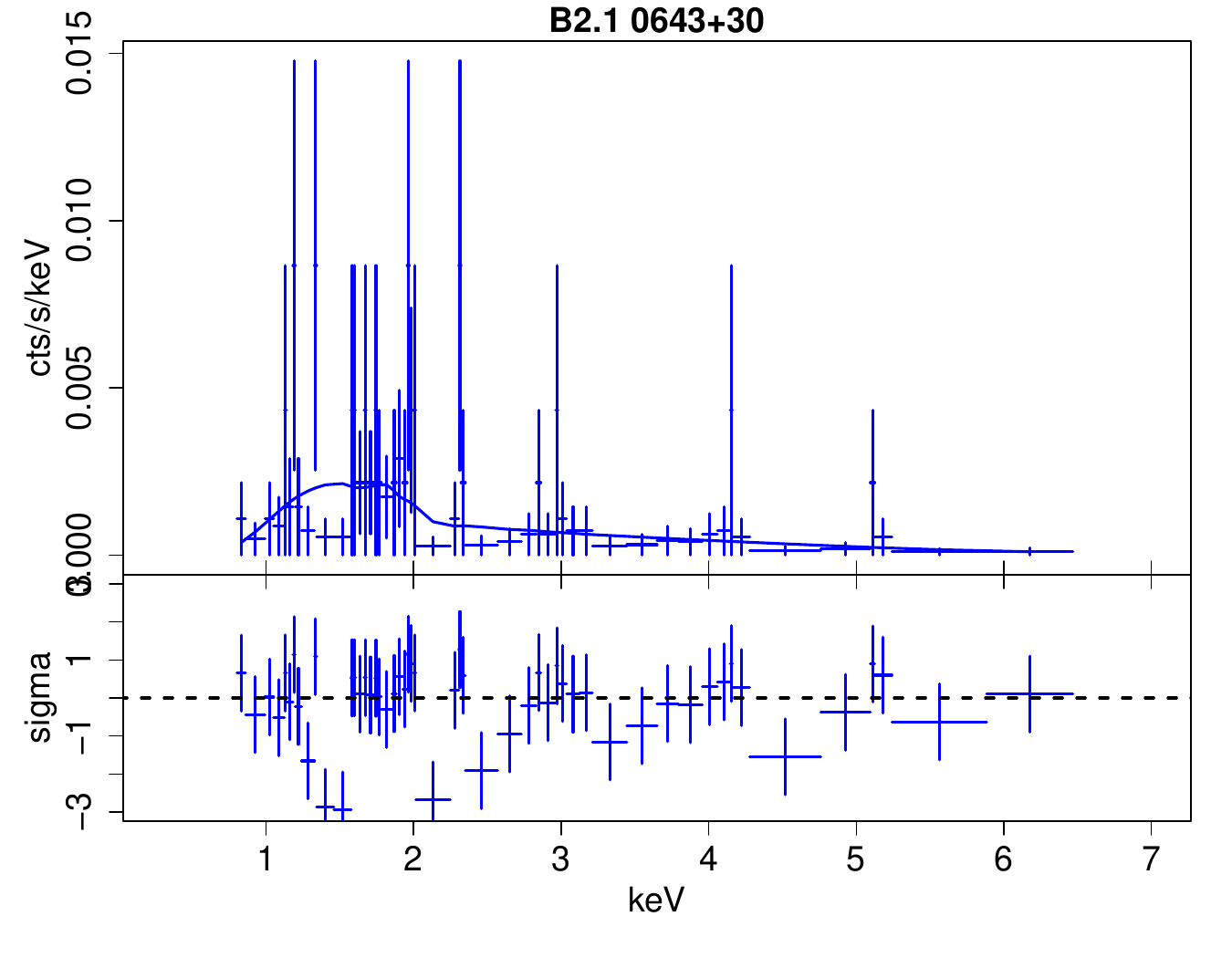}
\includegraphics[scale=0.29]{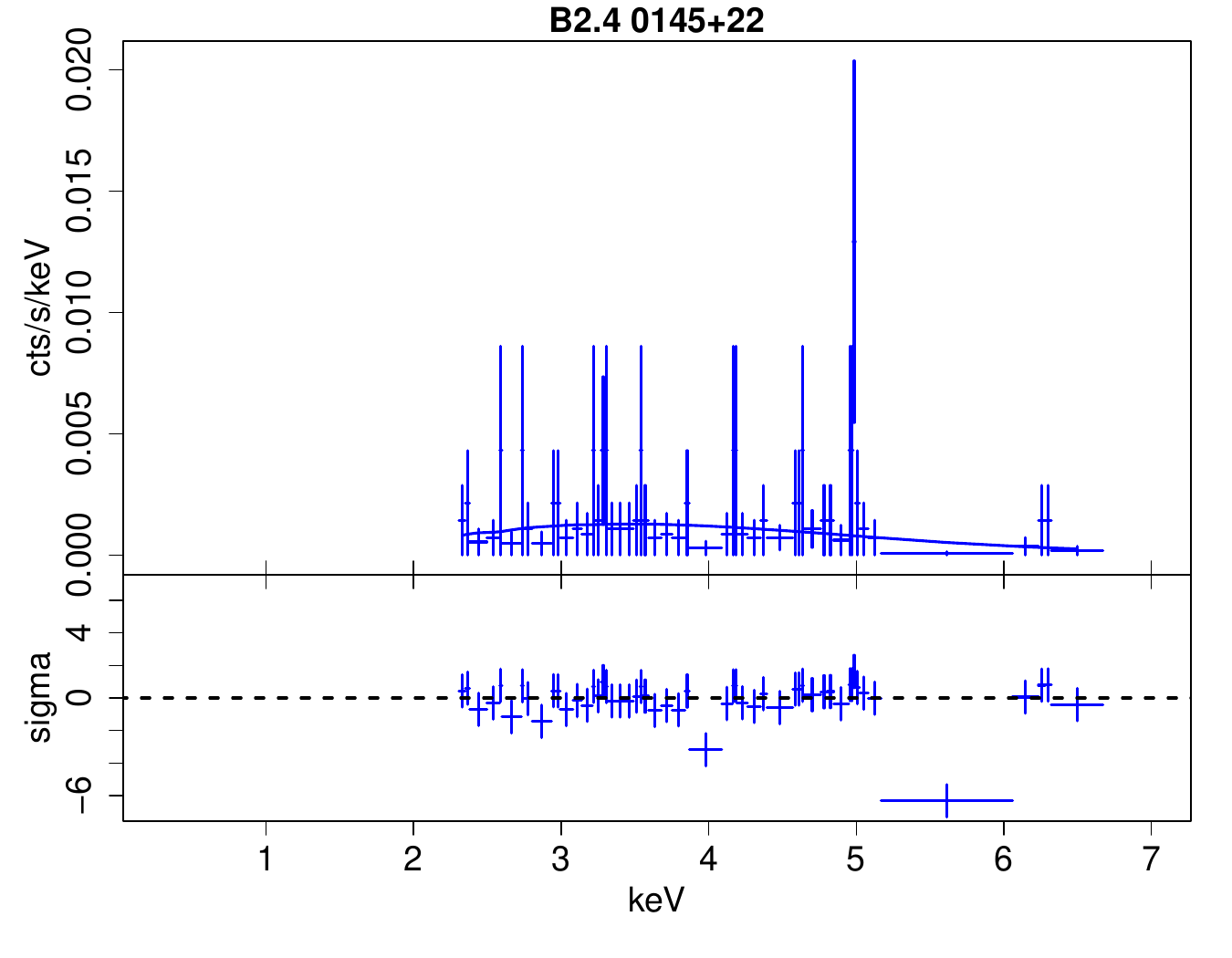}
\includegraphics[scale=0.29]{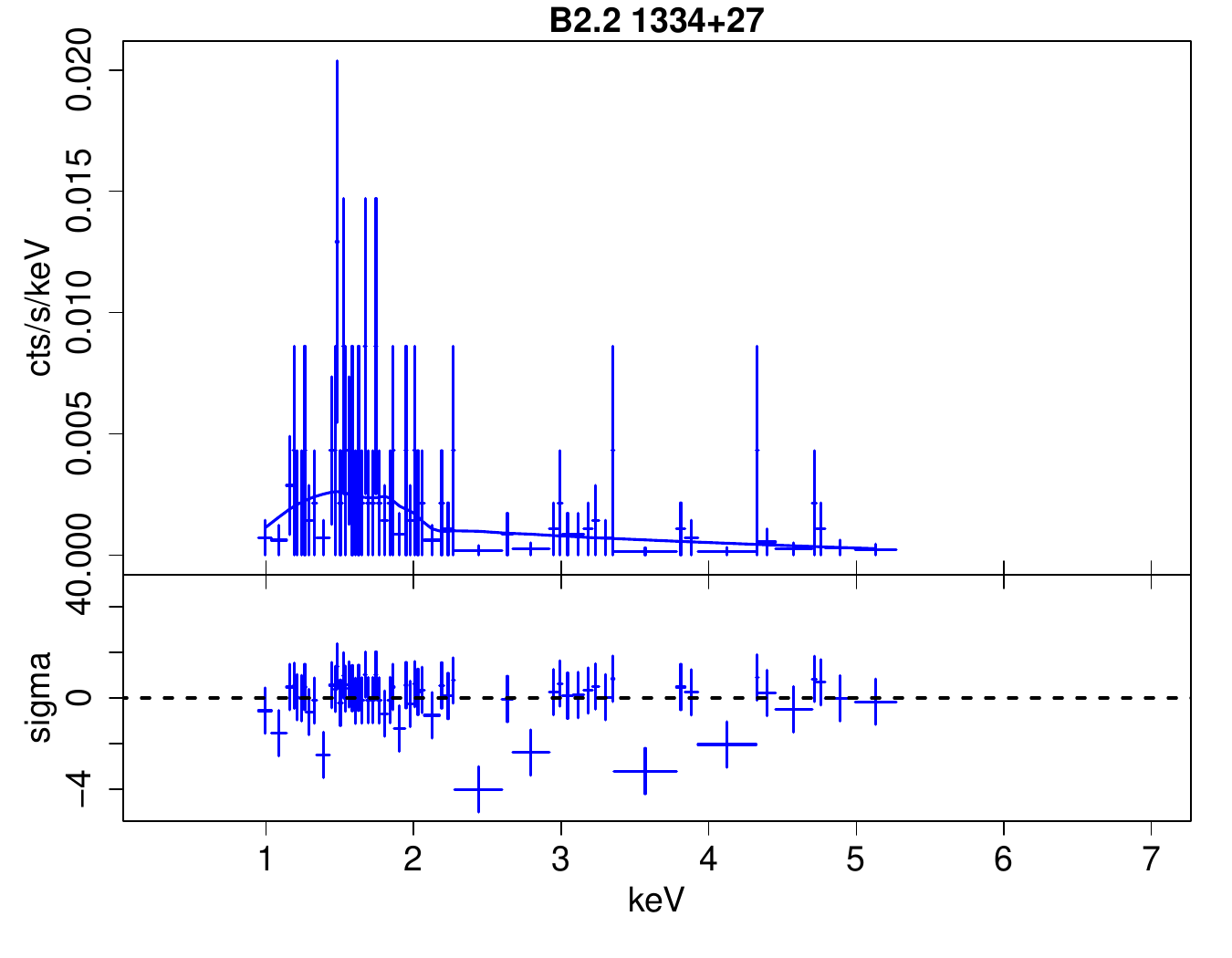}
	\caption{Continued.}
\end{figure*}

\begin{figure}
	\centering
	\includegraphics[scale=0.29]{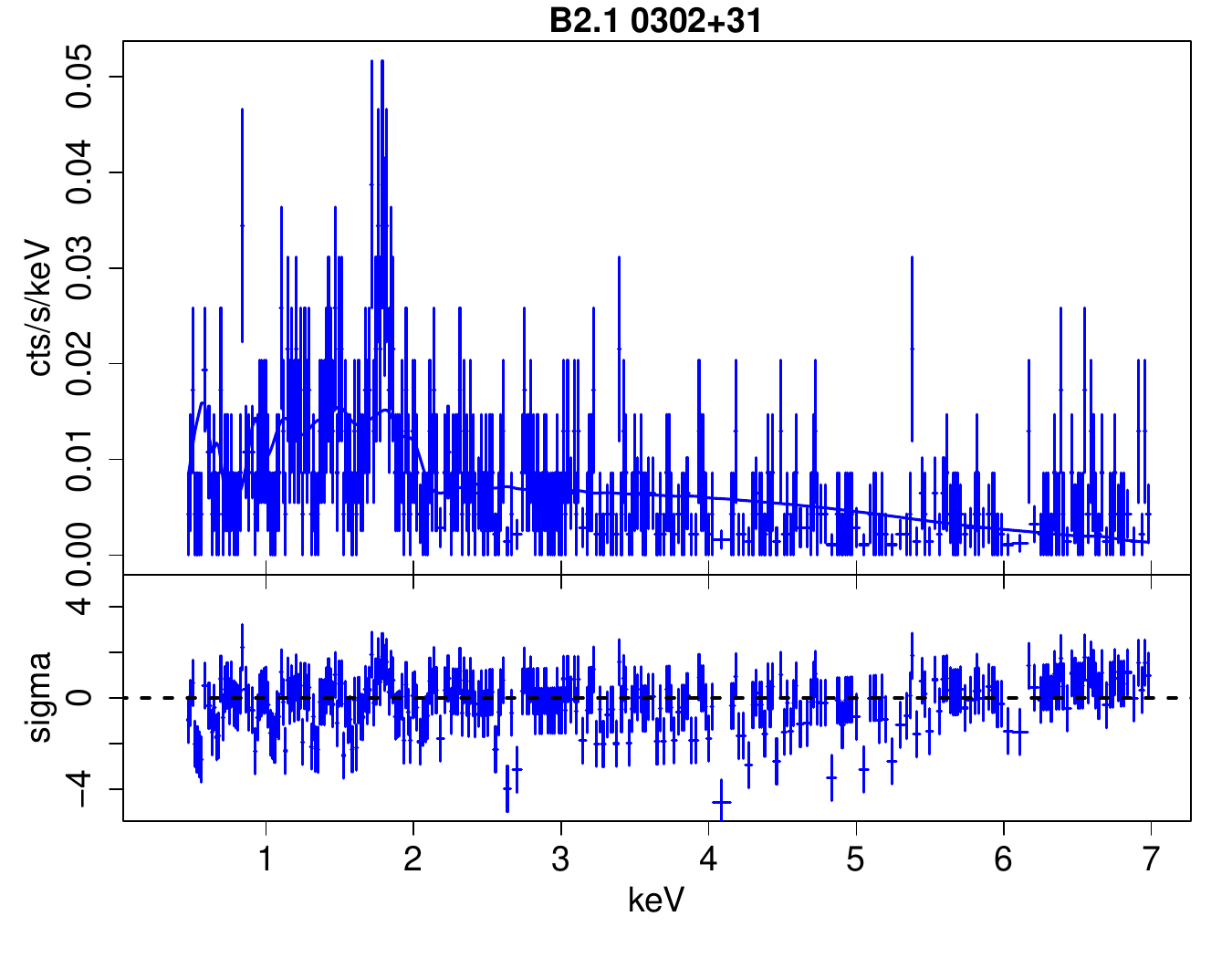}
	\includegraphics[scale=0.29]{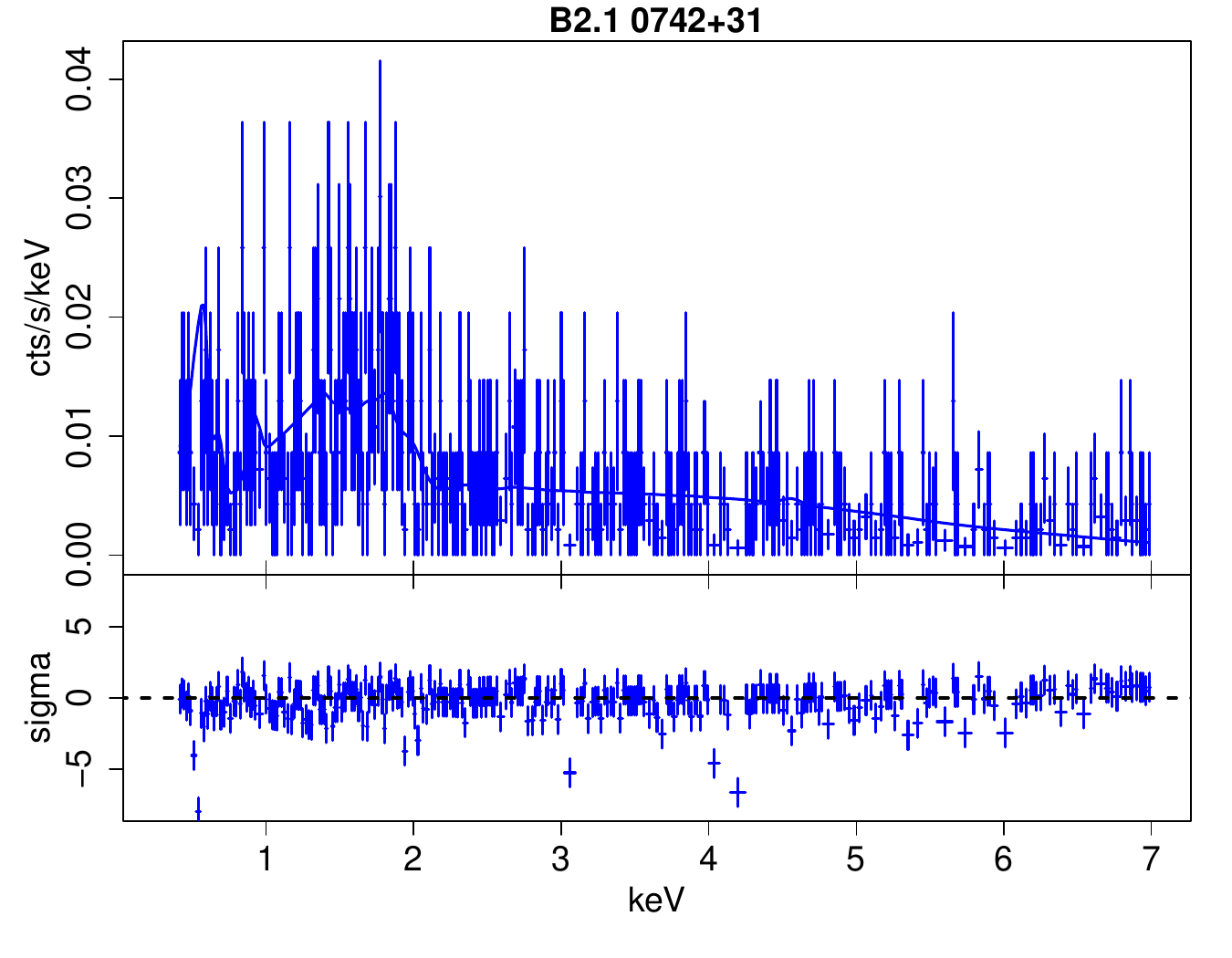}
	\includegraphics[scale=0.29]{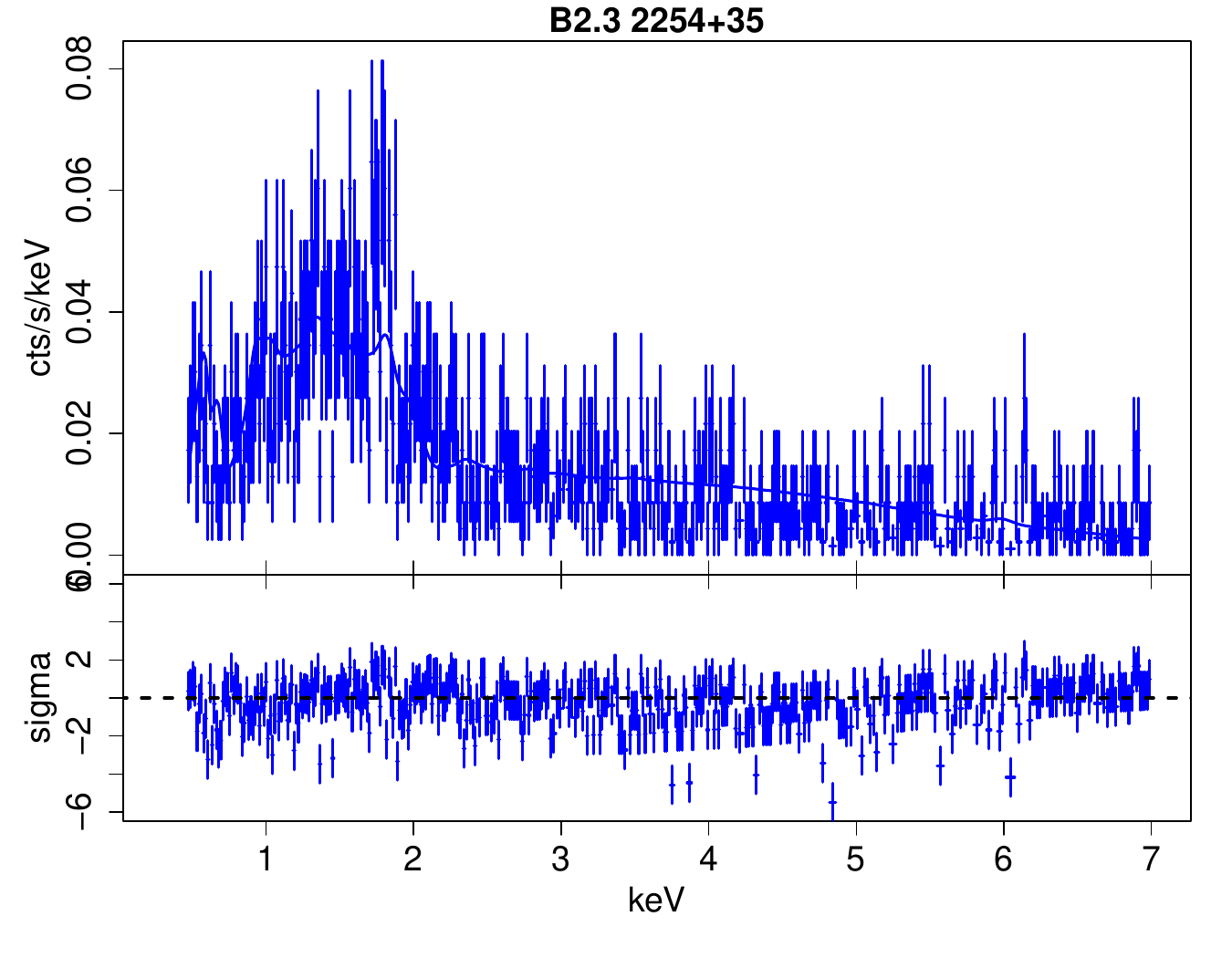}
	\caption{Spectral fits of the extended X-ray emission presented in Table \ref{tab:thermal_spectra_radio} with a model including a thermal plasma plus Galactic absorption. For each source, in the upper part of the panel we show the extracted spectrum with blue crosses and the best fit model with a blue line, while in lower part of the panel we show the residuals.}
	\label{fig:thermal_spectra_radio}
\end{figure}

\begin{figure}
	\centering
	\includegraphics[scale=0.29]{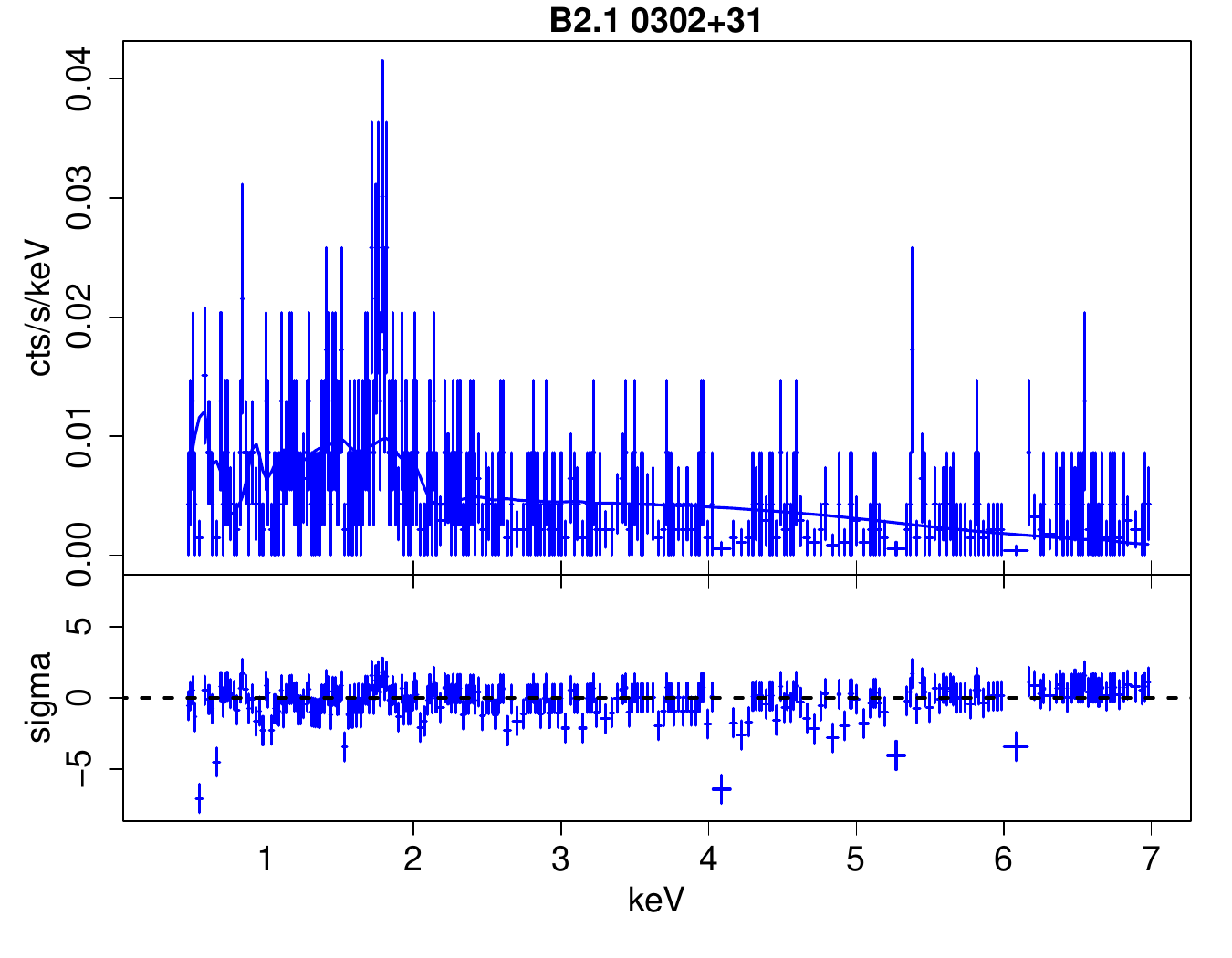}
	\includegraphics[scale=0.29]{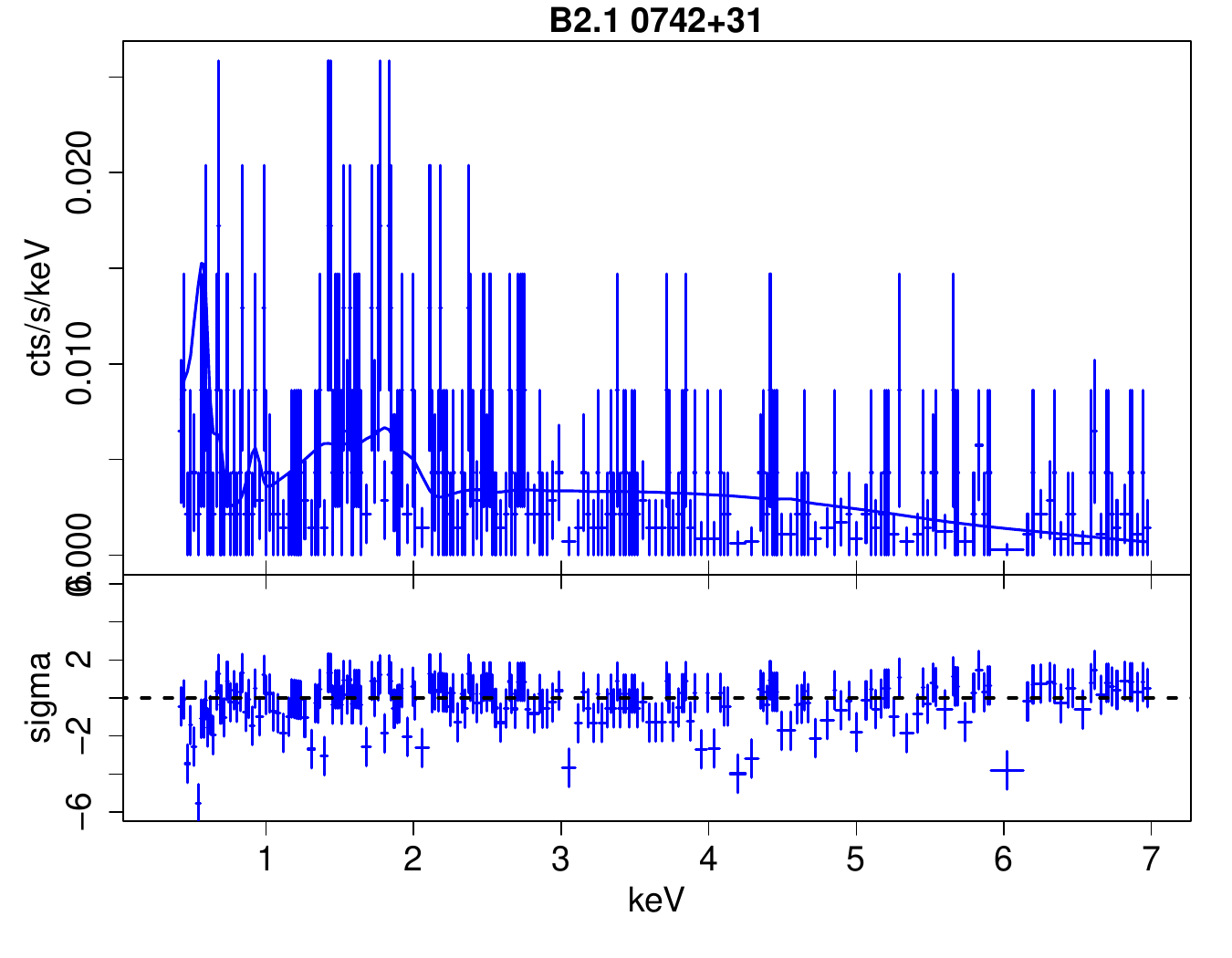}	
	\includegraphics[scale=0.29]{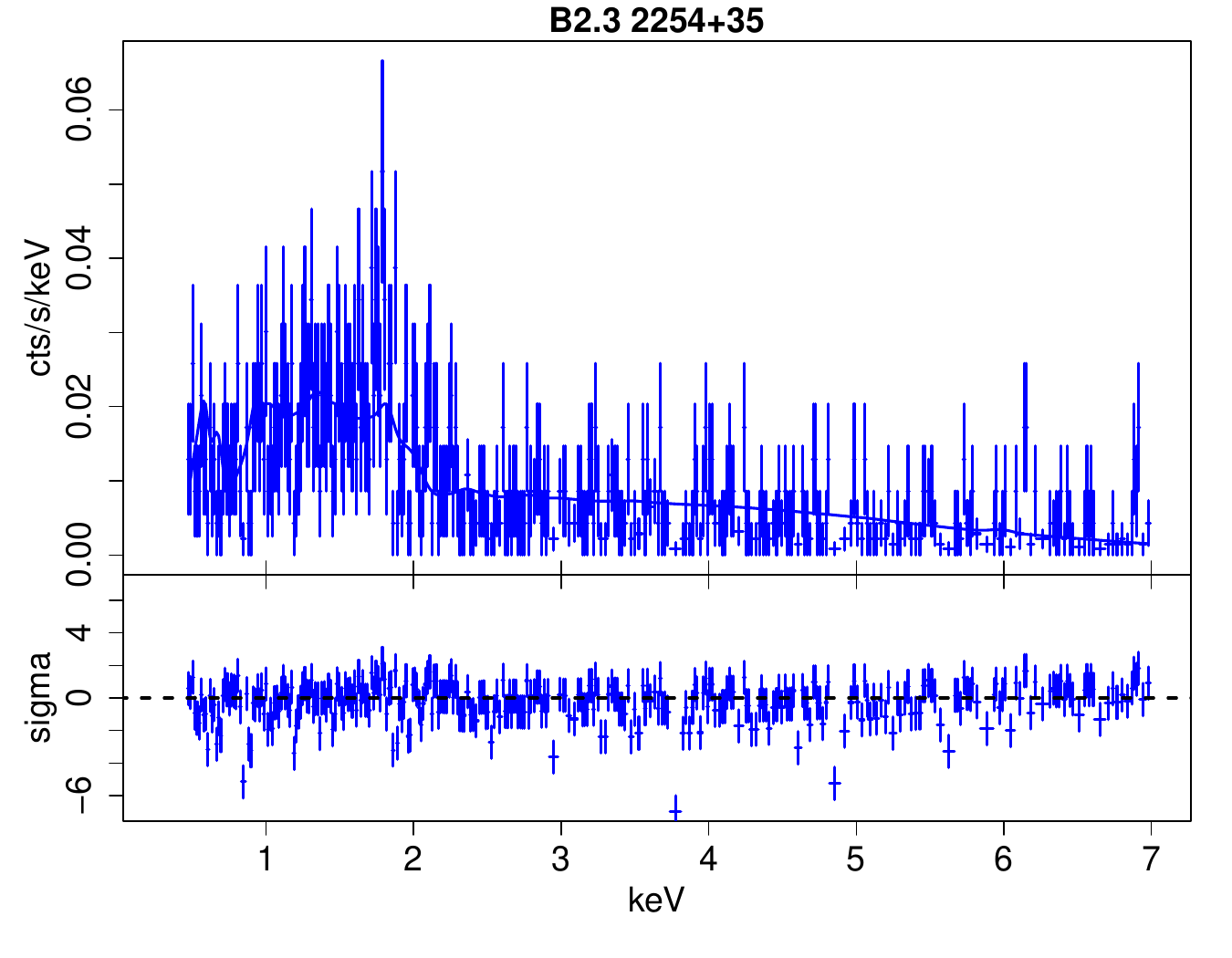}
	\caption{Spectral fits of the extended X-ray emission (excluding the regions connected with extended radio structures) presented in Table \ref{tab:thermal_spectra_noradio} with a model including a thermal plasma plus Galactic absorption. For each source, in the upper part of the panel we show the extracted spectrum with blue crosses and the best fit model with a blue line, while in lower part of the panel we show the residuals.}
	\label{fig:thermal_spectra_noradio}
\end{figure}

\begin{figure}
	\centering
	\includegraphics[scale=0.7]{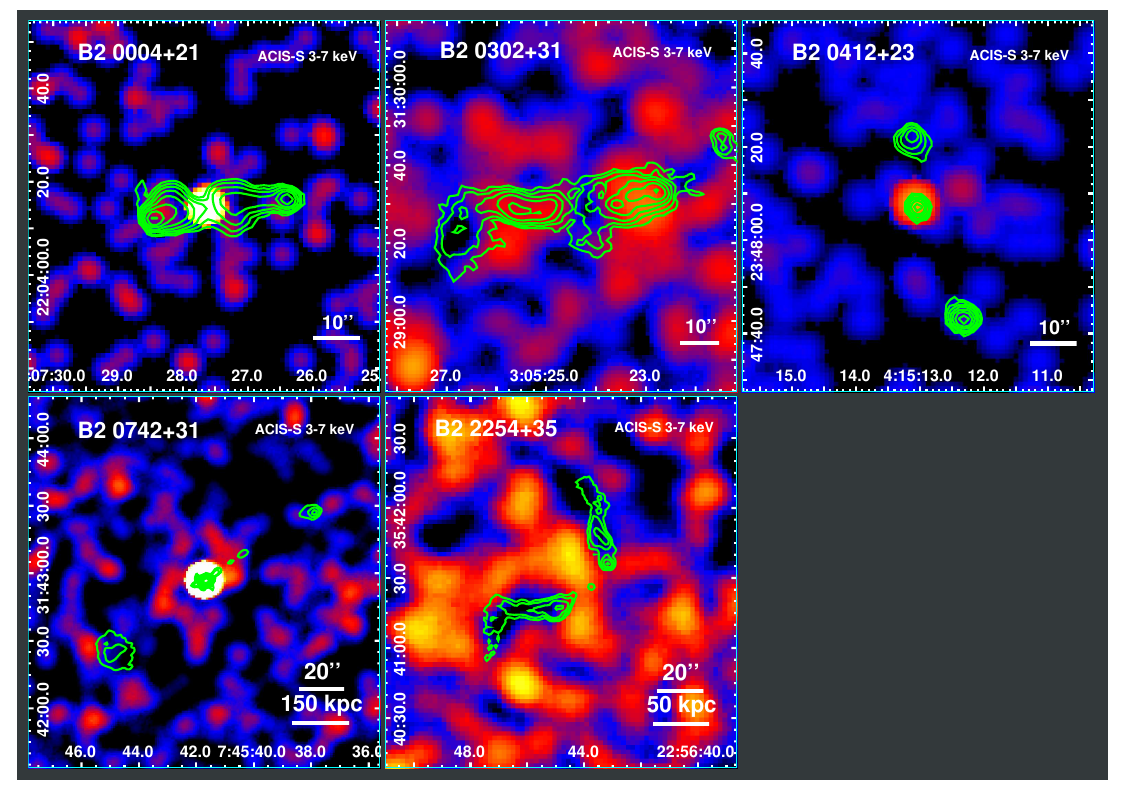}
	\caption{Hard band \(3-7 \text{ keV}\) \textit{Chandra} ACIS-S flux maps for the sources in the present sample that show evidence of extended X-ray emission (see Sect. \ref{sec:imaging}). VLASS \(3 \text{ GHz}\) contours overlaid in green are the same from Fig. \ref{fig:maps}.}
	\label{fig:hard}
\end{figure}

\begin{figure}
	\centering
	\includegraphics[scale=0.5]{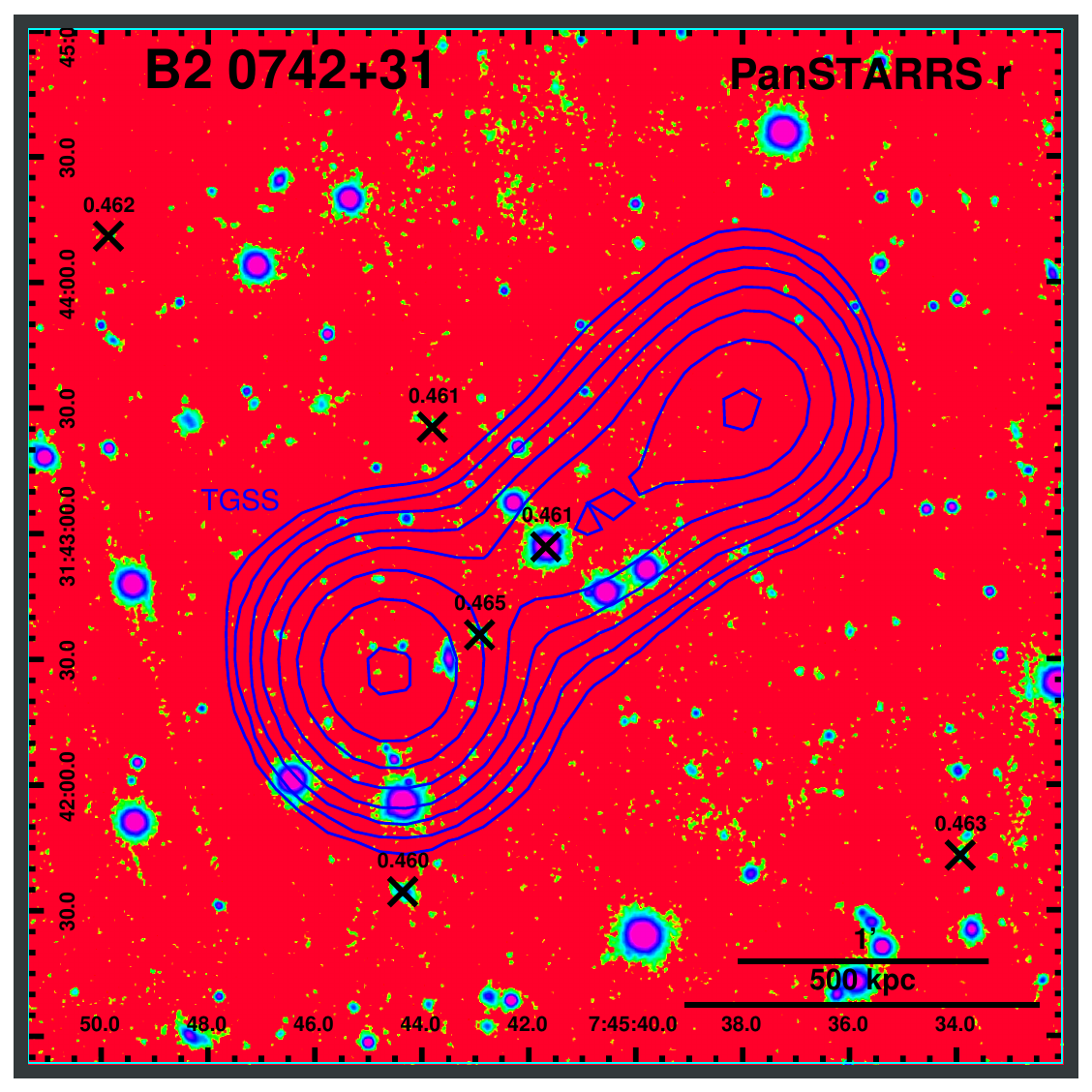}
	\caption{Pan-STARRS \citep{2016arXiv161205560C} \(r_{P1}\) filter image of B2.1 0742+31, with overlaid in blue the \(150 \text{ MHz}\) TGSS contours from Fig. \ref{fig:maps}. Black crosses indicate, in addition to the galaxy hosting the radio source, field sources with redshift measurement close to that of B2.1 0742+31 (0.461). The redshift value is indicated next to each cross.}
	\label{fig:0742_z}
\end{figure}

\begin{figure}
	\centering
	\includegraphics[scale=0.5]{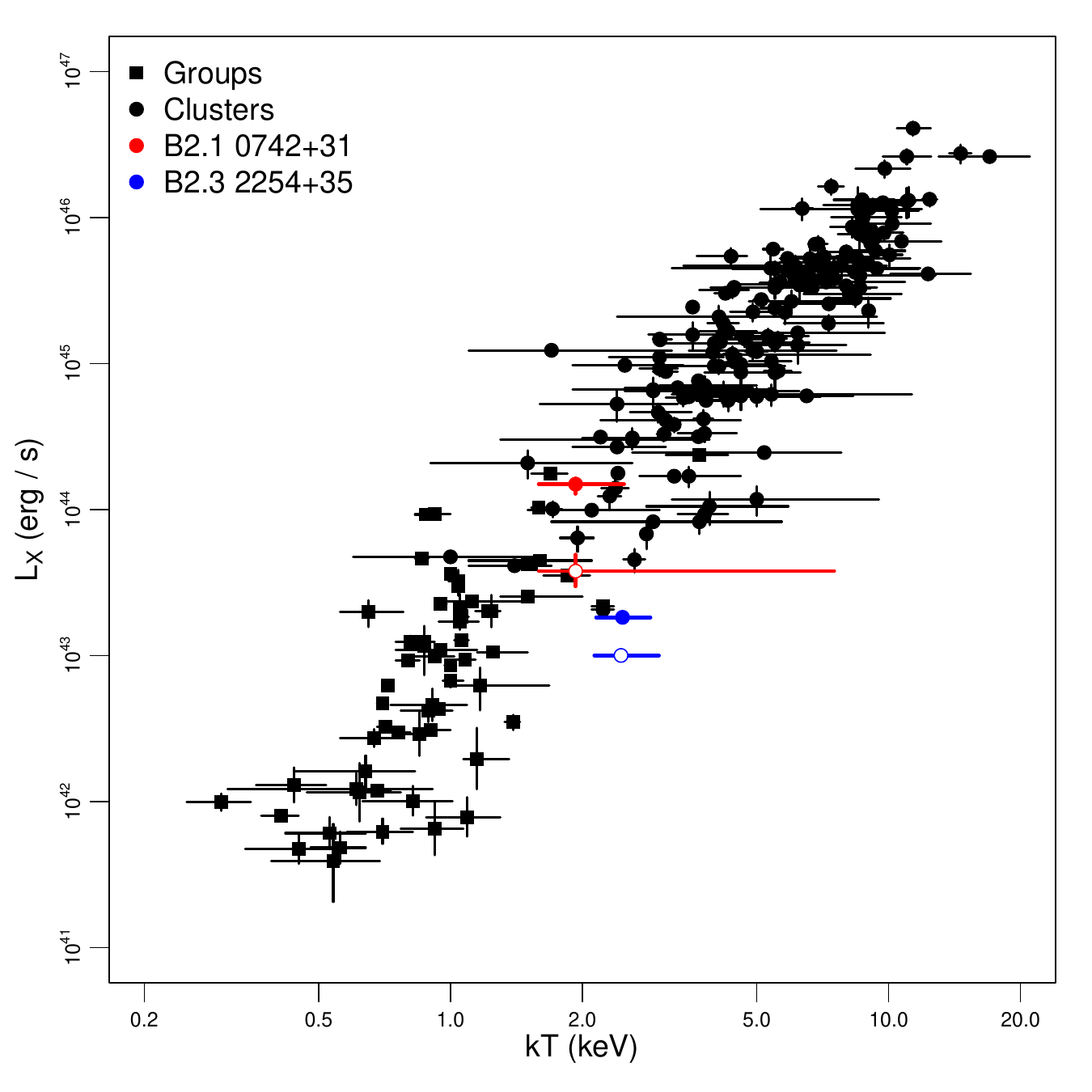}
	\caption{X-ray luminosity (\(L_X\)) vs. temperature (\(kT\)) of the X-ray emitting plasma in groups (black squares) and clusters (black circles) of galaxies \citep{2000ARA&A..38..289M}. Groups data are from \citet{2000ApJ...538...65X} and \citet{2000MNRAS.315..356H}, while data of clusters are from \citet{1999ApJ...524...22W}. The X-ray luminosities have been rescaled to the cosmology adopted in the present analysis. In this figure we represent with red and blue circles the properties of the X-ray emitting gas surrounding B2.1 0742+31 and B2.3 2254+35, respectively. Full colored circles indicate the results from the spectra extracted from the whole diffuse X-ray emission, while empty colored circles indicate the results from the spectra extracted from the regions that exclude the extended radio structures.}
	\label{fig:lx_vs_T}
\end{figure}

\appendix
\restartappendixnumbering

\section{Radio Maps}\label{app:a}

\begin{figure}
	\centering
	\includegraphics[scale=0.9]{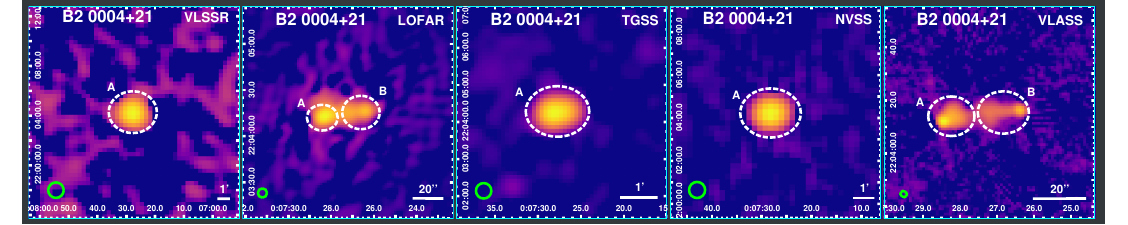}
	\caption{Radio maps for the sources considered in the present work. The white ellipses indicate the discernible radio structures, while the green ellipses on the lower-left of each panel represent the FWHM restoring beam. The complete figure set (33 images) is available in the online journal.}\label{fig:radiomaps}
\end{figure}

In this appendix we report the available radio images for the sources considered in this work, that is the \(74 \text{ MHz}\) VLSSR, \(145 \text{ MHz}\) LOFAR, \(150 \text{ MHz}\) GMRT TGSS, \(1.4 \text{ GHz}\) NVSS, and \(3 \text{ GHz}\) VLASS maps. The radio maps for each source are presented in Fig. \ref{fig:radiomaps}, where we overplot to the maps white dashed ellipses indicating the different radio structures, generally the two radio lobes (indicated with A and B) and radio core (indicated with N). When no radio structure appears discernible, only one ellipse (indicated with A) marks the bulk emission. In Table \ref{tab:radiotable} we report the specific flux estimates (in mJy) for the various structures observed in the radio maps. For each specific flux estimate we report an error that includes both the statistical and the systematic uncertainty, the latter ranging between \(10\%\) and \(15\%\) of the specific flux \citep[e.g.,][]{2014MNRAS.440..327L, 2017A&A...598A..78I, 2021A&A...648A...2S, 2020Symm...12..527K}.

\begin{table}
	\caption{Specific flux estimates for the radio structures indicated in Fig. \ref{fig:radiomaps}. In particular, \(F_{74 \text{ MHz}}\) represents the VLSSr flux, \(F_{145 \text{ MHz}}\) the LOFAR flux, \(F_{150 \text{ MHz}}\) the TGSS flux, \(F_{1.4 \text{ GHz}}\) the NVSS flux, and \(F_{3 \text{ GHz}}\) the VLASS flux. The quoted errors include both the statistical and the systematic uncertainty.}\label{tab:radiotable}
	\begin{center}
		\resizebox{\textwidth}{!}{
			\begin{tabular}{|l|c|c|c|c|c|c|c|c|c|c|c|c|c|c|c|}
				\hline
				\hline
				Source Name & \multicolumn{3}{c}{\(F_{74 \text{ MHz}}\) }& \multicolumn{3}{|c|}{\(F_{145 \text{ MHz}}\)} & \multicolumn{3}{|c|}{\(F_{150 \text{ MHz}}\)} & \multicolumn{3}{|c|}{\(F_{1.4 \text{ GHz}}\)} & \multicolumn{3}{|c|}{\(F_{3 \text{ GHz}}\)} \\
				\hline
				& \multicolumn{3}{|c|}{mJy} & \multicolumn{3}{|c|}{mJy} & \multicolumn{3}{|c|}{mJy} & \multicolumn{3}{|c|}{mJy} & \multicolumn{3}{|c|}{mJy} \\
				\hline
				& A & B & N & A & B & N & A & B & N & A & B & N & A & B & N \\
				\hline
B2.4 0004+21 & \(9472 \pm 1579\) & - & - & \(3196 \pm 320\) & \(1798 \pm 180\) & - & \(4824 \pm 731\) & - & - & \(815 \pm 83\) & - & - & \(270 \pm 41\) & \(125 \pm 20\) & - \\
B2.2 0038+25B & \(12086 \pm 1981\) & - & - & \(4933\pm 494\) & \(2981 \pm 298\) & - & \(5585 \pm 852\) & - & - & \(639 \pm 65\) & - & - & \(127 \pm 20\) & \(87 \pm 14\) & - \\
B2.2 0143+24 & \(4137 \pm 844 \) & - & - & \(896 \pm 91 \) & \(684 \pm 69 \) & - & \(992 \pm 153\) & \(679 \pm 105\) & - & \(187 \pm 20\) & - & - & \(49 \pm 8 \) & \(34 \pm 6 \) & - \\
B2.4 0145+22 & \(2796 \pm 580 \) & - & - & \(562 \pm 57 \) & \(1155 \pm 116\) & - & \(459 \pm 75 \) & \(1107 \pm 173\) & - & \(82 \pm 9 \) & \(195 \pm 20\) & - & \(61 \pm 10\) & \(116 \pm 19\) & - \\
B2.4 0229+23 & \(514 \pm 242 \) & - & - & \(469 \pm 48 \) & - & - & \(571 \pm 93 \) & - & - & \(459 \pm 47\) & - & - & \(277 \pm 42\) & - & - \\
B2.1 0241+30 & \(6123 \pm 1120\) & - & - & \(1229 \pm 124\) & \(2111 \pm 212\) & - & \(709 \pm 111\) & \(1345 \pm 207\) & - & \(173 \pm 18\) & - & - & \(22 \pm 4 \) & \(38 \pm 6 \) & \( 2 \pm 1 \) \\
B2.1 0302+31 & \(3435 \pm 698 \) & - & - & \(850 \pm 86 \) & \(856 \pm 87 \) & - & \(1679 \pm 260\) & - & - & \(373 \pm 38\) & - & - & \(60 \pm 10\) & \(83 \pm 13\) & - \\
B2.4 0401+23 & \(2529 \pm 587 \) & - & - & \(1179 \pm 118\) & \(371 \pm 37 \) & - & \(868 \pm 134\) & \(297 \pm 48 \) & - & \(198 \pm 21\) & - & - & \(66 \pm 10\) & \(20 \pm 3 \) & \(2 \pm 1 \) \\
B2.2 0410+26 & \(8379 \pm 1477\) & - & - & \(5335 \pm 534\) & - & - & \(5286 \pm 800\) & - & - & \(529 \pm 54\) & - & - & \(188 \pm 29\) & - & - \\
B2.4 0412+23 & \(2566 \pm 577 \) & - & - & \(797 \pm 80 \) & \(1622 \pm 163\) & - & \(502 \pm 81 \) & \(787 \pm 123\) & - & \(276 \pm 29\) & - & - & \(23 \pm 4 \) & \(96 \pm 15\) & \(36 \pm 6 \) \\
B2.3 0454+35 & \(1108 \pm 509 \) & - & - & - & - & - & \(418 \pm 70 \) & \(251 \pm 48 \) & - & \(97 \pm 11\) & \(66 \pm 7 \) & - & \(25 \pm 4 \) & \(24 \pm 5 \) & \(9 \pm 2 \) \\
B2.1 0455+32B & \(2185 \pm 542 \) & - & - & \(1355 \pm 139\) & - & - & \(1639 \pm 256\) & - & - & \(346 \pm 36\) & - & - & \(73 \pm 11\) & \(98 \pm 15\) & - \\
B2.1 0455+32C & \(5326 \pm 1019\) & - & - & \(857 \pm 88 \) & \(1460 \pm 149\) & - & \(3140 \pm 482\) & - & - & \(561 \pm 57\) & - & - & \(112 \pm 17\) & \(171 \pm 26\) & - \\
B2.3 0516+40 & \(1246 \pm 382 \) & - & - & \(1325 \pm 133\) & - & - & \(1700 \pm 262\) & - & - & \(265 \pm 28\) & - & - & \(123 \pm 19\) & - & - \\
B2.1 0536+33B & \(1141 \pm 390 \) & - & - & \(282 \pm 29 \) & \(535 \pm 54 \) & - & \(1129 \pm 178\) & - & - & \(192 \pm 20\) & - & - & \(34 \pm 6 \) & \(51 \pm 8 \) & \(5 \pm 1 \) \\
B2.1 0549+29 & \(1590 \pm 446 \) & - & - & \(1039 \pm 105\) & - & - & \(1453 \pm 229\) & - & - & \(218 \pm 23\) & - & - & \(50 \pm 8 \) & \(57 \pm 9 \) & - \\
B2.1 0643+30 & \(570 \pm 216 \) & - & - & - & - & - & \(870 \pm 144\) & - & - & \(869 \pm 88\) & - & - & \(581 \pm 88\) & - & - \\
B2.1 0742+31 & \(7098 \pm 1168\) & \(5018 \pm 859\) & - & \(3718 \pm 373\) & \(1818 \pm 182\) & \(325 \pm 33\) & \(3728 \pm 569\) & \(1702 \pm 263\) & \(450 \pm 72\) & \(642 \pm 65\) & \(475 \pm 48\) & - & \(221 \pm 35\) & \(142 \pm 24\) & \(696 \pm 105\) \\
B2.2 0755+24 & \(1239 \pm 364 \) & - & - & \(824 \pm 83 \) & - & - & \(1003 \pm 159\) & - & - & \(206 \pm 22\) & - & - & \(53 \pm 8 \) & \(44 \pm 7 \) & - \\
B2.3 0848+34 & \(2861 \pm 567 \) & - & - & \(2105 \pm 211 \) & - & - & - & - & - & \(341 \pm 35\) & - & - & \(77 \pm 12 \) & \(81 \pm 12 \) & - \\
B2.4 0939+22A & \(1754 \pm 404 \) & - & - & \(207 \pm 21 \) & \(810 \pm 81 \) & - & \(1126 \pm 175\) & - & - & \(207 \pm 22\) & - & - & \(22 \pm 4 \) & \(82 \pm 13\) & - \\
B2.4 1112+23 & \(2447 \pm 547 \) & - & - & \(1652 \pm 166\) & - & - & \(1448 \pm 224\) & - & - & \(375 \pm 39\) & - & - & \(199 \pm 30\) & - & - \\
B2.3 1234+37 & \(9736 \pm 1662\) & - & - & \(3184 \pm 319\) & \(2480 \pm 248\) & - & \(6700 \pm 102\) & - & - & \(923 \pm 94\) & - & - & \(262 \pm 40\) & \(161 \pm 25\) & - \\
B2.2 1334+27 & \(2732 \pm 584 \) & - & - & \(2086 \pm 209\) & \(257 \pm 26 \) & - & \(1390 \pm 217\) & - & - & \(188 \pm 20\) & - & - & \(24 \pm 4 \) & \(59 \pm 9 \) & - \\
B2.2 1338+27 & \(2695 \pm 588 \) & - & - & \(683 \pm 69 \) & \(2206 \pm 222\) & - & \(355 \pm 59 \) & \(1139 \pm 176\) & - & \(51 \pm 6 \) & \(120 \pm 13\) & - & \(19 \pm 3 \) & \(49 \pm 8 \) & - \\
B2.2 1439+25 & \(655 \pm 230 \) & \(1074 \pm 309\) & - & \(403 \pm 41 \) & \(755 \pm 76 \) & - & \(319 \pm 59 \) & \(707 \pm 118\) & - & \(73 \pm 8 \) & \(132 \pm 14\) & - & \(40 \pm 7 \) & \(70 \pm 12\) & - \\
B2.4 1512+23 & \(1280 \pm 424 \) & - & - & \(319 \pm 33 \) & \(365 \pm 37 \) & - & \(261 \pm 45 \) & \(278 \pm 48 \) & - & \(241 \pm 25\) & - & - & \(58 \pm 10\) & \(60 \pm 10\) & \(3 \pm 1 \) \\
B2.4 2054+22B & \(1928 \pm 485\) & - & - & - & - & - & \(1238 \pm 196\) & - & - & \(194 \pm 21\) & - & - & \(47 \pm 8\) & \(46 \pm 7\) & - \\
B2.2 2104+24 & \(6928 \pm 1232\) & - & - & \(2112 \pm 217\) & \(794 \pm 84 \) & - & \(3466 \pm 532\) & - & - & \(501 \pm 51\) & - & - & \(144 \pm 22\) & \(50 \pm 8 \) & \(17 \pm 3 \) \\
B2.2 2133+27 & \(3760 \pm 744 \) & - & - & \(1746 \pm 179\) & - & - & \(1447 \pm 224\) & - & - & \(281 \pm 29\) & - & - & \(72 \pm 12\) & \(47 \pm 8 \) & - \\
B2.3 2254+35 & \(5701 \pm 1138\) & - & - & \(1304 \pm 131\) & \(1911 \pm 193\) & - & \(852 \pm 140\) & \(1182 \pm 202\) & - & \(492 \pm 51\) & - & - & \(78 \pm 13\) & \(117 \pm 19\) & \(2 \pm 1 \) \\
B2.2 2328+26 & \(2860 \pm 650 \) & - & - & \(1790 \pm 179\) & - & - & \(1508 \pm 235\) & - & - & \(413 \pm 42\) & - & - & \(210 \pm 32\) & - & - \\
B2.3 2334+39 & \(2747 \pm 592 \) & - & - & \(917 \pm 94 \) & \(697 \pm 72 \) & \(119 \pm 12\) & \(815 \pm 130\) & \(719 \pm 117\) & - & \(469 \pm 48\) & - & - & \(102 \pm 18\) & \(76 \pm 14\) & \(9 \pm 2 \) \\
				\hline
				\hline
			\end{tabular}
		}
	\end{center}
\end{table}

\end{document}